\documentclass[12pt,letterpaper]{article}

\pdfoutput=1

\usepackage[left=2cm,top=1cm,right=3cm,nohead]{geometry}
\usepackage[dvipsnames]{xcolor}
\usepackage{colortbl}
\usepackage[utf8]{inputenc} 
\usepackage{color}
\usepackage{amsfonts}
\usepackage{amssymb}
\usepackage{amsthm}
\usepackage{amsmath}

\DeclareFontFamily{U}{mathx}{}
\DeclareFontShape{U}{mathx}{m}{n}{<-> mathx10}{}
\DeclareSymbolFont{mathx}{U}{mathx}{m}{n}
\DeclareMathAccent{\widehat}{0}{mathx}{"70}
\DeclareMathAccent{\widecheck}{0}{mathx}{"71}
\usepackage{epsfig,graphics,latexsym,mathtools,amsbsy,multirow,mathrsfs,wasysym,textcomp,subfigure,wrapfig,comment,bbold,array,longtable,setspace,pifont, enumitem}
\usepackage{titlesec}
\usepackage{tabularx}
\usepackage{graphicx}
\usepackage{graphbox}
\usepackage{setspace}
\usepackage{rotate}
\usepackage{accents}
\usepackage[bf]{caption}
\usepackage{braket}
\usepackage{fancyhdr}
\usepackage{subcaption}
\usepackage{tikz}
\usepackage{tikz-3dplot}
\usetikzlibrary{arrows}
\usetikzlibrary{3d}
\usepackage{comment}
\usepackage{float}
\usepackage{soul}
\usetikzlibrary{mindmap,trees,shadows}
\colorlet{color1}{gray!25}
\usetikzlibrary{positioning}
\usetikzlibrary{intersections} 
\usetikzlibrary{decorations.pathmorphing}
\usetikzlibrary{matrix,decorations.pathreplacing}
\usetikzlibrary{calc,angles,positioning,intersections,quotes,decorations.markings}
\usepackage{tkz-euclide}
\newlength{\PicScale}
\usepackage[UKenglish]{babel}
\usepackage[toc,page]{appendix}
\usepackage{hhline}
\usepackage{geometry}
\usepackage{cite}
\usepackage{datetime}
\usepackage{bbm} 
\usepackage{xcolor}
\usepackage{diagbox}
\usepackage{slashbox}
\usepackage{dynkin-diagrams}
\numberwithin{equation}{section}
\usepackage[colorlinks=true,
linkcolor=Blue, 
citecolor=Blue,
filecolor=Blue,
urlcolor=Blue,
linktoc=page,
pdfstartview=FitV,
bookmarksopen=true]{hyperref}

\hypersetup{pdftitle={},pdfauthor={},pdfsubject={}} 
\usepackage{subfiles}
\usepackage{simpler-wick}

\newlength{\dhatheight}

\theoremstyle{plain}

\newtheorem{thm}{Theorem}
\newtheorem*{thm*}{Theorem}
\newtheorem*{def*}{Definition}
\newtheorem{proposition}{Proposition}

\newcommand{\cgg}{{\mathfrak{g}}}

\newcommand{\ckk}{{\mathfrak{k}}}
\newcommand{\chh}{{\mathfrak{h}}}
\newcommand{\cpp}{{\mathfrak{p}}}
\newcommand{\cmm}{{\mathfrak m}}
\newcommand{\caa}{{\mathfrak{a}}}
\newcommand{\cnn}{{\mathfrak{n}}}

\newcommand{\dd}{\mathrm{d}}
\newcommand{\ee}{\mathrm{e}}
\newcommand{\ii}{\mathrm{i}}
\newcommand{\zz}{\vec{\alpha}}

\def\so{\mathfrak{so}}

\def\diag{{\rm diag \,}}
\def\cS{\mathcal{S}}

\newcommand{\stimes}{\!\times\!}

\DeclareMathOperator{\spa}{span}
\DeclareMathOperator{\Ad}{Ad}
\DeclareMathOperator{\rank}{rank}
\DeclareMathOperator{\ad}{ad}
\DeclareMathOperator{\Tr}{Tr}
\DeclareMathOperator{\class}{class}
\DeclareMathOperator{\vol}{vol}

\definecolor{applegreen}{rgb}{0.55, 0.71, 0.0}

\newcommand{\dext}{d}
\newcommand{\dt}{k}
\newcommand{\rt}{r}
\newcommand{\dimV}{N}
\newcommand{\dimVstar}{N_*}
\newcommand{\dimAbelian}{p}
\newcommand{\cartanelem}{p}

\begin{document}

\begin{titlepage}

\vspace*{-2cm} 
\begin{flushright}
{\tt \phantom{xxx}  MPP-2025-78} \qquad \qquad \\
{\tt \phantom{xxx}  Imperial/TP/2025/DW/1} \qquad \qquad \\
{\tt \phantom{xxx} IFT-UAM/CSIC-25-89} \qquad \qquad
\end{flushright}

\vspace*{0.8cm} 
\begin{center}
{\Huge The Boundary of Symmetric Moduli Spaces \\
\vspace{0.5cm}
 and the Swampland Distance Conjecture}\\

 \vspace*{1.5cm}
{\bf Stephanie Baines},$^{1}$ {\bf Veronica Collazuol},$^{2}$ {\bf Bernardo Fraiman},$^{3}$\\ 
{\bf Mariana Gra\~na}$^{4}$   and {\bf Daniel Waldram}$^{1}$\\

\begin{flushleft}
\begin{small}
 \vspace*{0.7cm} 
$^1$ {\it Abdus Salam Centre for Theoretical Physics, Imperial College London,
Prince Consort Road, London, SW7 2AZ, UK}\\[2mm]

$^2$ {\it Instituto de Física Téorica IFT-UAM/CSIC, C/ Nicolás Cabrera 13-15, Campus de
Cantoblanco, 28049 Madrid, Spain
}\\[2mm]

$^3$ {\it Max-Planck-Institut f\"ur Physik, Boltzmannstrasse 8, 85748 Garching bei München, Germany}\\[2mm]

$^4$ {\it Institut de Physique Th\'eorique, Universit\'e Paris Saclay, CEA, CNRS, 
Orme des Merisiers, 91191 Gif-sur-Yvette CEDEX, France}\\[2mm]

\end{small}
\end{flushleft}

\vspace{1cm}
\small{\bf Abstract} \\[3mm]\end{center}

\noindent For non-compact, locally symmetric moduli spaces ${\cal M}$, the set of geodesics and the geometry of the boundary can be completely characterised using group theory. In particular, geodesics that asymptote to a given infinite distance boundary point are characterised by a choice of rational parabolic subgroup $P(\mathbb{Q})$ of the local isometry group $G$ and an element of the Cartan subalgebra of $P(\mathbb{Q})$. Under the assumption that ${\cal M}$ satisfies the ``compactifiability'' constraint of \cite{Delgado:2024skw} and some mild conditions on the spectrum of states,  we use this formalism to prove the Swampland Distance Conjecture for essentially all locally symmetric spaces ${\cal M}$. We show that the states necessarily transform in some representation of $G$, and further that the convex hull encoding the exponential rate at which the leading tower of states becomes light is simply the convex hull of the weights of the representation. 
In a companion paper \cite{ourpaper}, we then use the formalism to classify all  locally symmetric spaces and irreducible representations that are consistent with the Emergent String Conjecture.

\vspace{2cm}

\end{titlepage}

\tableofcontents 
 \newpage

\section{Introduction}
The idea that quantum gravity leaves imprints on low-energy quantum field theories has sparked growing interest in recent years, leading to the now widely held view that not all consistent effective field theories are compatible with quantum gravity. This is formalized in the Swampland program \cite{Vafa:2005ui} (see also \cite{Brennan:2017rbf,Palti:2019pca,Agmon:2022thq,vanBeest:2021lhn,Grana:2021zvf} for reviews on the topic) by a series of conjectures telling apart  effective theories that are (the Landscape) from those that are not (the Swampland). 

Building on a recurring pattern observed in string theory, a notable conjecture was proposed in \cite{Ooguri:2006in}, known as the \emph{Swampland Distance Conjecture} (SDC). It states the following:
\begin{quote}
\textit{Let $d(p,q)$ be the minimum geodesic distance between two points $p$ and $q$ in the moduli space $\mathcal{M}$ of a quantum gravity theory. For fixed $p$, as $d(p, q)\to \infty$, an infinite tower of states in the theory at $q$ becomes exponentially light (in Planck units) as}
\begin{equation*} \label{eq:sdc} 
m(q) \sim m(p) \, e^{-\alpha \,  d(p, q)} \, ,
\end{equation*}
\textit{where $\alpha$ is some positive, order-one number.}
\end{quote} 
Recently, arguments giving a bottom-up motivation for this quantum gravity feature have been put forward, either from finiteness of black hole entropy \cite{Hamada:2021yxy}, or from the notion of information metric \cite{Stout:2021ubb,Stout:2022phm} or from the covariant entropy bound in backgrounds with dynamical cobordism \cite{Calderon-Infante:2023ler}. The {\it Emergent String Conjecture} (ESC) further posits that such a tower can only be made either of Kaluza--Klein states that signal a decompactification limit to a higher-dimensional theory, or of oscillator modes of a weakly coupled and tensionless critical string \cite{Lee:2018urn,Lee:2019wij} (see also \cite{Basile:2023blg,Bedroya:2024ubj} for recent bottom-up approaches).

The SDC has been extensively studied in many contexts \cite{Heidenreich:2015nta,Blumenhagen:2017cxt,Grimm:2018ohb,Heidenreich:2018kpg,Blumenhagen:2018nts,Lee:2018spm,Ooguri:2018wrx,Corvilain:2018lgw,Grimm:2018cpv,Buratti:2018xjt,Lust:2019zwm,Joshi:2019nzi,Erkinger:2019umg,Marchesano:2019ifh,Font:2019cxq,Baume:2020dqd,Perlmutter:2020buo,Klaewer:2020lfg,Lanza:2021udy,Lee:2021qkx,Lee:2021usk,Rudelius:2022gbz,Baume:2023msm,Rudelius:2023mjy,Alvarez-Garcia:2023gdd,Alvarez-Garcia:2023qqj,Castellano:2023jjt,Castellano:2023stg,Ooguri:2024ofs,Calderon-Infante:2024oed,Ashmore:2015joa,Aoufia:2024awo,Friedrich:2025gvs,Baume:2016psm,Klaewer:2016kiy,Gendler:2020dfp,Lanza:2020qmt,
Cribiori:2021gbf,Buratti:2021fiv,Basile:2023rvm,Castellano:2021yye,Castellano:2022bvr,Castellano:2021mmx,Scalisi:2018eaz}, verified in a variety of string examples and refined in several ways. For example, a more precise characterisation of the exponent $\alpha$ has been carried out in \cite{Gendler:2020dfp,Bedroya:2019snp,Andriot:2020lea,Lanza:2020qmt}, and currently is reflected in the lower bound \cite{Etheredge:2022opl} 
\begin{equation*}
\label{eqn:sharpeneddc}
	\alpha \geq \frac{1}{\sqrt{\dext-2}} \, ,
\end{equation*}
for a theory in $\dext$ (external) dimensions. The SDC with the addition of this lower bound on $\alpha$ is known as the \emph{Sharpened Distance Conjecture}, and it is widely believed that the equality is saturated along directions in moduli space that feature an emergent tensionless string limit. Refining $\alpha$ to be the exponent for the fastest decaying tower of states as one approaches a given point on the boundary $\partial\mathcal{M}$, one can plot it as a function of the unit normal to the boundary of the moduli space, to form the ``$\alpha$-hull''~\cite{Etheredge:2023odp}. It turns out that extrema of the $\alpha$-hull lie on a convex hull formed by charge-to-mass ratios (or ``tower-vectors'') $\zeta$ of the states in the theory~\cite{Calderon-Infante:2023ler}, which has been characterised both for particles and for membranes 
\cite{Etheredge:2023odp,Etheredge:2023zjk,Etheredge:2024tok,Etheredge:2023usk,Grieco:2025bjy,Etheredge:2024amg,Etheredge:2025ahf} and, for the Sharpened SDC to hold, should contain a ball of radius $\frac{1}{\sqrt{\dext-2}}$.

The goal of this paper and of the companion one \cite{ourpaper}, is to study in detail these conjectures in the context of locally symmetric moduli spaces using their underling Lie group structure, and to establish a dictionary between the physical and the group theoretic quantities. We will introduce the mathematical description of locally symmetric spaces and their boundary, and then discuss the SDC\footnote{For previous discussion on this see \cite{Cecotti:2021cvv}.}. In particular, we will prove that for essentially any locally symmetric space, the SDC holds for infinite-distance geodesics, assuming only the ``compactifiability condition'' of~\cite{Delgado:2024skw} and a fairly weak set of assumptions about the spectrum of states. Furthermore, we show that in this case, the convex hull of tower-vectors mentioned above is actually just the convex hull $C$ of the weights of a representation of a reductive Lie group, and prove that extrema of the $\alpha$-hull lie on extremal distance points to $C$. In the companion paper we will focus on the Sharpened SDC, and constrain the possible convex hulls (for particle and strings) that are compatible with the presence of a duality group. 

Although far from covering all cases, the analysis covers a fairly large class of spaces relevant to physics. The most straightforward examples are toroidal compactifications M- and string theory. The full range of applications is however much wider, including Kähler deformations in certain Calabi-Yau compactifications \cite{Farquet:2012cs}, orbifolds, and the moduli spaces arising from consistent truncations. Furthermore, it includes all theories with maximal or half-maximal supersymmetry irrespective of how they arise. As we discuss in the conclusions, much of our analysis can also apply under the much weaker assumption that the moduli space approximates a locally symmetric space asymptotically, near a boundary point. 

Locally symmetric spaces are diffeomorphic to cosets of the form 
\begin{equation*} \label{M}
    \mathcal{M}=\Gamma \backslash G / K \, , 
\end{equation*}
for some reductive Lie group $G$, its maximally compact subgroup $K$ and a discrete subgroup $\Gamma \subset G$. In toroidal string theory compactifications, $\Gamma$ is the T-duality group (or more generally, the U-duality group for maximal supergravities) that maps the lattice of charges of the string/M-theory spectrum to itself. In all known cases it has the property of being ``arithmetic'' (roughly meaning it can be thought of as a type of matrix Lie group where one restricts to integer-valued matrices). The group theoretical structure of $\mathcal{M}$ together with $\Gamma$ being arithmetic and the usual assumption that $\mathcal{M}$ has finite volume (up to flat factors) allows one to invoke a number of powerful results from mathematics and it is this that essentially allows one to prove the SDC. We should note that use of results on symmetric spaces to analyse the SDC has appeared before, most notably in \cite{Cecotti:2015wqa, Cecotti:2020rjq} in the context of $\mathcal{N}=2$ moduli spaces. Our analysis here is however very general, applying to any locally symmetric space and to all boundary points and geodesics. 

As mentioned, in the first part of the paper we review the structure of generic non-compact locally symmetric spaces $\mathcal{M}$ and their geodesics and boundaries, without imposing any restriction coming from string theory, or swampland criteria. The infinite-distance boundary $\partial\mathcal{M}$ is described by equivalence classes of geodesics, where two geodesics are in the same class if the distance between them remains finite as one approaches $\partial\mathcal{M}$. The class is invariant under the action of a parabolic subgroup of $G$, which one can think of the as the stability group of points on the boundary in the same way that $K$ is the stability group of points in $\mathcal{M}$. For a fixed parabolic group, each equivalence class has a canonical representative which is determined by a choice of unit-length element $H$ of the positive Weyl chamber of the Cartan subalgebra defined by $P$. Under the assumption that $\Gamma$ is arithmetic, after quotienting out by the discrete subgroup, one finds that the boundary parabolic subgroups have to be rational $P(\mathbb{Q})$, that is a type of matrix group with rational coefficients. Hence boundary points are labelled by a pair $(P(\mathbb{Q}),H)$. Two boundary points are equivalent if the corresponding parabolic subgroups are related by an action of $\Gamma$, and there are only a finite number of such equivalence classes. These define different ``cusps'' of the boundary. Since the rational parabolic subgroups are a set of measure zero among all real parabolic subgroups, only very special geodesics reach the boundary asymptotically. Most have instead an ergodic motion~\cite{Mautner-ergodic,Moore-ergodic}. For these geodesics, the distance ``as the crow flies'' $d(p,q)$ from the starting point $p$ scales only logarithmically with the distance along the geodesic~\cite{sullivan,Kleinbock-Margulis}.  

We then look at various simple string compactifications using this formalism.\footnote{Many aspects of $SL(2,\mathbb Z) \backslash SL(2,\mathbb{R}) / SO(2)$ and $O(1,17;\mathbb{Z})\backslash O(1,17) / (O(1) \times O(17) )$ cases were nicely worked out in \cite{Keurentjes:2006cw}.} In all cases the duality group $\Gamma$ is arithmetic and the charges $q$ of the particle states transform in a given representation $\rho$ of $\Gamma$, that is also a representation of $G$. The full set of states $\Pi$ is then typically a subspace of the $\Gamma$-invariant lattice $L$ in the representation space of $\rho$. (Since $G$ is reductive, the representation can always be recomposed into irreps. We note that the fact that the charges should be in a semi-simple representation of $\Gamma$ has also recently been argued for in the supersymmetric case in \cite{Delgado:2024skw}.) For toroidal compactifications, the canonical geodesics that reach the boundary  simply correspond to the radii of the torus growing or shrinking. In all cases, the mass is given by a simple BPS-like quadratic formula, and the tower-vectors $\zeta$ are simply the weights of the representation. 

We then turn to the general proof of the SDC, showing how these features are consequences of the compactifiability condition that the volume must not grow faster than that of Euclidean space ~\cite{Delgado:2024skw}, and some simple conditions on the spectrum. In particular, we show that compactifiability implies that the non-flat part of the moduli space $\mathcal{M}$ has finite volume, which crucially then implies that $\Gamma$ must be arithmetic. By considering the most general expression for the mass, invariant under $K$ and $\Gamma$, we show that if a single state becomes massless at some point on the boundary, then the leading order terms in the mass formula are quadratic, as they are in the simple string BPS-like examples. From this we can then deduce the relation between the $\alpha$-hull and the convex hull of the weights of the representation $\rho$. Since this applies to essentially any locally symmetric space and any representation $\rho$, we see that the SDC by itself is not particularly constraining from a bottom-up perspective. Refining our analysis, in the companion paper \cite{ourpaper}, we find all the irreducible representation and groups $G$ that are compatible with the known rates for emergent strings and Kaluza--Klein towers. The allowed cases turn out to be much more limited. The power of this formalism is that these constraints can be expressed purely in terms of the Lie algebra $\cgg$ of $G$, making the classification very easy and clean. 

This paper is structured as follows. In section~\ref{sec:geodesics}, we introduce globally symmetric spaces and describe their geodesics and boundaries. 
In section~\ref{sec:discrete}, we consider the discrete quotient, and how it affects the boundary. Section~\ref{sec:strmod} serves as a bridge from the mathematical definitions to the physics, as we explicitly apply the machinery built up in the previous sections to symmetric moduli spaces of string or M-theory compactifications. Finally, in section~\ref{sec:weight-polytopes} we show how the SDC is always satisfied at the infinite distance limits of these spaces under some assumptions on the spectrum, which we specify, and also show the relation between the $\alpha$-hull and the convex hull of the weights. We summarize our results and mention future directions in Section \ref{sec:conclusions}.

\section{Globally symmetric spaces: geodesics and boundaries}
\label{sec:geodesics}
The moduli in a $\dext$-dimensional quantum gravity theory appear in the action as scalar fields $\phi^i$ with a non-linear kinetic term 
\begin{equation} \label{eqn:dimredaction}
    S=\frac{M_{P}^{\dext-2}}{2} \int d^{\dext}x \sqrt{-g^E} \left(R^E - G_{ij}(\phi) \partial \phi^i \partial\phi^j\right) + \dots \, ,
\end{equation}
where $\phi^i$ can be viewed as coordinates on a moduli space $\mathcal{M}$ with metric $G_{ij}(\phi)$. The space-time metric is in the Einstein frame (denoted $g^E$) and $M_P$ is the $\dext$-dimensional Planck scale. 

In order to formalise the SDC, we will study geodesic motion and the various infinite distance limits in cases where $\mathcal{M}$ is a locally symmetric space. Non-compactness implies the metric on $\mathcal{M}$ has non-positive sectional curvature. Such spaces have the form 
\begin{equation}
\label{eq:M-product}
    \mathcal{M} = \Gamma\backslash G / K ,
\end{equation}
where $G=\mathbb{R}^{\dimAbelian}\times G_1\times \dots\times G_s$ is a reductive group with Abelian factor $\mathbb{R}^{\dimAbelian}$ and non-compact simple factors $G_a$, $K=K_1\times\dots\times K_s$ is its maximally compact subgroup and $\Gamma$ is a discrete group. In string theory examples $\Gamma$ is an arithmetic subgroup of $G$ possibly including outer automorphisms and an action on the flat factor $\mathbb{R}^{\dimAbelian}$. We assume $\mathcal{M}$ is non-compact and for the SDC we will be interested in geodesics on $\mathcal{M}$ that reach the boundary $\partial\mathcal{M}$. The group structure uniquely defines the metric on $\mathcal{M}$ up to some positive constants $\kappa_a$ so that
\begin{equation}
\label{eq:modspace-metric}
    G_{ij}(\phi) \dd \phi^i \dd \phi^j = \dd s_0^2 + \sum_a \kappa_a
    \dd \tilde{s}_a^2 \, ,
\end{equation}
where $\dd s_0^2$ is the flat metric on $\mathbb{R}^{\dimAbelian}$ and each $\dd\tilde{s}_a^2$ factor is an Einstein metric on $G_a/K_a$. The latter are normalised so that $\tilde{R}^{(a)}_{ij}=-\frac12\tilde{g}^{(a)}_{ij}$ where $\tilde{R}^{(a)}_{ij}$ and $\tilde{g}^{(a)}_{ij}$ are Ricci tensors and metrics respectively. The structure of equation \eqref{eq:modspace-metric} comes from the decomposition of the Lie algebra $\cgg$ of $G$ into an Abelian factor (giving the flat metric) and simple factors, and the normalisations of the $\tilde{g}^{(a)}_{ij}$ metrics are fixed by choosing the Killing form as the invariant metric on the simple factors in $\cgg$. 

Following mainly \cite{Link:2008,Erickson:2008,Borel}, in this section we will review the structure of globally symmetric spaces with non-positive sectional curvature. These are of the form $G/K$ and we will refer to them as $\mathcal{S}$ in order to distinguish them from the physical moduli spaces $\mathcal{M}\cong \Gamma\backslash \mathcal{S}$ which include the discrete quotient and are only locally symmetric. We will then review locally symmetric spaces in section~\ref{sec:discrete}. The central point we wish to convey is that geodesics and boundaries of both globally and locally symmetric spaces can be understood systematically using only the underlying group structure. In particular, while points in the interior of $\mathcal{S}$ or $\mathcal{M}$ are characterised by being stabilised by maximally compact subgroups $K\subset G$, points on the boundary are characterised in the same way by parabolic subgroups $P\subset G$. Throughout these sections, we will illustrate the concepts with the simplest example available. This corresponds to $SL(2,\mathbb{Z})\backslash SL(2)/SO(2)$, the moduli space of Type IIB supergravity in 10 dimensions.

Before doing so, let us note that moduli spaces of the form~\eqref{eq:M-product} are examples of a larger class of metric spaces with non-positive sectional curvature. In fact, given any such space that is both non-compact and of finite volume, if it has ``rank'' greater than one, it is necessarily a locally symmetric space, a result known as ``rank rigidity''~\cite{Ballman,BurnsSpatzier}. The notion of rank here is related to whether one can find subspaces that are flat in an infinitesimal sense around any given geodesic. Although we will restrict to locally symmetric spaces of non-compact type, it is very natural to wonder how the analysis might extend to non-positively curved spaces in general. It would also be interesting to ask the same question of spaces that are only locally symmetric of non-positive curvature near the boundary since, as we shall see, only the asymptotic behaviour of the geodesics is relevant to the distance conjectures. 

\subsection{Globally symmetric spaces of non-compact type}

The analysis of globally symmetric spaces goes back to Cartan, who showed that they can be classified using Lie theory. Perhaps the simplest way to define a symmetric space $\mathcal{M}$ is as a Riemannian space with constant Riemann tensor. It turns out that this is equivalent to requiring that, in the neighbourhood of each point $x\in\mathcal{M}$ one has an {\it inversion symmetry}, i.e.~an isometry that reverses geodesics through $x$ sending $\gamma(t)$ to $\gamma(-t)$. If the isometry exists globally then $\mathcal{M}$ is a globallly symmetric space. Any locally symmetric space has the form 
\begin{equation}
    \mathcal{M} = \Gamma \backslash \cS \, , 
\end{equation}
where $\cS$ is a simply-connected globally symmetric space and $\Gamma$ is a discrete subgroup of isometries without fixed points. The simply-connected globally symmetric spaces come in three types, related to their isometry group. Since our moduli space $\mathcal{M}$ is non-compact it is by definition locally the product of a flat space and a symmetric space of ``non-compact type'', meaning it has a non-compact and semi-simple isometry group\footnote{Spheres are also symmetric spaces, but of compact type since they are diffeomorphic to $S^k \cong SO(k+1)/SO(k)$ and have a compact isometry group. We will not consider such spaces in this work.}. 

To connect to the group-theoretic description, one notes that the inversion symmetry implies that $\cS$ is homogeneous and geodesically complete. Thus
\begin{equation}
\label{eq:coset}
\cS \cong G/K \, ,
\end{equation}
for some $G$ and $K$. Cartan classified globally symmetric spaces, that is the possible $G$ and $K$, by analysing the corresponding Lie algebras. The inversion symmetry guarantees that the real Lie algebra $\cgg$ of $G$ splits into 
\begin{equation}
\label{eqn:cartandecomposition}
    \cgg =   \cpp \oplus \ckk \, ,
\end{equation}
where $\ckk$ and $\cpp$ are respectively the positive and negative eigenspaces of the involution\footnote{Or equivalently, $\ckk$ is the Lie algebra of $K$ and, as we will see, $\cpp$ is its orthogonal complement with respect to the Killing form of $G$.} and one has
\begin{align}\label{lem:1}
    [\ckk,\ckk]\subseteq \ckk,\qquad 
    [\ckk,\cpp]\subseteq \cpp,\qquad 
    [\cpp,\cpp]\subseteq \ckk \, .
\end{align}
By definition, we the tangent space at any point on the manifold satisfies $T_x\mathcal{S}\simeq\cpp$. 

For a simply-connected globally symmetric space (with non-positive sectional curvature), $G$ is reductive and decomposes into irreducible pieces
\begin{equation}
    G = \mathbb{R}^{\dimAbelian} \times G_1 \times \dots \times G_s \, ,
\end{equation}
where $\mathbb{R}^{\dimAbelian}$ denotes the Abelian factor and $G_a$ are non-compact simple Lie groups. The space correspondingly decomposes as 
\begin{equation}
\label{eq:Sfactors}
    \cS = \mathbb{R}^{\dimAbelian} \times \cS_1 \times \dots \times \cS_s \, , 
\end{equation}
where $\cS_a=G_a/K_a$ and we can take $G_a$ to the be the connected component of the isometry group of $\cS_a$ and $K_a\subset G_a$ to be the maximally compact subgroup, also equal to the isotropy group that stabilises any point $x\in\cS_a$. Note that the group $G_a$ have zero centre. The $\mathbb{R}^{\dimAbelian}$ factor is flat and the other factors are of non-compact type. The Lie algebra decomposes as 
\begin{equation}
\label{eq:simple-decomp}
    \cgg = \mathbb{R}^{\dimAbelian} \oplus \cgg_1 \oplus \cgg_2 \oplus \dots \oplus \cgg_s \, ,
\end{equation}
and the invariant metric on $\mathcal{S}$, given $P=P_0 \oplus P_1\oplus\dots\oplus P_s\in\cpp=\mathbb{R}^{\dimAbelian} \oplus \bigoplus_a \cpp_a$, has the form
\begin{equation} \label{eqn:kf}
    \langle P, P \rangle = P_0^TP_0 + \sum_a \kappa_a \mathcal{K}(P_a,P_a) \, ,  
\end{equation}
where $P_0^TP_0$ is just the flat metric on $\mathbb{R}^{\dimAbelian}$, while $\mathcal{K}$ is the Killing form 
\begin{equation} \label{eqn:product}
    \mathcal{K}(A,B) = \Tr(\ad_A \ad_B) \,,
\end{equation}
where $A,B\in \cgg_a$, $\kappa_a$ are positive constants in agreement with~\eqref{eq:modspace-metric}. Cartan showed that for the non-compact case one has two possibilities
\begin{enumerate}
    \item $G_a$ is the complexification of a real simple compact Lie group $K_a$ (e.g.~$\cgg_a=\mathfrak{sl}(k,\mathbb{C})$, $\ckk_a=\mathfrak{su}(k)$)
    \item $G_a$ is a non-compact real simple Lie group and $K_a$ is its maximal compact subgroup (e.g.~$\cgg_a=\mathfrak{so}(p,q)$, $\ckk_a=\mathfrak{so}(p)\times \mathfrak{so}(q)$). 
\end{enumerate}
It is worth noting that variants of the groups $G$ with the same Lie algebra can lead to the same symmetric spaces, but in general $G$ will not act effectively on $\cS$, that is there may be elements of $G_i$ that leave $\cS$ invariant. The classic example is hyperbolic space $\cS=PSL(2,\mathbb{R})/SO(2)\simeq SL(2,\mathbb{R})/SO(2)$ where $-\mathbb{1}_2\in SL(2,\mathbb{R})$ acts trivially on $\cS$. For simplicity we will indeed sometimes use quotients $G/K$ where $G$ does not act effectively in the following.   

The Cartan decomposition of the algebra \eqref{eqn:cartandecomposition} also induces a decomposition of $G$~\cite{Link:2008,Erickson:2008}. One starts by identifying a maximal Abelian subalgebra $\caa\subset \cpp$ (any two such algebras are conjugate under the action of $K$). This is a real Cartan subalgebra of the real algebra $\cgg$, defined as the maximal set of elements $X\in\cgg$ such that $\ad_X$ can be simultaneously diagonalised over the reals. (If $\cgg$ includes an Abelian factor then this trivially becomes part of $\caa$.) In general the real Cartan subalgebra is smaller or equal to the usual Cartan subalgebra where the diagonalization is taken over the complex numbers. One defines the rank of the symmetric space by 
\begin{equation}
    r = \rank_{G/K} := \dim\caa = \rank_\mathbb{R}\cgg \leq \rank_\mathbb{C} \cgg \, .
\end{equation}
Defining the subgroup $A=\ee^\caa\subset G$, the Cartan decompositions then states that
\begin{align}
    \label{eqn:cartdec}
        \forall g\in G\;\, \exists \, k_1,k_2\in K, \; a\in A:\, g=k_1\,a\, k_2 \qquad\Leftrightarrow\qquad G=KAK \,  \, . 
\end{align}

One can also use the Cartan subalgebra $\caa$ to define a set of (restricted) roots $\alpha\in \caa^*$ in the usual way via the root spaces
\begin{equation}
    \cgg_\alpha = \left\{ X \in \cgg : [H,X] = \alpha(H) X , \forall H\in \caa \right\} \neq 0 \, , 
\end{equation}
for $\alpha\neq 0$. The corresponding restricted root system $\Phi$ has a Weyl group $W\subset K$ and one can fix a positive Weyl chamber $\caa^+$, its closure $\overline{\caa^+}$ and sets of positive and negative roots $\Phi^+$ and $\Phi^-$. Since the Weyl group $W \subset K$ maps one Weyl chamber to another, one can restrict $a$ in equation \eqref{eqn:cartdec} to an element of $\ee^{\overline{\caa^+}}$ giving 
$G=K\, e^{\overline{\caa^+}}\, K$ and hence 
\begin{equation}
\label{eq:decompositionpositiveWc}
    \mathcal{S} \cong K\, \ee^{\overline{\caa^+}}\cdot o \, , 
\end{equation}
for the corresponding symmetric space, where $o$ is a point in $\mathcal{S}$ stabilized by $K$. It also implies that $\cpp=\Ad_K \overline{\caa^+}$ where $\Ad_K$ is the adjoint action of the group $K$. 

Note that in general the structure of the restricted root system $\Phi$ for a non-compact simple Lie algebra can be read off from the ``Tits--Sasaki'' diagram, a refinement of the standard Dynkin diagram of the complexified algebra (for a review see for example~\cite{Henneaux:2007ej}). One can again introduce a set of fundamental roots $\{\alpha_i\}$, but unlike the complex case, one can have $\dim \cgg_{\alpha_i} >1$ and also sometimes $\cgg_{2\alpha_i}\neq0$. Finally, recall that for each complex simple Lie algebra there is always a real simple Lie algebra $\cgg$, known as the ``split form'' such that $\rank_\mathbb{R}\cgg = \rank_\mathbb{C} \cgg$ and for which Tits--Sasaki and Dynkin diagrams, and hence the restricted and conventional complex root systems, are the same. 

\subsubsection*{Example: $SL(2,\mathbb{R})/SO(2)$}
Let us illustrate these definitions on the upper half plane $\mathbb{H}^2$ 
\begin{equation}
    \mathbb{H}^2 = \{ \tau=\tau_1 + i \tau_2 : \tau_2 \geq 0 \} \, ,
\end{equation}
with hyperbolic metric 
\begin{equation}
\label{eq:H2-metric}
    \dd s^2 = \frac{1}{\tau_2^2}\left( \dd \tau_1^2 + \dd \tau_2^2 \right) \, . 
\end{equation}
It is easy to see that the group
\begin{equation}
    G=SL(2,\mathbb{R}) = \Bigg{\{} \begin{pmatrix}
        a & b\\
        c & d \\
    \end{pmatrix} , \,  a,b,c,d \in \mathbb{R}, \, ad-bc=1  \Bigg{\}} \, ,
\end{equation}
acting on $\tau \in \mathbb{H}^2$ in the usual way
\begin{equation}
    \tau \to \frac{a \tau +b}{c\tau+d} \, ,
\end{equation}
leaves the metric invariant, although is not effective since $-\mathbb{1}_2\in G$ acts trivially. (Thus the actual isometry group is $PSL(2,\mathbb{R})=SL(2,\mathbb{R})/\mathbb{Z}_2$.) Since the space is homogeneous we can identify the isotropy group by choosing any fixed point $o\in\mathbb{H}^2$. Taking $o = i$, it is easy to see that $K=SO(2)\subset SL(2,\mathbb{R})$ fixes $o$ since for 
\begin{align}
\label{eq:so2angcoord}
    k_\theta =  \begin{pmatrix}
        \cos\theta & \sin\theta\\
        -\sin\theta & \cos\theta
    \end{pmatrix} \qquad \text{we have} \qquad
    k_\theta \cdot i = \frac{i\cos\theta + \sin\theta}{-i(\sin\theta+i\cos\theta)}=i \, ,
\end{align}
and hence $\mathbb{H}^2 \cong SL(2) /SO(2)$. (Again the actual isotropy group is $SO(2)/\mathbb{Z}_2\simeq SO(2)$ since $k_\pi=\mathbb{1}_2$ acts trivially.) The algebra $\mathfrak{sl}(2, \mathbb{R})$ is generated by
\begin{align}\label{eq:slalg}
    h=\begin{pmatrix}
 1 & 0 \\
 0 & -1 \\
 \end{pmatrix},\;\;e_1=\begin{pmatrix}
 0 & 1 \\
 1 & 0 \\
\end{pmatrix},\;\;e_2=\begin{pmatrix}
 0 & -1 \\
 1 & 0 \\
\end{pmatrix}.
\end{align}
where $h$ generates the Cartan subalgebra $\caa=\spa_{\mathbb{R}}(\{h\})$, $\mathfrak{so}(2)=\spa_{\mathbb{R}}(\{e_2\})$ generates the set of antisymmetric matrices  and $\cpp=\spa_{\mathbb{R}}(\{h,e_1\})$ is the set of the symmetric matrices. Thus the involution defining the decomposition~\eqref{eqn:cartandecomposition} is simply $X\to -X^T$. Furthermore $\cpp$ is orthogonal to $\mathfrak{so}(2)$ with respect to the Killing form $\mathcal{K}(A,B)=4\Tr(AB)$, and is isomorphic to $T_o\mathbb{H}^2$, the tangent space to $\mathbb{H}^2$ at $o=i$. The Cartan decomposition is given in terms of $\theta,\phi\in [0,2\pi]$ and $t\geq 0$ by
\begin{align}
    g = k_\theta \,\ee^{th} k_\phi \in SL(2) \, ,
\end{align}
and one can write a general point in $\mathbb{H}^2$ as 
\begin{align}
\label{eq:SL2cartan}
    \tau = k_\theta\, \ee^{th} \cdot i = \frac{i-e^{2t}\cot{\theta}}{e^{2t}+i\cot{\theta}} \, ,  
\end{align}
where taking $0\leq\theta<\pi$ with $t\geq0$, it covers the whole of $\mathbb{H}^2$. 
Finally, we note that $\mathfrak{sl}(2,\mathbb{R})$ is the split form and so $\rank_\mathbb{R}\cgg = \rank_\mathbb{C} \cgg=1$. 

\subsubsection*{Example: $SO(1,3)/SO(3)$ and $SO(2,2)/SO(2)^2$} 

Consider a basis for $\mathfrak{so}(1,3)$ that stabilises $\eta=\diag(1,-1,-1,-1)$. We can define a commuting subalgebra by $\chh=\spa_{\mathbb{C}}(\{h_1,h_2\})$ where 
\begin{equation}\arraycolsep=2pt\def\arraystretch{0.8} \label{CartansSO13}
	h_1=\left(
\begin{array}{@{}cccc@{}}
 0 & 1 & 0 & 0 \\
 1 & 0 & 0 & 0 \\
 0 & 0 & 0 & 0 \\
 0 & 0 & 0 & 0 \\
\end{array}
\right),\;\; 
h_2=\left(
\begin{array}{@{}cccc@{}}
 0 & 0 & 0 & 0 \\
 0 & 0 & 0 & 0 \\
 0 & 0 & 0 & -1 \\
 0 & 0 & 1 & 0 \\
\end{array}
\right)
\, .
\end{equation}
Over the complex numbers we can simultaneously diagonalise both $h_1$ and $h_2$ and so \mbox{$\rank_\mathbb{C}\mathfrak{so}(1,3)=2$}. However, while $h_1$ has real eigenvalues, $h_2$ does not, and hence over the reals \mbox{$\rank_\mathbb{R}\mathfrak{so}(1,3)=1$} and we have a one-dimensional Cartan subalgebra $\caa=\spa_\mathbb{R}(\{h_1\})$. 

The rest of the algebra is given by 
\begin{equation}
\label{rootsSO13}
\begin{aligned} 
	b_1 &=\left(
\arraycolsep=2pt\def\arraystretch{0.8}\begin{array}{@{}cccc@{}}
 0 & 0 & 1 & 0 \\
 0 & 0 & 0 & 0 \\
 1 & 0 & 0 & 0 \\
 0 & 0 & 0 & 0 \\
\end{array}
\right), &
b_2 &=\left(
\arraycolsep=2pt\def\arraystretch{0.8}\begin{array}{@{}cccc@{}}
 0 & 0 & 0 & 1 \\
 0 & 0 & 0 & 0 \\
 0 & 0 & 0 & 0 \\
 1 & 0 & 0 & 0 \\
\end{array}
\right),\\
r_1 &=\left(
\arraycolsep=2pt\def\arraystretch{0.8}\begin{array}{@{}cccc@{}}
 0 & 0 & 0 & 0 \\
 0 & 0 & -1 & 0 \\
 0 & 1 & 0 & 0 \\
 0 & 0 & 0 & 0 \\
\end{array}
\right), &
r_2 &=\left(
\arraycolsep=2pt\def\arraystretch{0.8}\begin{array}{@{}cccc@{}}
 0 & 0 & 0 & 0 \\
 0 & 0 & 0 & -1 \\
 0 & 0 & 0 & 0 \\
 0 & 1 & 0 & 0 \\
\end{array} 
\right),
\end{aligned}
\end{equation}
and we have 
\begin{align}
    \ckk &= \so(3)=\spa_{\mathbb{R}}(\{h_2,r_1,r_2\}) , &
    \cpp &= \spa_{\mathbb{R}}(\{h_1,b_1,b_2\}) \, ,
\end{align}
corresponding to antisymmetric and symmetric elements respectively, and so the involution is $X\to -X^T$ as above. One can identify the complex root spaces for $\chh$ as 
\begin{equation}
\begin{aligned}
    \cgg_{\pm\beta_1} &= \spa_{\mathbb{C}}(\{b_2\mp r_2\pm i b_1 - ir_1\}) , & 
    \beta_1 &= (1,i) \, , \\
    \cgg_{\pm\beta_2} &= \spa_{\mathbb{C}}(\{-b_2\mp r_2\mp i b_1 - ir_1\}) , & 
    \beta_2 &=(1,-i) \, , 
\end{aligned}
\end{equation}
noting that the involution exchanges the fundamental roots $\beta_1\leftrightarrow \beta_2$. 

The real root spaces on the other hand are eigenspaces of $h_1$ alone, and are given by 
\begin{align}
    \cgg_{\pm\alpha} = \spa_{\mathbb{R}}(\{b_1\mp r_1, b_2\mp r_2\}) , \qquad
    \dim_\mathbb{R} \cgg_\alpha = 2 \, , 
\end{align}
which can be regarded as a projection of the two pairs of roots $(\beta_1,\beta_2)$ and $(-\beta_1,-\beta_2)$ onto the one dimensional space spanned by $h_1$. The corresponding Tits--Satake diagram for the restricted root system is 
\begin{equation*}
    \dynkin[scale=2.5,edge/.style={draw=none},labels*={\beta_1,\beta_2},involutions={[out=-80,in=-100,relative]12}]{D}{oo}
\end{equation*}
the arrow indicating that the two roots are exchanged under the involution. 

Repeating the analysis for the split case of $\mathfrak{so}(2,2)$, we take a basis for the algebra that stabilises $\eta=\diag(1,-1,1,-1)$ with the Cartan subalgebra $\caa=\spa_{\mathbb{R}}(\{h_1,h_2\})$ where 
\begin{align} \label{CartansSO22}
	h_1=\left(
\arraycolsep=2pt\def\arraystretch{0.8}\begin{array}{@{}cccc@{}}
 0 & 1 & 0 & 0 \\
 1 & 0 & 0 & 0 \\
 0 & 0 & 0 & 0 \\
 0 & 0 & 0 & 0 \\
\end{array}
\right),\;\; h_2=\left(
\arraycolsep=2pt\def\arraystretch{0.8}\begin{array}{@{}cccc@{}}
 0 & 0 & 0 & 0 \\
 0 & 0 & 0 & 0 \\
 0 & 0 & 0 & 1 \\
 0 & 0 & 1 & 0 \\
\end{array}
\right) \, .
\end{align}
$h_1$ and $h_2$ can be simultaneous diagonalised over the reals and so $\rank_\mathbb{R}\mathfrak{so}(2,2)=2$. The rest of the algebra is given by 
\begin{equation}
\label{rootsSO22}
\begin{aligned} 
b_1 &=\left(
\arraycolsep=2pt\def\arraystretch{0.8}\begin{array}{@{}cccc@{}}
 0 & 0 & 0 & 1 \\
 0 & 0 & 0 & 0 \\
 0 & 0 & 0 & 0 \\
 1 & 0 & 0 & 0 \\
\end{array}
\right), &
b_2 &=\left(
\arraycolsep=2pt\def\arraystretch{0.8}\begin{array}{@{}cccc@{}}
 0 & 0 & 0 & 0 \\
 0 & 0 & 1 & 0 \\
 0 & 1 & 0 & 0 \\
 0 & 0 & 0 & 0 \\
\end{array}
\right) \, , \\
r_1 &=\left(
\arraycolsep=2pt\def\arraystretch{0.8}\begin{array}{@{}cccc@{}}
 0 & 0 & -1 & 0 \\
 0 & 0 & 0 & 0 \\
 1 & 0 & 0 & 0 \\
 0 & 0 & 0 & 0 \\
\end{array}
\right), &
r_2 &=\left(
\arraycolsep=2pt\def\arraystretch{0.8}\begin{array}{@{}cccc@{}}
 0 & 0 & 0 & 0 \\
 0 & 0 & 0 & -1 \\
 0 & 0 & 0 & 0 \\
 0 & 1 & 0 & 0 \\
\end{array}
\right) ,
\end{aligned}
\end{equation}
and we have 
\begin{align}
    \ckk &= \so(2)\oplus \so(2) =\spa_{\mathbb{R}}(\{r_1,r_2\}) , &
    \cpp &= \spa_{\mathbb{R}}(\{h_1,h_2, b_1,b_2\}) , 
\end{align}
and so the involution is again $X\to -X^T$. One can identify the real root spaces as 
\begin{equation}
\label{rootspacesSO22}
\begin{aligned}
    \cgg_{\pm\alpha_1} &= \spa_{\mathbb{R}}(\{r_1 - r_2 \pm b_1\mp b_2\}) , & 
    \alpha_1 &= (1,1) \, , \\
    \cgg_{\pm\alpha_2} &= \spa_{\mathbb{R}}(\{r_1 + r_2 \pm b_1 \pm b_2,\}) , &
    \alpha_2 &=(-1,1) \, ,
\end{aligned}
\end{equation}
and we note that the involution exchanges $\alpha_i\leftrightarrow -\alpha_i$. Given $H=\lambda_1h_1+\lambda_2h_2$, the positive Weyl chamber corresponds to $\lambda_1+\lambda_2>0$ and $\lambda_2-\lambda_1>0$. 

\subsection{Geodesics and boundaries on globally symmetric spaces}
\label{sec:geodbund}

Viewing $\cS$ as a quotient of Lie groups makes it easy to characterise geodesics. Let $o\in\cS$ be the point stabilized by $K\subset G$ and write $x=g\cdot o$ for some $g\in G$ for a generic point $x\in\cS$. A general geodesic passing through $x$ then has the form 
\begin{equation}  \label{eqn:genericgeodesic}
    \gamma (t)= g \, e^{tX} \cdot o \, ,
\end{equation}
where $X\in\cpp$. Furthermore, if we fix the normalization such that $\left< X,X\right>=1$ then $t$ is an affine parameter giving the geodesic distance away from $x$. The physical picture is the following: for geodesics passing through $o$, we can identify $X$ with a choice of the initial velocity $X\in T_o\mathcal{S}=\cpp$. The action of $g$ then maps the geodesic from one passing through $o$ to one passing through $x$. Finally, we note that the geodesic distance between any two points $x_1=g_1 \cdot o$ and $x_2=g_2 \cdot o$ in $\cS$ is then given by 
\begin{equation} \label{distancegeod}
    d(x_1,x_2)=\left<X,X\right>^{1/2} \, ,   \quad \text{where} \ \ \text{exp}(X) \cdot o=g_2^{-1} g_1 \cdot o \, .
\end{equation}

For the SDC, we would like to consider geodesics that go to infinite distance, that is, all the way to the boundary. This naturally defines the boundary $\partial\cS$ as the ``geodesic compactification'' as follows. Consider an equivalence class of {\it asymptotic geodesics}
\begin{equation}
    \gamma_1(t) \sim \gamma_2(t) \quad \text{iff} \quad  \limsup_{t\to \infty} d(\gamma_1(t),\gamma_2(t))< \infty \, , 
\end{equation} 
that is, the distance between them is bounded from above as $t\to\infty$. The boundary is then given by equivalence classes of asymptotic geodesics
\begin{align}
    \partial\cS = \{ \text{all geodesics in $\cS$} \} / \sim \, .
\end{align}
The fact that $\cS$ is simply-connected and non-positively curved means that, fixing any point $x\in\cS$, there is precisely one geodesic through $x$ for each equivalence class. Taking $x=o$ from~\eqref{eqn:genericgeodesic} we can then associate 
\begin{equation} \label{bdy}
    \partial\cS\cong \cpp_1 \cong S^{n-1} \, , 
\end{equation}
where the subscript~1 denotes the restriction to norm-one vectors in $T_o\cS=\cpp$ and $n=\dim \cS$. 

We can make this more explicit using the Cartan decomposition \eqref{eq:decompositionpositiveWc}. Each geodesic through $o\in\cS$, given $k\in K$ and $h\in\overline{\caa^+_1}$ takes the form 
\begin{align}
    \gamma(t) = k \,\ee^{th} \in \partial\cS\, .  
\end{align}
Furthermore, each choice of $h$ corresponds to a different class of asymptotic geodesics. This gives the boundary the following structure. We have a projection map $\pi:\partial\cS\to\overline{\caa^+_1}$. Above each point $h\in\overline{\caa^+_1}$ the preimage $\pi^{-1}(h)$ is diffeomorphic to $K/M_h$ where $M_h$ is the centraliser of $h$ in $K$. 

\subsubsection*{Example}
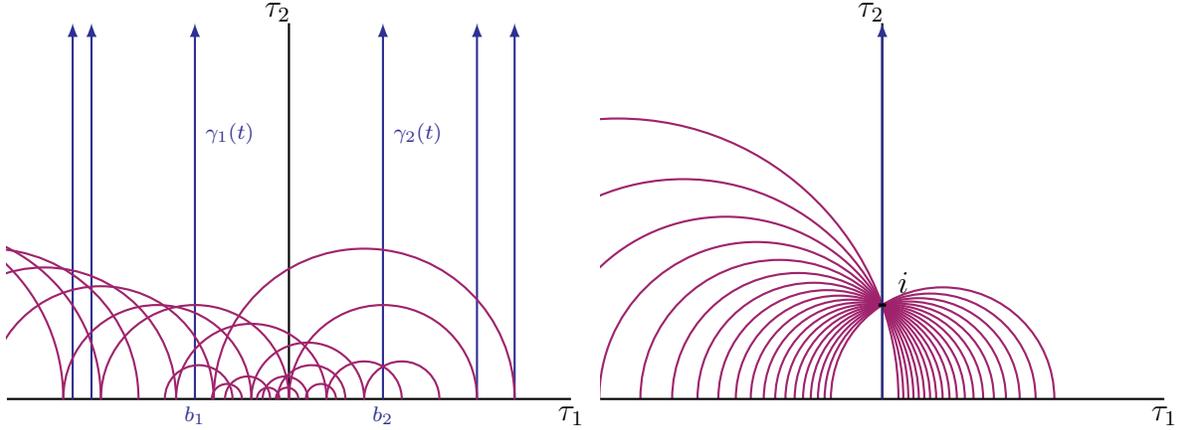
\begin{figure}[t]
	\begin{center}
	\subfigure{\begin{tikzpicture}[scale=2.5]
  \def\xmin{-1.5} \def\xmax{1.5}
  \def\ymin{-0.1} \def\ymax{2}
  \clip (-1.5,-0.2) rectangle (1.6,2.4);

  \draw [thick] (\xmin,0) -- (\xmax,0);
  \draw [thick] (0,0) -- (0,\ymax);
  \node[xshift=0.0cm, yshift=-0.22cm] at (\xmax,0) {$\tau_1$};
  \node[xshift=-0.15cm, yshift=0.15cm] at (0,\ymax) {$\tau_2$};

  \foreach \x/\y/\z in {
    -0.5/\scriptsize$b_1$/\scriptsize$\gamma_1(t)$,
     0.5/\scriptsize$b_2$/\scriptsize$\gamma_2(t)$
  } {
    \draw [thick,Blue,-latex] (\x,0) -- (\x,\ymax);
    \node[xshift=0.0cm, yshift=-0.22cm, Blue] at (\x,0) {\y};
    \node[above right, xshift=0.0cm, yshift=0.0cm, Blue] at (\x,1.3) {\z};
  }

  \foreach \P in {-1.15,-1.05,1.0, 1.2} 
    {\draw [thick,Blue,-latex] (\P,0) -- (\P,\ymax);}

  \foreach \x/\y in 
  {    -1.2/1.0, -1.0/0.9, -0.8/0.8, -0.6/0.7,  -0.4/0.6, -0.2/0.5, 0/0.5,
     0.2/0.4, 0.4/0.3, 0.6/0.2,
     0.8/0.2, 1.0/0.5, 1.2/0.8,
       -0.1/0.12, 0.1/0.12, 
  -0.3/0.18, 0.3/0.18, 
  -0.05/0.06, 0.05/0.06,
  -0.25/0.08, 0.25/0.08
     }
    {\draw[thick,RedViolet] (\x,0) arc (0:180:\y);}
\end{tikzpicture}}
\subfigure{\begin{tikzpicture}[scale=2.5]
  \def\xmin{-1.5} \def\xmax{1.5}
  \def\ymin{-0.1} \def\ymax{2}
  \clip (-1.5,-0.2) rectangle (1.6,2.4);

  \foreach \x/\y in 
{0.915244/0.594198,0.815521/0.561037,0.730848/0.536458,0.657718/0.51891,0.593621/0.507383,0.536713/0.501256,0.485607/0.500213,0.439239/0.504203,0.396776/0.513427,0.357556/0.528374,0.321046/0.549875,0.28681/0.579234,0.254484/0.618432,0.223764/0.670506,0.19439/0.740233,0.166137/0.835461,0.138808/0.96993,0.112228/1.16992,0.0862384/1.49259}
    {\draw[thick,RedViolet] (\x,0) arc (0:180:\y);}

 \draw [thick] (\xmin,0) -- (\xmax,0);
  \draw [thick] (0,0) -- (0,\ymax);
  \node[xshift=0.0cm, yshift=-0.22cm] at (\xmax,0) {$\tau_1$};
  \node[xshift=-0.15cm, yshift=0.15cm] at (0,\ymax) {$\tau_2$};

  \foreach \P in {0} 
    {\draw [thick,Blue,-latex] (\P,0) -- (\P,\ymax);}
  
    \draw [very thick] (-0.02,0.5) -- (0.02,0.5) node [above right] {$i$};
\end{tikzpicture}}
					\caption{Geodesics of $\mathbb{H}^2$. These are straight vertical lines or semi-circles. On the right we show the geodesics passing through the point $\tau=i$.}			
		    \label{fig:straightgeod}
	\end{center}
\end{figure}

From the hyperbolic metric~\eqref{eq:H2-metric} on $\mathbb{H}^2$ one finds that the geodesics are either straight lines or semicircles centred on the real line, as shown in Figure \ref{fig:straightgeod}. In particular, the two ``straight line'' geodesics
\begin{equation}
\label{eq:straightgeod}
    \gamma_1(t) = b_1 + i \, \ee^{t} \, , \quad \gamma_2(t) = b_2 + i \, \ee^{t} \, , 
\end{equation}
with $b_1 \ne b_2$ have $ \lim_{t \to + \infty} d(\gamma_1(t),\gamma_2(t)) \to 0$, so they define the same boundary point $\tau = i \infty$. On the other hand, $ \lim_{t \to - \infty} d(\gamma_1(t),\gamma_2(t)) \to + \infty$, corresponding to two distinct boundary points $\tau = b_1$ and $\tau=b_2$. The semi-circular geodesics on the other hand connect two points on the $\tau_2=0$ line in the infinite affine parameter limit. So it appears that the boundary of the symmetric space includes one point at $i\infty$ and all points $b\in \mathbb{R}$. Hence, we have 
\begin{equation}
\label{H2-bdry}
    \partial \mathbb{H}^2 = \mathbb{R} \cup \{ \infty \} \cong S^1 \, ,
\end{equation}
which is indeed diffeomorphic to $\cpp_1$. 

To see this equivalently from the Cartan decomposition, we note that $\overline{\caa^+_1}$ contains a single element 
\begin{align}
    h=\begin{pmatrix}
 1 & 0 \\
 0 & -1 \\
 \end{pmatrix} \, ,
\end{align}
and so from~\eqref{eq:SL2cartan} the geodesics through $i\in\mathbb{H}^2$ have the form 
\begin{align}
    \gamma(t) = \frac{i-e^{2t}\cot{\theta}}{e^{2t}+i\cot{\theta}} \,  ,
\end{align}
and are illustrated on the right of Figure \ref{fig:straightgeod}. 
Taking the limit we have 
\begin{align}
    \lim_{t\to\infty} \gamma(t)  
       = \begin{cases} 
        -\cot\theta & \text{ if $0<\theta < \pi$} \\
        i\infty & \text{ if $\theta = 0$} 
    \end{cases} \, , 
\end{align}
confirming~\eqref{H2-bdry}. The centraliser of $h$ is $\mathbb{Z}_2$ (the rotation by $\pi$) and hence the preimage of the projection map $\pi:\partial\cS\to\overline{\caa^+_1}$ is $\pi^{-1}(h)=\partial\cS=SO(2)/\mathbf{Z}_2\cong S^1$. In particular, we see that, for $\theta\neq 0$, 
\begin{align}
\label{eq:bdry-SO2}
    k^{-1}_{\theta}\cdot i\infty = \frac{i\cos\theta}{i\sin\theta} = \cot\theta\, ,
\end{align}
or $0\leq\theta\leq\pi$ so that the boundary points are indeed all related by the action of $SO(2)/\mathbb{Z}_2$ where the $\mathbb{Z}_2$ action sends $\theta\to\theta+\pi$. 

Another simple example is the boundary of $\cS\cong SO(1,3)/SO(3)$. The Cartan generator and roots are given in \eqref{CartansSO13}--\eqref{rootsSO13}. Here, $\overline{\caa_{1}^+}=h_1$, is once again a point. The centraliser of $h_1$ is again non-trivial, $M_h=SO(2)\subset SO(3)$ and hence $\pi^{-1}(h)=\partial\cS=SO(3)/SO(2)\cong S^2$ is a 2-sphere, as expected.

\subsection{Boundaries and parabolic subgroups} \label{sec:parabsub}

A very useful refinement of the description of the boundary of symmetric spaces is through the notion of \textit{parabolic subgroups} of $G$. In particular, each point on the boundary, as equivalence classes of geodesics, is fixed under the action of a parabolic subgroup of $G$. For example, for $SL(2,\mathbb{R})/SO(2)$, we can easily check that the elements of the form
\begin{equation}
\label{eq:parabolicsl2}
 p=   \begin{pmatrix}
 a & b \\
 0 & a^{-1} \\
 \end{pmatrix} \, , 
\end{equation}
for $a, \, b \in \mathbb{R}$ form a parabolic subgroup $P_\infty\subset SL(2, \mathbb{R})$ that leaves the point $\tau = i \infty$ fixed. Given the other points in $\partial\mathbb{H}^2$ are related to $\tau = i \infty$ by the action of $K=SO(2)$, as in~\eqref{eq:bdry-SO2}, we see that each point $\tau=-\cot\theta$ on the boundary can be associated to the action of a different parabolic group given by $P_\theta=k_\theta P_\infty k^{-1}_\theta\subset SL(2, \mathbb{R})$.  

To see how this works more generally we start by introducing  parabolic subgroups. Consider a reductive real Lie algebra $\cgg$ with a choice of Cartan subalgebra $\caa$ and positive and negative roots $\Phi^\pm$. The latter define two nilpotent subalgebras
\begin{align}
    \cnn^+=\sum_{\alpha\in\Phi^+}\cgg_{\alpha}\, , \quad \cnn^-=\sum_{\alpha\in\Phi^+}\cgg_{-\alpha}=\sum_{\alpha\in\Phi^-}\cgg_{\alpha} \, ,
\end{align}
with  corresponding unipotent groups $N^+=\ee^{\cnn^+}$ and $N^-=\ee^{\cnn^-}$. The remaining part of $\cgg$ is the centraliser of $\caa$, i.e.~the subalgebra that commutes with $\caa$. It has the form 
\begin{equation}
    \mathfrak{z}(\caa) = \cmm \oplus \caa \, ,
\end{equation}
where $\cmm=\mathfrak{z}(\caa)\cap\ckk$ and hence we have the full decomposition
\begin{align}
    \cgg = \caa \oplus \cmm \oplus \cnn^+ \oplus \cnn^- \, . 
\end{align}
(Note that in the split case $\cmm=0$). The {\it minimal standard parabolic subgroup} $P_\emptyset\subset G$ is defined as the subgroup of $G$ whose algebra has the form 
\begin{align}
    \cpp_\emptyset = \caa \oplus \cmm \oplus \cnn^+ \, . 
\end{align}
To get the global structure correct one then defines the corresponding parabolic group $P_\emptyset\subset G$ as the normaliser of $\cpp_\emptyset$ in $\cgg$, that is the set of elements $g\in G$ such that $\Ad_g p\in\cpp_\emptyset$ for all $p\in\cpp_\emptyset$. One furthermore has the Langlands decomposition 
\begin{equation} \label{MAN}
P_\emptyset=N^+AM \, ,     
\end{equation}
where $A=\exp(\caa)$ is the Cartan subgroup in $G$ and $M$ is the centraliser of $A$ in $K$. This in turn leads to the standard \emph{horospherical coordinates} on $\cS$
\begin{equation}
\label{P0horosperical}
    \cS \cong N^+ A \cdot o\, ,
\end{equation}
which we will regularly use in the following. 

One can define other larger \emph{standard parabolic subgroups} as follows. Let $\Delta$ be the set of fundamental roots and choose a subset $I\subseteq \Delta$. Then if $\Phi^-_I\subset\Phi^-$ is the subset of negative roots generated by $I$, the standard parabolic Lie algebra is given by 
\begin{align}
    \cpp_I = \cpp_\emptyset \oplus \sum_{\alpha\in\Phi^-_I} \cgg_\alpha \, , 
\end{align}
and again the group $P_I$ is defined as the normaliser of $\cpp_I$. Note that if $I=\emptyset$ then $p_I$ is the minimal parabolic, while if $I=\Delta$ then $\cpp_I=\cgg$. The parabolic group also picks out a natural subalgebra of the Cartan subalgebra $\caa$ defined by
\begin{align}
    \caa_I = \left\{ h\in\caa : \alpha(h) = 0 , \forall \alpha\in I \right\} \, ,
\end{align}
in other words elements of $\caa_I$ commute with the root spaces $g_\alpha$ for all $\alpha\in I$. One can then define a corresponding positive Weyl chamber $\caa^+_I$. For each $I$, the closure $\overline{\caa_I^+}$ is a different face of the closure of the positive Weyl chamber $\overline{\caa^+}$ of the full Cartan subalgebra. For completeness, note that decomposing a parabolic algebra as 
\begin{align}
    \cpp_I = \caa_I \oplus \cmm_I \oplus \cnn_I \, ,
\end{align}
where $\cnn_I\subset \cnn^+$ is the sum over the positive root spaces not in $\Phi^+_I$, again gives a Langlands decomposition of the parabolic group 
\begin{align}
\label{eq:LanglangsI}
    P_I = N_I A_I M_I \, , 
\end{align}
where $A=\exp(\caa_I)$ is the Cartan subgroup, $N_I=\ee^{\cnn_I}$ is the unipotent component, $M_I$ has Lie algebra $\cmm_I$ and $A_IM_I$ is the centraliser of $A_I$ in $G$. The corresponding horospherical coordinates give 
\begin{align}
\label{eq:horospherical}
    \mathcal{S} \cong N_I A_I S_I \, , 
\end{align}
where now the groups act on a subspace $S_I=M_I/K_I$ (where $K_I=M_I\cap K$) rather than just the special point $o$ stabilised by $K$. 

To see how the standard parabolic groups appear in a particular example, consider $\mathfrak{sl}(k,\mathbb{R})$ and take $\caa$ to be diagonal traceless matrices. $\cpp_\emptyset$ is then the set of upper-triangular matrices, while $\cpp_I$ adds in elements in the lower triangle such that the extra elements generate one or more $\mathfrak{sl}(p,\mathbb{R})\subset\mathfrak{sl}(k,\mathbb{R})$ subalgebra. Concretely for $\mathfrak{sl}(4,\mathbb{R})$, with the natural ordering of the roots, we have
\begin{equation}
\begin{aligned}
    \cpp_\emptyset &\ni \left(\begin{smallmatrix}
        * & * & * & * \\  & * & * & * \\ && * & * \\ &&& * 
    \end{smallmatrix}\right) , &
    \cpp_{\{\alpha_1\}} &\ni \left(\begin{smallmatrix}
        * & * & * & * \\  * & * & * & * \\ && * & * \\ &&& * 
    \end{smallmatrix}\right) , &
    \cpp_{\{\alpha_2\}} &\ni \left(\begin{smallmatrix}
        * & * & * & * \\  & * & * & * \\ & * & * & * \\ &&& * 
    \end{smallmatrix}\right) , &
    \cpp_{\{\alpha_3\}} &\ni \left(\begin{smallmatrix}
        * & * & * & * \\  & * & * & * \\ & & * & * \\ && * & * 
    \end{smallmatrix}\right) , \\
    \cpp_{\{\alpha_1,\alpha_2\}} &\ni \left(\begin{smallmatrix}
        * & * & * & * \\  * & * & * & * \\ * & * & * & * \\ &&& * 
    \end{smallmatrix}\right) , &
    \cpp_{\{\alpha_1,\alpha_3\}} &\ni \left(\begin{smallmatrix}
        * & * & * & * \\  * & * & * & * \\ & & * & * \\ && * & * 
    \end{smallmatrix}\right) , &
    \cpp_{\{\alpha_2, \alpha_3\}} &\ni \left(\begin{smallmatrix}
        * & * & * & * \\  & * & * & * \\ & * & * & * \\ & * & * & * 
    \end{smallmatrix}\right) , &
    \cpp_\Delta &\ni \left(\begin{smallmatrix}
        * & * & * & * \\  *  & * & * & * \\ * & * & * & * \\ * & * & * & * 
    \end{smallmatrix}\right) .  
\end{aligned}    
\end{equation}

Crucially, any parabolic subgroup $P\subset G$ is conjugate, under the action of $K$ (and hence also $G$), to a unique standard parabolic subgroup $P_I$, that is $P=kP_Ik^{-1}$ for some $k\in K$. One then refers to $I$ as the type of the parabolic group $P$. In particular, one can also define the corresponding ``split component'' of the Cartan subalgebra as $\caa_P = \Ad_k \caa_I$, and similarly $N_P=kN_Ik^{-1}$, $M_P=kM_Ik^{-1}$, etc. in the Langlands decomposition~\eqref{eq:LanglangsI} and in the definition of the horospherical coordinates~\eqref{eq:horospherical}. 

The relation to the boundary $\partial\cS$ is then as follows. One can show that the different geodesics in an equivalence class of asymptotic geodesics are all related by the action of some (proper) parabolic group $P$ (proper meaning $P\neq G$). In particular, we can write each member of the class as
\begin{align}
\label{eq:gen-geo-class}
    \gamma(t) = n\, a\,\ee^{tH}z \, , 
\end{align}
where $n\in N_P$, $a\in A_P$ and $z\in S_P$ are horospherical coordinates parameterising the different elements of the class, and $H\in\caa^+_{P,1}$ where $\caa^+_{P,1}$ is the space of unit length vectors in $\caa^+_P$. Changing $H$ changes the equivalence class, so that one can also define a canonical geodesic for each class
\begin{align}
    \gamma_H(t) = \ee^{tH}\cdot o \, . 
\end{align}
Thus, points on the boundary correspond to specifying the pair $(P,H)$ where $P$ is a proper parabolic subgroup and $H\in\caa^+_{P,1}$. More formally we have a disjoint decomposition 
\begin{align}
\label{tits-building}
    \partial\cS = \coprod_P \caa^+_{P,1} \, .  
\end{align}
This structure is also known as the ``Tits building'' of the group $G$. 

\subsubsection*{Examples}

In the case of $SL(2)$ there is a single root, and thus a single class of parabolic subgroup, the minimal one. We have (see \eqref{eq:slalg})
\begin{align}
    \caa &= \spa_{\mathbb{R}}(\{ h\})\, , &
    \cnn^+ &= \spa_{\mathbb{R}}(\{e_1-e_2\})\, , &
    \cnn^- &= \spa_{\mathbb{R}}(\{e_1+e_2\})\, ,
\end{align}
with the root $\alpha=2$. The centraliser $M$ of $\caa$ is trivial, so the standard minimal parabolic subgroup is generated by $\mathfrak{a}$ and $\mathfrak{n}^+$, and is given by
\begin{align}
\label{eq:parabolicsplitsl2}
   P_{\emptyset} =  \ee^{\cnn^+}\ee^{\caa}\, \ni \begin{pmatrix}
       1 & n\\
       0 & 1
   \end{pmatrix}\begin{pmatrix}
       \ee^t & 0\\
       0 & \ee^{-t}
   \end{pmatrix} = \begin{pmatrix}
       \ee^t & \ee^{-t} n\\
       0 & \ee^{-t}
   \end{pmatrix} \, ,
\end{align}
and we see that $P_{\emptyset}$ is the parabolic group $P_\infty$ in~\eqref{eq:parabolicsl2} that fixes the point $\tau = i \infty$. Note that the corresponding horospherical coordinates are 
\begin{align}
    \label{horospherical}
    \tau = \begin{pmatrix}
       \ee^t & \ee^{-t} n\\
       0 & \ee^{-t}
   \end{pmatrix} \cdot i = n + i \ee^{2t} \, , 
\end{align}
which indeed cover $\mathbb{H}^2$. All other parabolic groups are of the form 
\begin{equation}
    P_{\theta} = k_\theta  P_\emptyset  k_\theta^{-1} \, , 
\end{equation}
for $0<\theta<\pi$ and fix the point $\tau=-\cot\theta$. We have 
\begin{align}
    \caa^+_{P_{\theta},1}=\{ \Ad_{k_\theta}h \} = \left\{ \begin{pmatrix}
       \cos2\theta & -\sin2\theta \\
       \sin2\theta & \cos2\theta
   \end{pmatrix} \right\}
\end{align}
is a single point and so from~\eqref{tits-building}
\begin{align}
    \partial\cS = \coprod_P \caa^+_{P,1} 
        = \coprod_{0\leq\theta<\pi} \begin{pmatrix}
       \cos2\theta & -\sin2\theta \\
       \sin2\theta & \cos2\theta
   \end{pmatrix}      
        = S^1 \, ,  
\end{align}
as expected. 

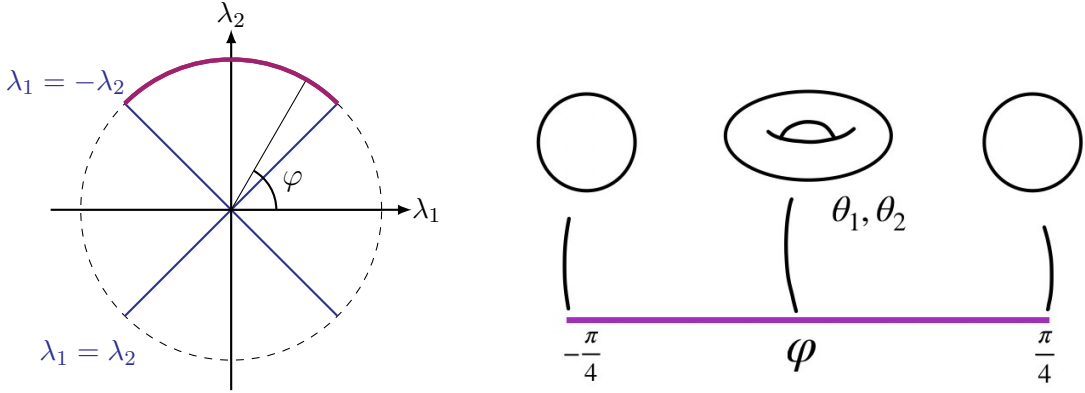
\begin{figure}[t]
\begin{center}	
\subfigure{\begin{tikzpicture}[scale=2]
  \draw[dashed] (0,0) circle(1);
  \foreach \angle in {45, 225} {
    \draw[ thick,Blue] (0,0) -- ({cos(\angle)}, {sin(\angle)});}
    \foreach \angle in {135, 315} {
    \draw[ thick,Blue] (0,0) -- ({cos(\angle)}, {sin(\angle)});}
    \foreach \angle in {0, 90, 180, 270} {
    \draw[] (0,0) -- ({cos(\angle)}, {sin(\angle)});}
  \draw[ultra thick, RedViolet] (135:1) arc (135:45:1);
\draw [thick,-latex] (-1.2,0) -- (1.2,0);
\draw [thick,-latex] (0,-1.2) -- (0,1.2);
  \node at (1.3,0) {\small$\lambda_1$};
  \node at (0,1.3) {\small$\lambda_2$};
  \node[Blue] at (-1.1,0.85) {\small$\lambda_1 = -\lambda_2$};
  \node[Blue] at (-0.95,-0.95) {\small$\lambda_1 = \lambda_2$};
  \draw[thick] (0:0.3) arc (0:60:0.3);
  \node at (25:0.45) {$\varphi$};
 \foreach \angle in {0, 60} {
    \draw[] (0,0) -- ({cos(\angle)}, {sin(\angle)});}
  \draw[ultra thick, RedViolet] (135:1) arc (135:45:1);
\end{tikzpicture}}\quad\quad
\subfigure{\includegraphics[height=120pt]{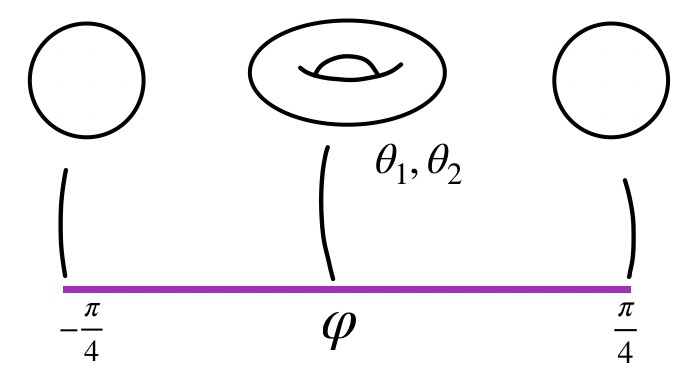}}
					\caption{Left: in purple, one choice of $\overline{\mathfrak{a}_1^+}$ for the Cartan subalgebra of $SO(2,2)$. Right: the structure of the boundary of $\cS=SO(2,2)/SO(2)\times SO(2)$ under the projection map $\pi\to\overline{\mathfrak{a}_1^+}$. }			
		    \label{fig:o22}
	\end{center}
\end{figure}

For a slightly more involved example, let us consider the group $SO(2,2)$. We have three proper standard parabolic algebras 
\begin{align}
    \cpp_\emptyset &= \caa \oplus \cgg_{\alpha_1} \oplus \cgg_{\alpha_2} , &
    \cpp_{\{\alpha_1\}} &= \caa \oplus \cgg_{\alpha_1} \oplus \cgg_{-\alpha_1} \oplus \cgg_{\alpha_2} , &
    \cpp_{\{\alpha_2\}} &= \caa \oplus \cgg_{\alpha_1} \oplus \cgg_{\alpha_2} \oplus \cgg_{-\alpha_2}, 
\end{align}
where the root spaces are given in~\eqref{rootspacesSO22}. The corresponding split components are
\begin{equation}
\begin{aligned}
    \caa^+_\emptyset &= \left\{ \lambda_1h_1+\lambda_2h_2 : \lambda_2> |\lambda_1| , \lambda_1^2+\lambda_2^2 = 1 \right\} , \\
    \caa^+_{\{\alpha_1\}} &= \left\{ \tfrac{1}{\sqrt{2}}(h_2-h_1) \right\} , \\
    \caa^+_{\{\alpha_2\}} &= \left\{ \tfrac{1}{\sqrt{2}}(h_2+h_1) \right\} , 
\end{aligned}
\end{equation}
and we see that $\caa^+_{\{\alpha_1\}}$ and $\caa^+_{\{\alpha_2\}}$ lie at the boundary points of $\caa^+_\emptyset$ where $\lambda_2=\pm\lambda_1$. We note that the centralisers in $K=SO(2)\times SO(2)$ of $\caa^+_\emptyset$ and the two $\caa^+_{\{\alpha_i\}}$ and $\caa^+_{\{\alpha_2\}}$ are $\mathbb{Z}_2$ and $SO(2)$ respectively. Since $SO(2)\times SO(2)/\mathbb{Z}_2=T^2$, the boundary then splits into three pieces
\begin{equation}
\begin{aligned}
    \partial\cS &= \coprod_{\class{P}=\{\alpha_1\}} \caa^+_{P,1} \ \amalg \
        \coprod_{\class{P}=\emptyset} \caa^+_{P,1} \ \amalg \
        \coprod_{\class{P}=\{\alpha_2\}} \caa^+_{P,1} \\ 
    &= S^1 \times \{\varphi=-\tfrac14\pi\} \ \amalg \
        T^2\times \{-\tfrac14\pi < \varphi < \tfrac14 \pi\} \ \amalg \
        S^1 \times \{\varphi=\tfrac14\pi\} \\
    &= S^3 \, ,  
\end{aligned}
\end{equation}
where we have parametrised $\lambda_1=\sin\varphi$ and $\lambda_2=\cos\varphi$. This strucure is depicted in Figure \ref{fig:o22}. 

\section{Locally symmetric spaces and their boundaries}
\label{sec:discrete}
The symmetric spaces that appear as moduli spaces in string compactifications are generically only locally symmetric. That is, they have the form 
\begin{equation}
\label{eqn:modulispace}
    \mathcal{M} = \Gamma \backslash G/K \, ,
\end{equation}
where $\Gamma$ is a discrete subgroup of $G$, and in physical examples corresponds to some kind of duality group. They are still non-compact, and so have a boundary, but are argued to have finite volume\footnote{Or more refinedly, the argument is that finiteness of quantum gravity amplitudes requires moduli spaces of finite volume, or at worst whose volume grows no faster than that of flat space \cite{Delgado:2024skw}. As we show in section~\ref{sec:weight-polytopes} for locally symmetric space the latter condition implies finite volume for semi-simple $G$.}. In fact, any finite-volume locally symmetric space takes the form~\eqref{eqn:modulispace}, in which case $\Gamma$ is called a lattice in $G$ and is called non-uniform if $\mathcal{M}$ is non-compact. A famous theorem of Margulis~\cite{Margulis} states that if $G$ is a linear semi-simple group of real rank at least two then any lattice $\Gamma$ is necessarily a particular kind of discrete group known as an \emph{arithmetic group}\footnote{In fact this result was extended by Corlette~\cite{Corlette} to include the rank-one groups $Sp(\dt,1)$ and $F_{4(-20)}$ leaving only $G=SO(\dt,1)$ and $G=SU(\dt,1)$ as the excluded cases.}, a notion that we will define below. 

Remarkably, we can then still understand the geodesics and boundary of $\mathcal{M}$ using parabolic subgroups, but now the parabolic group has to be \emph{rational}. As we will see in the later sections, it is this rationality that is key to understanding the SDC. It is the appearance of these rational groups that explains, for example, why in~\cite{Keurentjes:2006cw,Israel:2025ouq} the small radius limit of the nine-dimensional heterotic string is a decompactification if and only if the Wilson line is rational, and is the origin of, among others, the conjecture about the denseness of the boundary geodesics in~\cite{Etheredge:2023usk}.

\subsection{Discrete subgroups and how to choose them}
\label{sec:funddom}

The reductive groups $G$ that define $\cS=G/K$ are all real linear algebraic groups. That means they can be defined as subgroups $G\subset GL(\dt,\mathbb{R})$ satisfying a set of equations polynomial in the matrix elements and with real coefficients, for example, $\det M$=1 for $M \in SL(\dt,{\mathbb{R}})$, or $M^TM = \mathbb{1}$ for $ M \in O(\dt)$. To define an \emph{arithmetic group}, we require that the defining polynomials must have rational coefficients, noting that this will of course be possible in many different ways, defining equivalent reductive groups $G$. One then defines the corresponding \emph{rational group} $G(\mathbb{Q})$ and  \emph{arithmetic group} $\Gamma$ as the sets of matrices with rational or integer entries, that is, 
\begin{align}
\label{eq:GQ-Gamma}
    G(\mathbb{Q}) = G \cap GL(\dt,{\mathbb Q}) , \qquad
    \Gamma = G \cap GL(\dt,{\mathbb Z}) ,
\end{align}
the classic example being the modular group $SL(2,\mathbb{Z})\subset SL(2,\mathbb{R})$. In general, different sets of rational polynomials that define isomorphic $G(\mathbb{Q})$ groups will define non-isomorphic arithmetic groups $\Gamma$ as we see in the example below. In fact there are an infinite number of such distinct arithmetic subgroups. Rather than changing the polynomials, one can start with a fixed $G(\mathbb{Q})$ and a corresponding fixed arithmetic group $\Gamma$. A second discrete subgroup $\Gamma' \subset G(\mathbb{Q})$ is arithmetic if it is commensurate to $\Gamma$, that is, if $\Gamma /(\Gamma \cap \Gamma')$ and $\Gamma' / (\Gamma \cap \Gamma')$ are finite. 

As is familiar from $SL(2,\mathbb{Z})\subset SL(2,\mathbb{R})$, quotienting a symmetric space $\cS$ by an arithmetic subgroup divides it into fundamental domains. These are the largest subsets $\mathcal{F}\subset\cS$ such that no two points in $\mathcal{F}$ are related by the action of $\Gamma$.\footnote{From the string theory point of view, the closure of a fundamental domain determines the region on the coset space that represents the physical moduli space.} Or more precisely, the orbit $\Gamma\cdot x$ of any point $x\in\cS$ under the action of $\Gamma$ contains at least one point in the closure $\overline{\mathcal{F}}$ and no two interior points lie in the same orbit. The boundary of $\mathcal{M}=\Gamma\backslash\cS$ is thus closely related to the boundary of the fundamental domain. As we will discuss, this is in turn related to a choice of rational parabolic group. 

\subsubsection*{Example}
Consider $SL(2,\mathbb{R})/SO(2)$ with two different arithmetic subgroups $\Gamma_1$ and $\Gamma_2$ generated by
\begin{equation}
\label{Gammas}
\begin{aligned}
    \Gamma_1&: \quad T_1=\begin{pmatrix}
        1 & 1\\
        0 & 1 \\
    \end{pmatrix} \text{ and }  S=\begin{pmatrix}
        0 & 1\\
        -1 & 0 \\
    \end{pmatrix} \, , \\
    \Gamma_2&: \quad T_2 = \begin{pmatrix}
        1 & 2 \\
        0 & 1 \\
    \end{pmatrix} \text{ and }  S=\begin{pmatrix}
        0 & 1\\
        -1 & 0 \\
    \end{pmatrix} \, .
\end{aligned}   
\end{equation}
The former corresponds to the usual version of $SL(2,{\mathbb Z})$ as defined by
\begin{equation} \label{SL2Z}
    \Gamma_1 = SL(2,\mathbb{Z}) = \Bigg{\{} \begin{pmatrix}
        a & b\\
        c & d \\
    \end{pmatrix} , \,  a,b,c,d \in \mathbb{Z}, \, ad-bc=1  \Bigg{\}} \,.
\end{equation}
The latter includes a subset of these matrices where $a,d$ are even (and hence $b,c$ are odd) or the other way around, that is 
\begin{equation} 
    \Gamma_2 = \Bigg{\{} \begin{pmatrix}
        a & b\\
        c & d \\
    \end{pmatrix} \in SL(2,\mathbb{Z}) :  \text{$a,d$ even or $b,c$ even}\Bigg{\}} \,.
\end{equation}
A choice of fundamental domain $\mathcal{F}$ for the two quotients are shown in Figure \ref{fig:fdsl2}, in blue and violet respectively, where 
\begin{equation}
\label{fund-doms}
\begin{aligned}
    \mathcal{F}_{1} &= \{ -\tfrac12 <\tau_1 < \tfrac12, |\tau| > 1 \} \, , \\
    \mathcal{F}_{2} &= \{ 0 < \tau_1 < 2 , |\tau| > 1, |\tau-2| > 1 \} \, . 
\end{aligned}
\end{equation}
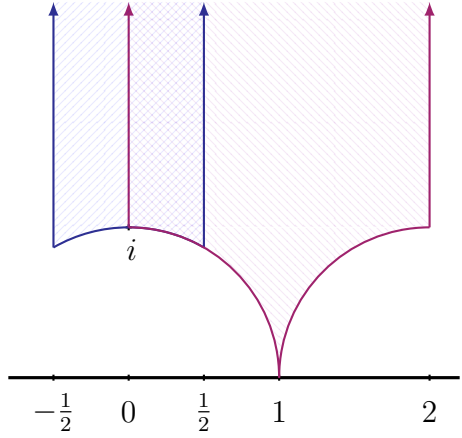
\begin{figure}[t]
\begin{center}
\begin{tikzpicture}[scale=2]
  \def\xmin{-0.8} \def\xmax{2.2}
  \def\ymin{-0.1} \def\ymax{2.5}
  \def\shift{{sqrt(3)/2}}
  \draw [very thick] (\xmin,0) -- (\xmax,0);
  \draw [thick] (-0.5,-0.02) -- (-0.5,0.02) node [below=0.1] {$-\frac{1}{2}$};
  \draw [thick] (0.5,-0.02) -- (0.5,0.02) node [below=0.1] {$\frac{1}{2}$};
  \draw [thick] (-0.02,1) -- (0.02,1) node [below] {$i$};
  \draw [thick] (1,-0.02) -- (1,0.02) node [below=0.2] {$1$};
  \draw [thick] (0,-0.02) -- (0,0.02) node [below=0.2] {$0$};
  \draw [thick] (2,-0.02) -- (2,0.02) node [below=0.2] {$2$};
  \draw [thick] (0,0.98) -- (0,1.02); 
  \draw [thick,Blue,-latex] (-0.5,\shift) -- (-0.5,\ymax);
  \draw [thick,Blue,-latex] (0.5,\shift) -- (0.5,\ymax);
  \draw [thick,Blue] (0.5,\shift) arc (60:120:1);
  \draw [thick,RedViolet,-latex] (2,1) -- (2,\ymax);
  \draw [thick,RedViolet,-latex] (0,1) -- (0,\ymax);
  \draw [thick,RedViolet] (2,1) arc (90:180:1);
  \draw [thick,RedViolet] (1,0) arc (0:90:1);
\path [pattern=north east lines, pattern color=blue, opacity=0.3](-0.5,\shift) -- (-0.5,\ymax)-- (0.5,\ymax) -- (0.5,\shift)  arc (60:120:1)    -- cycle;
\path [pattern=north west lines, pattern color=RedViolet, opacity=0.3](2,1) -- (2,\ymax)-- (0,\ymax) -- (0,1) --  (2,1) arc (90:180:1) --  (1,0) arc (0:90:1)  -- cycle;
\end{tikzpicture}
\caption{Fundamental domains for $\Gamma_1 \backslash \mathbb{H}^2$ in blue and $\Gamma_2 \backslash \mathbb{H}^2$ in violet. }			
\label{fig:fdsl2}
\end{center}
\end{figure}
The first fundamental domain has only one point at infinite distance $\tau=i\infty$, while the second has two, $\tau=i\infty$ and $\tau=1$. Since the boundary of the space before the quotient by the arithmetic group is $\tau=i \infty$ together with the real axis $\tau\in \mathbb{R}$, we see that the boundary of $\Gamma_1\backslash\mathbb{H}^2$ is a single point, whereas the boundary of $\Gamma_2\backslash\mathbb{H}^2$ is two distinct points. 

To see how the two arithmetic groups can be viewed as defined by different polynomials we use the isomorphism $SL(2,\mathbb{R})\cong SO(1,2)$. Let $g\in SL(2,\mathbb{R})$ and $u$ be the symmetric matrix 
\begin{equation}
   u = \begin{pmatrix}
        x & z \\
        z & y  \\
           \end{pmatrix} \, .  
\end{equation}
Under the isomorphism, $g$ defines an element $r\in SO(1,2)$ via\footnote{Strictly this is a double cover $SL(2,\mathbb{R})\to SO(1,2)$ but since the central element $-\mathbb{1}$ is in $K=SO(2)$ it is an isomorphism of their actions on $\cS$.}
\begin{equation}
    v' = r\cdot v \Leftrightarrow u'=gug^T \, ,
\end{equation}
where $r\in SL(3,\mathbb{R})$ acts linearly on the vector $v=(x,y,z)\in \mathbb{R}^3$ and, since $\det u'=\det gug^T = \det u$, preserves the $SO(1,2)$ metric 
\begin{align}
    \eta_1(v,v) = \det u = xy - z^2 \, .
\end{align}
$\Gamma_1$ is then the subgroup of $GL(3,\mathbb{Z})$ preserving $\eta_1$. Alternatively we could define $SO(1,2)$ as preserving the metric
\begin{align}
    \eta_2(\tilde{v},\tilde{v}) = \det \tilde{u} = \tilde{x}\tilde{y} - 4\tilde{z}^2 \, ,
\end{align}
in which case the isomorphism would be defined by 
\begin{equation}
   \tilde{u} = \begin{pmatrix}
        \tilde{x} & 2\tilde{z} \\
        2\tilde{z} & \tilde{y}  \\
           \end{pmatrix} \ ,  
\end{equation}
and one finds that $\Gamma_2$ is the subgroup of $GL(3,\mathbb{Z})$ preserving $\eta_2$. In particular, the transformation $T_1$ on $\tilde{v}$ sends $\tilde{z}\to \tilde{z}+\tfrac12 \tilde{y}$, and so is not an element of $GL(3,\mathbb{Z})$. On the other hand, $T_2$ sends $\tilde{z}\to \tilde{z}+\tilde{y}$ and is in the arithmetic group. We note that the two arithmetic groups are not equivalent because there is no $GL(3,\mathbb{Z})$ transformation that takes $\eta_1$ to $\eta_2$, although there is such a $GL(3,\mathbb{Q})$ transformation and so the corresponding rational groups are equivalent. Alternatively, the two arithmetic groups can be defined using different lattices with a fixed definition of the rational group. For example, one can take the metric $\eta_1$ to define the $SO(2,1)$ group acting on the vector $(\tilde{x},\tilde{y},2\tilde{z})\in\mathbb{R}^3$ and define $\Gamma_2$ as the arithmetic group that preserves the lattice with $\tilde{x},\tilde{y},\tilde{z} \in\mathbb{Z}$. Or even, as the subgroup $\Gamma_2 \subset GL(2,\mathbb{Z})$ acting on the vector $(a+b,a-b)\in\mathbb{R}^2$ and preserving the lattice $a,b\in\mathbb{Z}$. 

\subsubsection*{Chevalley arithmetic subgroups}

For generic semi-simple groups $G$ there is not a canonical or natural way to choose the arithmetic group, and so no natural definition of $\mathcal{M}$. The exception is when $G$ is split in which case one can define the so-called \emph{Chevalley arithmetic subgroup}, and it turns out that it is exactly these groups that appear in the moduli space of toroidally compactified bosonic and type II string theory and M-theory. 

The groups are constructed as follows~\cite{steinberg}. Given a split simple real Lie algebra $\cgg$ one can chose a \emph{Chevalley basis} of generators $\{h_i\}$ of the Cartan subalgebra $\caa$ and $e_\alpha$ of the root spaces $\cgg_\alpha$ such that 
\begin{equation}
    \begin{aligned}
        [h_i, h_j] &= 0 , &
        [h_i,e_\alpha ] &= n^j A_{ji} e_\alpha , \\
        [e_\alpha,e_{-\alpha}] &= k^i_\alpha h_i , &
        [e_\alpha,e_\beta] &= 
            \begin{cases} \pm(p+1)e_{\alpha+\beta} & \text{if $\alpha+\beta\in\Phi$} \\ 0 & \text{otherwise} \end{cases}, 
    \end{aligned}
\end{equation}
where we write $\alpha$ as a sum of fundamental roots $\alpha=n^i\alpha_i$, $A_{ij}$ is the Cartan matrix, $k^i_\alpha\in\mathbb{Z}$ and $p$ is the largest integer such that $\beta-n\alpha\in\Phi$ for all $0\leq n\leq p$ but not for $n=p+1$. The first key point is that all the structure constants are integers. 

Now consider some faithful $p$-dimensional real representation $\rho$ of $\cgg$. The second remarkable fact is that the matrices $\ee^{\lambda\rho(e_\alpha)}\subset GL(p,\mathbb{R})$ are integer valued for any $e_\alpha$ and $\lambda\in\mathbb{Z}$. The Chevalley group $G_\rho(\mathbb{Z})$ is then defined as the subgroup $G_{\rho}\subset GL(p,\mathbb{Z})$ generated by the $\ee^{\lambda\rho(e_\alpha)}$.\footnote{In fact, taking $\lambda\in F$ defines a Chevalley group for any field $F$. In particular, $G_\rho(\mathbb{Z})\subset G_\rho(\mathbb{Q})\subset G_\rho(\mathbb{R})$. Depending on the representation, $G_\rho(\mathbb{R})$ can take a different global form, for example $SO(\dt,\dt)$ versus $Spin(\dt,\dt)$.} The form of $G_\rho(\mathbb{Z})$ does not depend much on the choice of representation. There are two extreme cases: the adjoint group $G_0(\mathbb{Z})$ and the universal group $G_1(\mathbb{Z})$, corresponding to the adjoint representation and the sum of the fundamental representations respectively. The lattices in $\caa^*$ generated by the sets of weights of these representations are the root lattice $L_0$ and the weight lattice $L_1$. Any other faithful representation will correspond to a lattice $L$ contained between these two extremes $L_0 \subseteq L \subseteq L_1$ and representations $\rho$ with the same $L$ define the same Chevalley group $G_\rho(\mathbb{Z})$.

For the classical Lie algebras one finds the universal Chevalley groups 
\begin{equation}
    SL(\dt,\mathbb{Z}), \quad  Spin(\dt,\dt;\mathbb{Z}), \quad
    Spin(\dt+1,\dt;\mathbb{Z}), \quad Sp(2k,\mathbb{Z})\, , 
\end{equation}
where, as algebraic groups, the orthogonal and symplectic groups preserve bilinears of the forms
\begin{equation}
\label{Chevalley-bilinears}
    \eta = \begin{pmatrix}
        \mathbb{0} & \mathbb{1} \\ \mathbb{1} & \mathbb{0} 
    \end{pmatrix} , \qquad
    \eta = \begin{pmatrix}
        1 & 0 & 0 \\
        0 & \mathbb{0} & \mathbb{1} \\ 
        0 & \mathbb{1} & \mathbb{0} 
    \end{pmatrix} , \qquad
    \Omega = \begin{pmatrix}
        \mathbb{0} & \mathbb{1} \\ - \mathbb{1} & \mathbb{0} 
    \end{pmatrix} \, ,  
\end{equation}
respectively. For $E_{6(6)}$ and $E_{7(7)}$ the universal Chevalley groups can be defined using the $\mathbf{27}$ and $\mathbf{56}$ representations respectively, while for $E_{8(8)}$, $F_{4(4)}$ and $G_{2(2)}$ the root and weight lattices agree and so every representation gives the same Chevalley group. In all cases, the adjoint Chevalley group is given by  $G_0(\mathbb{Z})=G_1(\mathbb{Z})/C$ where $C$ is the centre. For $SO(\dt,\dt)$ choosing the vector representation $\mathbf{2k}$ gives a lattice $L$ strictly between $L_0$ and $L_1$ and defines the Chevalley group $SO(\dt,\dt;\mathbb{Z})$ where the metric again takes the form in~\eqref{Chevalley-bilinears}. 

As mentioned, it is precisely Chevalley groups that appear as duality groups for toroidally compactified bosonic and type II string theory and M-theory. They appear very directly (see for example~\cite{Obers:1998fb}), the generating elements of the T- or U-duality taking the form $\ee^{\rho(e_\alpha)}$ acting on some space of charged states that transform in a particular representation, such as the winding and momentum states forming the $\mathbf{2k}$ vector representation of $SO(\dt,\dt)$. 

In Margulis's theorem the only requirement was that $G$ is semi-simple and linear, and so it allows for groups $G$ with various different global structures for a given Lie algebra. On the other hand the Chevalley groups (over $\mathbb{R}$) are of particular  types: for example for $\so(\dt,\dt)$ they can give $Spin(\dt,\dt)$, $SO(\dt,\dt)$ and $Spin(\dt,\dt)/C$ but not $O(\dt,\dt)$, although the latter is linear. Thus the bosonic string duality group $O(\dt,\dt;\mathbb{Z})$ is not strictly a Chevalley group. However we can view it as semi-direct product of the Chevalley group $SO(\dt,\dt;\mathbb{Z})$ with an outer automorphism of the Lie algebra. For split Lie algebras, the inner automorphisms of $\cgg$ are those that are elements of $G_0(\mathbb{Z})$ while the outer automorphisms correspond to symmetries of the Dynkin diagram. Thus for the T-duality symmetry $O(\dt,\dt;\mathbb{Z})$ of the bosonic string we take the semi-direct product with the $\mathbb{Z}_2$ outer automorphism that exchanges the spinor nodes of the Dynkin diagram
\begin{equation}
\label{eq:onn-dynk}
\dynkin[scale=2.5,involutions={[out=80,in=100,relative]56}]{D}{oo.oooo} . 
\end{equation}
The corresponding fundamental domain will then, of course, be a $\mathbb{Z}_2$ quotient of the one for $SO(\dt,\dt;\mathbb{Z})$.

\subsection{Geodesics, boundaries and rational parabolic groups}
\label{sec:geo}

We saw previously that for globally symmetric spaces we could associate the boundary of $\cS$ with the Tits building $\partial\cS = \coprod_P \caa^+_{P,1}$ associated to the proper parabolic subgroups $P\subset G$. For each point on the boundary, elements of the equivalence class of asymptotic geodesics were related by the action of a parabolic subgroup $P$, and the class had a canonical representative of the form $\gamma_H(t)=\ee^{tH}\cdot o$ where $H\in\caa_{P,1}^+$ is a unit length element of the split component of the Lie algebra of $P$. 

Remarkably, when $\Gamma$ is arithmetic and $\mathcal{M}$ is non-compact but of finite volume, the boundary of the locally symmetric space has the same kind of structure. One finds~\cite{Borel} 
\begin{align}
\label{eq:boundaryM}
    \partial\mathcal{M} = \Gamma \mathbin{\big\backslash} \coprod_{P(\mathbb{Q})} \caa^+_{P(\mathbb{Q}),1} \, ,
\end{align}
where now the disjoint union runs over the proper \emph{rational} parabolic subgroups $P(\mathbb{Q})$, and we also have to quotient by the action of $\Gamma$. Elements of the equivalence class of geodesics going to a particular boundary point are related by the  parabolic group $P$ that is the real locus of $P(\mathbb{Q})$, that is defined as an algebraic group over $\mathbb{R}$ rather than $\mathbb{Q}$. And the class again has a canonical representative of the form 
\begin{align}
\label{eq:rational-geodesic}
    \gamma_H(t)=\ee^{tH}\cdot o \, , 
\end{align}
where 
$H\in\caa_{P(\mathbb{Q}),1}^+$ is a unit length element of the split component of the  Lie algebra of the real locus of $P(\mathbb{Q})$. Note that different $(P(\mathbb{Q}),H)$ pairs can be related by an element of $\Gamma$ and so a given point on the boundary will be associated to a set of rational parabolics $P_\mathbb{Q}$. In fact (see Prop. III.2.19 of~\cite{Borel}) there are only a finite number of such equivalence classes. 

To unpack this construction a little bit, recall first that a rational algebraic group $G(\mathbb{Q})$ is defined by the intersection $G(\mathbb{Q})=G\cap GL(\dt,\mathbb{Q})$ as in~\eqref{eq:GQ-Gamma}. A rational parabolic subgroup $P(\mathbb{Q})$ is then one of the form $P\cap G(\mathbb{Q})$ where as an algebraic group $P$ is defined by polynomials with rational coefficients. Equivalently one requires $P\cap G(\mathbb{Q})$ to be dense in $P$ in the same way that the rational numbers are dense in the reals. Thus the real parabolic subgroups that define rational parabolic subgroups form a countably infinite subset of the continuous set of all parabolic subgroups $P\subset G$. The real locus of $P(\mathbb{Q})$ is thus just the parabolic group $P$ used to define $P(\mathbb{Q})$. 

The reason rational parabolics appear is because of the \emph{reduction theory} for arithmetic groups (see for example~\cite{Borel}). The idea is to find a description of the fundamental domain using bits of parabolic groups and goes roughly as follows. To identify a fundamental domain $\mathcal{F}$ we need to assign a unique point in $\mathcal{F}$ to each orbit of $\Gamma\cdot x$ of points $x\in\cS$. We start by recalling that, given any minimal parabolic group, the Langlands decomposition allows one to write $\cS$ using horospherical coordinates $\cS\cong N^+A\cdot o$. If we choose $P$ to be the real locus of a minimal\footnote{One subtlety is that the rank of $G$ over $\mathbb{Q}$ may be smaller than the rank over $\mathbb{R}$, that is the dimension of maximal sets of elements $X\in\cgg$ such that $\ad_X$ can be simultaneously diagonalised over the rationals and reals are not the same, and  $\rank_\mathbb{Q}\cgg \leq \rank_\mathbb{R}\cgg \leq \rank_\mathbb{C}\cgg$ with equality if the algebra is split. For the rational parabolics one then uses the root spaces defined by the smaller rational Cartan subalgebra, but the definitions all go through as in the real case.} rational parabolic $P(\mathbb{Q})$ then the arithmetic unipotent subgroup $N^+(\mathbb{Z})=N^+\cap \Gamma$ will be non-zero. We also know that $P$ defines a set of points on the boundary $\partial\cS$. To assign a unique point to an orbit, we first choose the point in the orbit closest to the boundary defined by $P$, that is, writing each point in horospherical coordinates $\gamma\cdot x= n\, \ee^H\cdot o$ we maximise the norm of $H$. This maximizing point is not unique since the maximum is unchanged by the action of $n\in N^+(\mathbb{Z})\subset\Gamma$. Thus we can choose the maximum point to lie in a region of the form of a \emph{Siegel set}
\begin{equation}
    S_{P(\mathbb{Q}),T} = \left\{ n\,\ee^{H+T}\cdot o : n\in U, H\in \caa^+_{P(\mathbb{Q})}, \right\}  \, , 
\end{equation}
for some fixed $T\in \caa^+_{P(\mathbb{Q})}$ and a subset of the identity $U\subset N$, chosen such that the $N^+(\mathbb{Z})$ translations of $U$ generate the whole of $N^+$, that is $N^+(\mathbb{Z})\times U\cong N^+$. Essentially, $S_{P(\mathbb{Q}),T}$ is describing the part of the fundamental domain that is at least as close as the point $\ee^T\cdot o$ to the part of the boundary defined by $P(\mathbb{Q})$. Repeating this for each rational parabolic, since there are only a finite number of equivalence classes of $P(\mathbb{Q})$ under $\Gamma$ one gets the \emph{precise reduction} for the fundamental domain as a disjoint union for some suitably chosen set of $T_i \in \caa^+_{P(\mathbb{Q})}$ 
\begin{align}
    \mathcal{F} \cong \Omega_0 \ \amalg \ \coprod_i S_{P_i,T_i} \, , 
\end{align}
where $i$ runs over the equivalence classes, $P_i$ is a representative in each class and $\Omega_0$ is a bounded set corresponding to the region of the fundamental domain away from all the boundaries. (Note that one needs to slightly enlarge the notion of Siegel set to include non-minimal parabolics in the decomposition.) 

In summary, each Siegel set can be thought of as describing a different ``cusp'' of the boundary so that we have a disjoint union 
\begin{align}
    \partial\mathcal{M} = \coprod_{i} \mathcal{B}_i \, , \qquad \mathcal{B}_i \cong \caa^+_{P_i,1} \, ,
\end{align}
where the way the different $\mathcal{B}_i$ pieces are glued together follows from~\eqref{eq:boundaryM}. Furthermore, for each cusp, we have canonical representatives of boundary geodesics
\begin{align}
    \gamma_{H}(t) = \ee^{tH}\cdot o \, , \qquad H\in \caa^+_{P_i,1} \, . 
\end{align}
Finally, each representative $P_i$ will be conjugate under $G(\mathbb{Q})$ to a standard rational parabolic subgroup $P_I(\mathbb{Q})$ defined using a fixed rational Cartan subalgebra $\caa$. 

For the special case of Chevalley groups, there is only one ``large'' cusp, corresponding to a single minimal standard parabolic subgroup $P_\emptyset$, together with smaller cusps corresponding to each of the other standard parabolics, that is we can identify each representative $P_i$ with a different $P_I$. Since each $\overline{\caa^+_I}$ corresponds to a different facet of the Weyl polytope $\overline{\caa^+}$ one has the simple result that 
\begin{align}
\label{eq:Chevalley-boundary}
    \partial\mathcal{M} \cong \overline{\caa^+_{1}} \, , \qquad \text{for Chevalley groups $\Gamma$} \, ,
\end{align}
where $\caa$ is a fixed rational Cartan subalgebra for $G(\mathbb{Q})$. 

It is natural to ask what a typical geodesic in $\mathcal{M}$ looks like. Recall that the space of geodesics radiating from a point $x$ is parametrised by a unit velocity vector in $T_x\mathcal{M}\cong\cpp$. In the globally symmetric space $\cS$ every geodesic reaches the boundary, asymptoting to a point stabilized by some parabolic group $P$. In contrast, since the space of rational parabolic groups $P(\mathbb{Q})$ is measure zero in the space of all parabolic groups, a typical geodesic in $\mathcal{M}$ must fail to reach the boundary. Instead it has an ergodic behaviour, gradually passing through more and more of the space. In fact, it is a famous result of Sullivan~\cite{sullivan} for $\Gamma\backslash\mathbb{H}^2$, then generalised to generic locally symmetric spaces in~\cite{Kleinbock-Margulis}, that the largest distance between a point on the geodesic $\gamma(t)$ and its starting point $\gamma(0)$ grows only logarithmically, that is 
\begin{align}
    \limsup_{t\to\infty} \frac{d(\gamma(t),\gamma(0))}{\log t} = \text{const.}\, .
\end{align}
The denseness of the rational parabolic subgroups $P(\mathbb{Q})$ in the set of real parabolic subgroup also proves that the geodesics that reach the boundary are dense in the set of all geodesics. This proves the ``Dense Direction Conjecture'' of~\cite{Etheredge:2023usk} for any local symmetric space, and hence for maximally supersymmetric theories in particular. 

\subsubsection*{Example}

Continuing with our standard example, consider $\cS=SL(2,\mathbb{R})/SO(2)=\mathbb{H}^2$ with the two different arithmetic subgroups $\Gamma_1$ and $\Gamma_2$ given in~\eqref{Gammas}. Note that the geodesics and boundary in the first case with $\Gamma_1=SL(2,\mathbb{Z})$ is very nicely described using the explicit action of the arithmetic group in~\cite{Keurentjes:2006cw}.

By construction, in each case we have the same rational group $G(\mathbb{Q}) = SL(2,\mathbb{Q})$. To characterise the boundary of $\Gamma_1\backslash\mathbb{H}^2$ and $\Gamma_2\backslash\mathbb{H}^2$ we first need to identify the set of rational parabolic groups $P(\mathbb{Q})$ and the points on the boundary $\partial\mathbb{H}^2$ that they stabilise. Since $\rank_\mathbb{Q}SL(2,\mathbb{Q})=1$ all the parabolic groups are minimal and each $\caa^+_{P(\mathbb{Q}),1}$ is a point. 

Explicitly, the standard minimal rational parabolic group is given by (cf Eq. \eqref{eq:parabolicsl2}) 
\begin{equation}
    P_\emptyset(\mathbb{Q}) = \left\{  \begin{pmatrix}
   a & b \\ 0 & a^{-1} \end{pmatrix} : a, b \in \mathbb{Q}, a\neq 0 \right\} \,  , 
\end{equation}
stabilising the boundary point $\tau=i\infty$. All other rational parabolics are conjugate to $P_\emptyset$ 
\begin{equation}
    P(\mathbb{Q}) = g P_\emptyset(\mathbb{Q}) g^{-1} \quad \text{for some $g\in SL(2,\mathbb{Q})$} \, . 
\end{equation}
We note that  
\begin{align}
    g\cdot i\infty = \begin{pmatrix}
   a & b \\ c & d \end{pmatrix} \cdot i\infty = a/c \in \mathbb{Q}\, , 
\end{align}
and there is clearly enough freedom in $g$ to set $a/c$ to be any rational. Hence the set of rational parabolic groups $P(\mathbb{Q})$ corresponds to the rational points on real $\tau$ axis together with $i\infty$, that is  
\begin{equation}
    \coprod_{P(\mathbb{Q})} \caa^+_{P(\mathbb{Q}),1} \cong \mathbb{Q} \cup \{i\infty\} \, . 
\end{equation}
To understand the boundary of $\partial\mathcal{M}$ we then need to see which of these points are equivalent under the action of the arithmetic group $\Gamma$. 

In the first case, acting on $\tau=i\infty$ with $g\in \Gamma_1=SL(2,\mathbb{Z})$ gives
\begin{align}
    g\cdot i\infty = a/c \quad \text{with $a,c\in\mathbb{Z}$} \, . 
\end{align}
For any coprime $a$ and $c$, one can always find\footnote{$b/d$ can be taken to be the larger neighbour to $a/c$ in any Farey sequence~\cite{hardy}.} $b,d\in\mathbb{Z}$ such that $ad-bc=1$. Hence we see that all the rational parabolic groups are in fact equivalent under the action of $\Gamma_1$ and we have a single equivalence class containing the whole of $\mathbb{Q} \cup \{i\infty\}$  and so indeed the boundary is a single point. Choosing $P_\emptyset$ as the representative of the class, we get 
\begin{align}
    \partial(\Gamma_1\backslash\mathbb{H}^2) \cong \{i\infty\} \, , 
\end{align}
just as we expected from the fundamental domain given in Figure~\ref{fig:fdsl2}. 

For $\Gamma_2$ we consider its action on $\tau=i\infty$ and $\tau=1$ noting that the values of $a$, $b$, $c$ and $d$ are now restricted
\begin{align}
    \left.\begin{aligned}
       g\cdot i\infty &= a/c \\
       g\cdot 1 & = \frac{a+b}{c+d}
    \end{aligned}\quad \right\}
       \quad \text{with $a,d$ even, $b,c$ odd or $a,d$ odd, $b,c$ even} \, . 
\end{align}
In particular, we see that $1$ is not in the same equivalence class as $i\infty$. Instead the rational Tits building decomposes into two equivalence classes
\begin{equation}
\begin{aligned}
    U_1: &\left\{ p/2q, 2p/q : \text{$p$, $q$ are odd} \right\} \cup \{ i\infty \} \, , \\  
    U_2: &\left\{ p/q : \text{$p$, $q$ are odd} \right\} \, .  
\end{aligned}
\end{equation}
One can represent the classes by the parabolic groups that stabilise $i\infty$ and $1$ respectively, 
\begin{equation}
\label{Gamma2-paras}
    P_1 = P_\emptyset \, , \qquad 
    P_2 = C P_\emptyset C^{-1} \quad \text{with} \quad C=\begin{pmatrix}
      1 & 0 \\ 1 & 1 \end{pmatrix} \, , 
\end{equation}
where $C$ is an $SL(2,\mathbb{Z})$ transformation that maps $\tau=i\infty$ to $\tau=1$ which, crucially, is not in $\Gamma_2$. This gives
\begin{align}
    \partial(\Gamma_2\backslash\mathbb{H}^2) \cong \{i\infty\} \cup \{1\} \, , 
\end{align}
again just as  expected from the form of the fundamental domain given in Figure~\ref{fig:fdsl2}. 

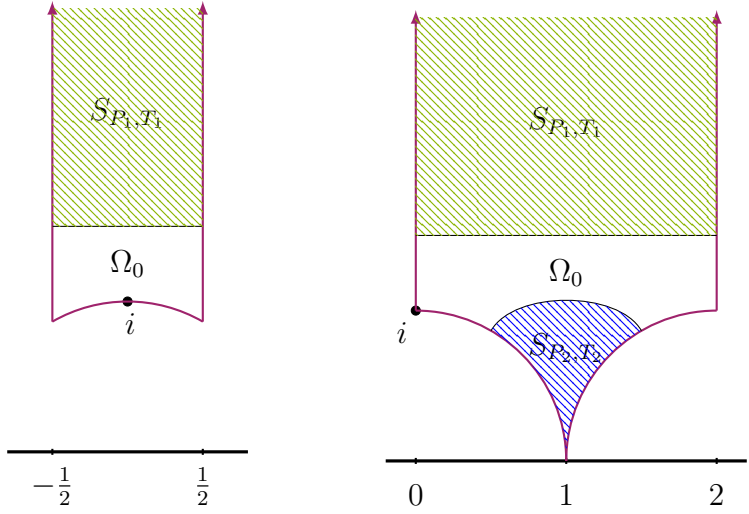
\begin{figure}[t]
\def\cola{RedViolet}
\def\colb{applegreen}
\def\colc{Blue}
	\begin{center}
\begin{tikzpicture}[scale=2]
  \def\xmin{-0.8} \def\xmax{0.8}
  \def\ymin{-0.1} \def\ymax{3}
  \def\shift{{sqrt(3)/2}}
  \draw [very thick] (\xmin,0) -- (\xmax,0);
  \draw [thick] (-0.5,-0.02) -- (-0.5,0.02) node [below=0.1] {$-\frac{1}{2}$};
  \draw [thick] (0.5,-0.02) -- (0.5,0.02) node [below=0.1] {$\frac{1}{2}$};
  \draw [very thick] (-0.02,1) -- (0.02,1) node [below] {$i$};
  \draw [very thick] (0,0.98) -- (0,1.02); 
  \filldraw[black] (0,1) circle(0.03);
  \node at (0,3*\ymax/4) {$S_{P_1,T_1}$};
  \node at (0,1.25) {$\Omega_0$};
  \draw [thick,\cola,-latex] (-0.5,\shift) -- (-0.5,\ymax);
  \draw [thick,\cola,-latex] (0.5,\shift) -- (0.5,\ymax);
  \draw [thick,\cola] (0.5,\shift) arc (60:120:1);
  \draw (-0.5,1.5) -- (0.5,1.5);
  \fill[pattern=north west lines, pattern color=\colb, opacity=1] (-0.5,1.5) rectangle (0.5,\ymax-0.05);
\end{tikzpicture}
\qquad \qquad
\begin{tikzpicture}[scale=2]
  \def\xmin{-0.2} \def\xmax{2.2}
  \def\ymin{-0.1} \def\ymax{3}
  \def\shift{{sqrt(3)/2}}
  \draw [very thick] (\xmin,0) -- (\xmax,0);
  \draw [thick] (1,-0.02) -- (1,0.02) node [below=0.2] {$1$};
  \draw [thick] (0,-0.02) -- (0,0.02) node [below=0.2] {$0$};
  \draw [thick] (2,-0.02) -- (2,0.02) node [below=0.2] {$2$};
  \draw [thick] (-0.02,1) -- (0.02,1) node [below left] {$i$};
  \filldraw[black] (0,1) circle(0.03);
  \draw [very thick] (0,0.98) -- (0,1.02); 
  \node at (1,3*\ymax/4) {$S_{P_1,T_1}$};
  \node at (1,0.75) {$S_{P_2,T_2}$};
  \node at (1,1.25) {$\Omega_0$};
  \draw [thick,\cola,-latex] (2,1) -- (2,\ymax);
  \draw [thick,\cola,-latex] (0,1) -- (0,\ymax);
  \draw [thick,\cola] (2,1) arc (90:180:1);
  \draw [thick,\cola] (1,0) arc (0:90:1);
  \draw (0,1.5) -- (2,1.5);
  \draw (0.5,\shift) to [out=60, in=120, looseness=0.8] (1.5,\shift);
  \fill[pattern=north west lines, pattern color=\colc, opacity=1] (0.5,\shift) to [out=60, in=120, looseness=0.8] (1.5,\shift) -- (1.5,\shift) arc (120:180:1) -- (1,0) arc (0:60:1) -- cycle;
\path [pattern=north west lines, pattern color=\colb, opacity=1](0,1.5) rectangle (2,\ymax-0.05);
\end{tikzpicture}
    \caption{Precise reduction for $\Gamma_1 \backslash \mathbb{H}^2$ and $\Gamma_2 \backslash \mathbb{H}^2$.}			
	\label{fig:Siegel}
	\end{center}
\end{figure}

For completeness, note that the corresponding precise reductions of the fundamental domains using Siegel sets are shown in Figure~\ref{fig:Siegel}. 
For $\Gamma_1$ the parabolic group $P_1$ is the standard minimal parabolic $P_\emptyset$ and the Siegel set is, for some $T_1>1$, 
\begin{equation}
    S_{P_1,T_1} = \left\{ \begin{pmatrix}
       \ee^t & \ee^{-t} n\\
       0 & \ee^{-t} \end{pmatrix}\cdot i : -\tfrac12 < n < \tfrac12, t > T_1 \right\} \, . 
\end{equation}
For $\Gamma_2$, we have $P_1$ and $P_2$ given in~\eqref{Gamma2-paras} with the corresponding Siegel sets, given $T_1,T_2>1$,
\begin{equation}
    \begin{aligned}
    S_{P_1,T_1} &= \left\{ \begin{pmatrix}
    \ee^t & \ee^{-t} n\\
       0 & \ee^{-t} \end{pmatrix}\cdot i : 0 < n < 2, t > T_1 \right\} \,  , \\
    S_{P_2,T_2} &= \left\{ C\begin{pmatrix}
    \ee^t & \ee^{-t} n\\
       0 & \ee^{-t} \end{pmatrix}C^{-1} \cdot i : -1 < n < 1, t > T_2 \right\} \,  .
    \end{aligned}
\end{equation}

\section{Applications to string compactifications}
\label{sec:strmod}
\label{sec:ex}

In Section \ref{sec:discrete}, we discussed how  to explicitly parametrise the infinite-distance boundary of a locally symmetric space and the geodesics that reach it using a group-theoretic approach. The goal of this section is to show how, for standard string compactifications, the SDC \eqref{eq:sdc} can be understood in a unified way using this group-theoretic language, emphasising several specific examples. 

\subsection{General set up}

We are interested in string compactifications that lead to an action of the form~\eqref{eqn:dimredaction} where the moduli space is a locally symmetric space $\mathcal{M}=\Gamma\backslash G/K$. The metric on $\mathcal{M}$ has the form
\begin{equation}
\label{eq:modspace-metric-2}
    G_{ij}(\phi) \dd \phi^i \dd \phi^j = \dd s_0^2 + \sum_a \kappa_a
    \dd \tilde{s}_a^2 \, , 
\end{equation}
as in equation~\eqref{eq:modspace-metric} where $\dd s_0^2$ is the flat metric on $\mathbb{R}^p$. Each $\dd\tilde{s}_a^2$ is an Einstein metric on a component of the decomposition of $G/K$ into irreducible pieces $\mathbb{R}^p\times G_1/K_1\times\dots\times G_s/K_s$. It is normalised so that $\tilde{R}^{(a)}_{ij}=-\frac12\tilde{g}^{(a)}_{ij}$. (In most of our examples the sum in~\eqref{eq:modspace-metric-2} runs over only one factor, and we also will always have at most a one-dimensional flat direction corresponding to the dilaton.) Furthermore, in all our examples $\Gamma$ is arithmetic, and other than a possible $\mathbb{R}$ factor corresponding to the dilaton, $\mathcal{M}$ has finite volume. In discussing the SDC the normalisation factors $\kappa_a$ are crucial since they fix the scale of the distance along the geodesic and hence the value of the exponential rate $\alpha$ in the SDC expression $M \sim \ee^{-\alpha t}$. In toroidal reductions of string theory, one has
\begin{equation} \label{eqn:hvee}
    \kappa_a = \frac{1}{2h_a^{\vee}} \, , 
\end{equation}
where $h_a^{\vee}$ is the dual Coxeter number of the group $G_a$ in the relevant irreducible factor. This corresponds to the longest root $\beta_{\text{long}}$ in the lie algebra $\cgg_a$ being normalised to $\langle \beta_{\text{long}} , \, \beta_{\text{long}}\rangle =2$. For eight or more supercharges, this is also a necessary condition for supersymmetry of the dimensionally reduced theory (except for a few cases of eight-supercharge vector multiplet moduli spaces, see table~\ref{tab:sk} in the appendix). 

The second ingredient in the SDC is the set of states of the theory, in which each state has a mass. In the string compactifications we consider, the discrete duality group $\Gamma$ is arithmetic and the states form a lattice $\Pi\cong \mathbb{Z}^{\dt}\subset V\cong \mathbb{R}^{\dt}$ corresponding to a particular representation of the duality group $\rho:\Gamma\to GL(\dt,\mathbb{R})$. Physically $\Pi$ is the lattice of charges under the $U(1)^{\dt}$ gauge symmetry of the perturbative states of the theory.\footnote{In \cite{Delgado:2024skw} it was argued for a generic supersymmetric compactification that the charged states should transform in a semi-simple representation of $\Gamma$.} At least for a subset of charges, typically those satisfying a BPS condition, the mass formula, as a function of $q\in\Pi$ and the moduli $\phi$, has a simple quadratic form, as follows. The $\Gamma$-representation extends to a real representation of the Lie group $\mathbb{R}^p\times G$. If we parametrise a point $\phi\in\mathcal{M}$ in the moduli space by a representative element $g_\phi\in \mathbb{R}^p\times G$ one can then construct a ``dressed'' charge vector 
\begin{equation}
\label{eq:dressed-q}
    v(q,\phi) = \rho(g_\phi^{-1})\cdot q \, , 
\end{equation}
which is invariant under the action of $\Gamma$. The reason for taking the inverse element $g_\phi^{-1}$ is because $\Gamma$ acts on the left in $\mathcal{M}$, that is, given $\gamma\in\Gamma$ one has $g_\phi\to\gamma g_\phi$, and so $\rho(g_\phi^{-1})\to \rho(g_\phi^{-1})\rho(\gamma^{-1})$ while $q\to \rho(\gamma)\cdot q$. However, it still transforms under the right action of $K$, namely under $g_\phi\to g_\phi k$ we have $v(q,\phi)\to \rho(k^{-1})\cdot v(q,\phi)$. Since $K$ is compact, its representations are unitary, or, since in this case $\rho$ is a real representation, they must be orthogonal. Thus we have a $K$-invariant positive-definite metric $\tilde\kappa$ on the representation space $\mathbb{R}^{\dt}$ and so one can construct the mass-squared of the state as a scalar that depends only on the charge $q$ and the point $\phi^i$ on the moduli space (i.e.~is independent of the choice of representative $g_\phi$ under that action of $K$) as\footnote{In Proposition \ref{prop:mass} in Section \ref{sec:weight-polytopes} we will show that this is the generic form for the mass of the states.}  
\begin{equation}
\label{eq:string-mass}
    m^2(q,\phi) = \left|v(q,\phi)\right|^2 = v^T \cdot \tilde\kappa \cdot v 
        = q^T \cdot \mathcal{H}^{-1} \cdot q \, ,
\end{equation}
where we define the moduli-dependent ``generalised metric''\footnote{The definition of the generalised metric and its inverse is conventional, chosen to match the standard expressions in, for example,~\cite{Giveon:1994fu,Coimbra:2011ky}.}
\begin{align}
\label{eq:gen-metric}
    \mathcal{H}= \rho(g_\phi)\tilde\kappa^{-1}\rho(g_\phi)^T \, . 
\end{align}
Note that, in many cases, the representation $\rho$ is actually the one used to define the algebraic group $\mathbb{R}^p\times G$, and $\tilde\kappa$ is the identity, so one simply has $\mathcal{H}=g_{\phi}g^T_{\phi}$. Whenever $\rho$ is faithful and $G$ is irreducible (and there is no $\mathbb{R}^p$ factor) one can also use $\mathcal{H}$ to write the metric on moduli space in the familiar form 
\begin{equation}
    G_{ij}(\phi) \dd \phi^i \dd \phi^j \propto \Tr \left( \dd \mathcal{H} \; \dd\mathcal{H}^{-1} \right) \, . 
\end{equation}
As mentioned above, the mass-formula~\eqref{eq:string-mass} typically applies to a subset of states satisfying a BPS-type condition. This means that the relevant set of states is not quite complete, in other words there is not necessarily a suitable state for every charge $q\in\Pi$. However, it turns out that there are always enough states to have a tower satisfying the SDC. 

\subsection{Explicit examples from string theory}  

As already mentioned, the moduli spaces of maximal and half-maximal supergravity are of symmetric type.  Let us recall their structure and possible ways to obtain them. 
The following is not an exhaustive list; there are for example other half-maximal theories with rank reduction \cite{deBoer:2001wca,Fraiman:2021hma}. The $\mathbb{R}$ factor is parametrised by the $\dext$-dimensional dilaton $\phi$.
\begin{itemize}
    \item M-theory on $T^{\dt}$ (or Type II on $T^{\dt-1}$): $E_{\dt(\dt)}(\mathbb{Z})\backslash E_{\dt(\dt)}/K_{\dt}$ as in Table \ref{tab:G/K}.
    \item Heterotic theory or Type I on $T^{\dt}$:  $\mathbb{R} \times O(\dt,\dt+16;\mathbb{Z}) \backslash O(\dt,\dt+16)/ (O(\dt)\times O(\dt+16))$.
    \item CHL string on $T^{\dt}$: $\mathbb{R} \times O(\dt,\dt+8;\mathbb{Z}) \backslash O(\dt,\dt+8)/ (O(\dt)\times O(\dt+8))$.
    \item M-theory on a M{\"o}bius strip: $\mathbb{R} \times O(1,1;\mathbb{Z})\backslash O(1,1)/(O(1)\times O(1))$. 
\end{itemize}
As mentioned previously, $E_{\dt(\dt)}(\mathbb{Z})$ is the universal Chevalley group, while in the other cases $\Gamma$ is defined using some fixed form of the $O(\dt,\dt+k)$ metric (as we will describe, where appropriate). 

There are also other instances of symmetric spaces that arise from compactifications with less or even no supersymmetry. When there is no (or small amounts of) supersymmetry, these are exact moduli spaces only at the classical level because they can receive quantum corrections. Nevertheless, it is expected that quantum corrections are exponentially suppressed close to infinite distance limits so that the statements of this paper are true asymptotically.\footnote{It is also plausible that some limits that seem allowed classically are actually obstructed by quantum corrections.} Some examples are
\begin{itemize}
    \item Bosonic theory on $T^{\dt}$: $\mathbb{R} \times O(\dt,\dt;\mathbb{Z}) \backslash O(\dt,\dt)/ (O(\dt)\times O(\dt))$.
        \item $ O(16) \times O(16)$ heterotic theory on $T^{\dt}$:  $\mathbb{R} \times \,\Gamma_{(\dt,\dt+16)} \backslash O(\dt,\dt+16)/ (O(\dt)\times O(\dt+16))$.
    \item $E_8$ heterotic string on $T^{\dt}$: $\mathbb{R} \times \Gamma_{(\dt,\dt+8)} \backslash O(\dt,\dt+8)/ (O(\dt)\times O(\dt+8))$\footnote{or the other non-supersymmetric rank reduced theories recently discovered \cite{Nakajima:2023zsh,DeFreitas:2024ztt}}. 
  \item Type II or heterotic theory on $T^{2\dt}/\mathbb{Z}_{\dt}$ with ${\dt}\, {\rt}$ matter fields:\\
    $\Gamma_{(\dt;r)} \backslash SU(\dt,\dt+\rt)/(U(1) \times SU(\dt) \times SU(\dt+\rt))$.  
    \item Type II or heterotic theory on Calabi-Yau (orientifolds), under certain conditions on the triple intersection numbers and gauge bundle topology~\cite{Farquet:2012cs}. The local moduli spaces should be in the list of symmetric special K\"ahler and quaternionic spaces summarised in the Appendix. 
\end{itemize}
In the non-supersymmetric heterotic cases listed above, despite the fact that the moduli spaces have the same form as their supersymmetric counterparts, the arithmetic groups $\Gamma_{\dt,\dt+16}$ and $\Gamma_{\dt,\dt+8}$ are different. For $\dt=1$, they include the automorphisms of the Generalized Dynkin Diagrams~\cite{Fraiman:2023cpa,DeFreitas:2024yzr,
Fraiman:2025wip}.

For the orthogonal cases $G=O(\dt,\dt+\rt)$, the lattice of particle states lives in the representation space of the vector representation $\mathbf{2\dt+\rt}$. Their charges come from the $U(1)^{2\dt+\rt}$ group of diffeomorphisms, gauge symmetries and residual $B$-field symmetries that survive the compactification on $T^{\dt}$. (In the case of bosonic and type II string compactifications, $\rt=0$.) Similarly, for $G=E_{\dt(\dt)}$ in the M-theory embedding, states are charged with respect to diffeomorphisms and residual form-field symmetries on the torus, corresponding to momentum, M2 and M5 brane charges\footnote{For $\dt\geq 7$, one should also account for diffeomorphisms of the dual graviton, and their associated charges.}. For particle states, the representations spaces are the familiar ones listed under $p=1$ in Table~\ref{tab:G/K}. A separate lattice of states also appears for external string-like states. These correspond to M2 and M5 branes with one space-like dimension in the external spacetime but wrapping respectively one and four directions on the compact $T^{\dt}$. Analogously, there are lattices of higher-dimensional objects. We list the corresponding representation spaces in Table \ref{tab:G/K}, denoted by their external spacetime dimension $p$.\footnote{Actually, all irreducible representations of 1/2 BPS objects can be found from branching rules of $E_{11}$  \cite{Riccioni_2007,Bergshoeff:2007qi,Riccioni_2009,Marrani:2019jvd}.}

\begin{table}\renewcommand{\arraystretch}{1.3}
\centering \small
\begin{tabular}{|>{$}c<{$}||>{$}c<{$}|>{$}c<{$}|>{$}c<{$}|>{$}c<{$}|>{$}c<{$}|>{$}c<{$}|>{$}c<{$}|>{$}c<{$}|>{$}c<{$}|} \hline
\dt & G=E_{\dt(\dt)} & K_{\dt} & p=1 & p=2 & p=3 & p=4 & p=5 & p=6 & p=7
\\ \hline\hline
2 & GL(2) & SO(2) & \mathbf{2}+\mathbf{1} & \mathbf{2} & \mathbf{1} & \mathbf{1} & \mathbf{2} & \mathbf{2}+\mathbf{1} & \mathbf{3}+\mathbf{1}  \\\cline{1-10}
3 & SL(2)\times SL(3) & SO(2) \times SO(3) & (\mathbf{2},\mathbf{\overline{3}})  & (\mathbf{1},\mathbf{3}) & (\mathbf{2},\mathbf{1}) & (\mathbf{1},\mathbf{\bar{3}}) & (\mathbf{2},\mathbf{3}) & \begin{matrix} (\mathbf{1},\mathbf{8})\\ (\mathbf{3},\mathbf{1})\end{matrix}
\\\cline{1-9}
 4 & SL(5) & SO(5) & \mathbf{\overline{10}} & \mathbf{5} & \mathbf{\bar{5}} & \mathbf{10} & \mathbf{24}  \\\cline{1-8}
 5 & SO(5,5) & SO(5)\times SO(5) & \mathbf{16} & \mathbf{10} & \mathbf{\overline{16}} & \mathbf{45}\\\cline{1-7}
 6 & E_{6(6)} & USp(8) & \mathbf{27} & \mathbf{\overline{27}} & \mathbf{78} \\\cline{1-6}
7 & E_{7(7)} & SU(8) & \mathbf{56} & \mathbf{133}\\\cline{1-5}
 8 & E_{8(8)} & SO(16) &\mathbf{248} \\\cline{1-4}
\end{tabular}
\caption[caption]{Cosets $G/K$ of maximal supergravities in $11$-$\dt$ dimensions, and corresponding $E_{\dt(\dt)}$ representations of objects of $p-1$ spatial dimensions in the external space (see for instance \cite{Obers:1998fb}). We give only the representations that will be relevant here, which are for $p\leq 9-\dt$. 
}  
\label{tab:G/K}
\end{table}

\bigskip
\noindent
In the following subsections, we will look at some of the examples we cited above in more depth. We will explicitly show how the moduli of the theories are embedded in the coset spaces and discuss the boundary structure and geodesic motion, and identify the corresponding massless towers. In particular, for toroidal reductions, consistent with the Emergent String Conjecture, we will find that the possible exponential decay rates are  
\begin{equation} 
\label{eqn:alphakkstring}
    \alpha_{\text{KK};n} = \sqrt{\frac{1}{\dext-2}+\frac{1}{n}}\, ,\qquad     
    \alpha_{\text{St}} = \frac{1}{\sqrt{\dext-2}} \, ,
\end{equation}
where $\alpha_{\text{KK};n}$ is the exponent for decompactifying $n$ dimensions (see for example~\cite{Etheredge:2022opl}) and $\alpha_{\text{St}}$ is the exponent for an emergent string limit.

\subsubsection{Toroidal compactifications of the bosonic theory} 
\label{sec:bosonicTd} 

We first consider the simple case of the bosonic string, or equivalently, the NSNS sector of type II theories, reduced on $T^{\dt}$. As noted above the moduli space is locally symmetric of the form 
\begin{equation}
\label{eq:Mbosonic}
    \mathcal{M} = \mathbb{R} \times O(\dt,\dt;\mathbb{Z}) \backslash O(\dt,\dt)/ (O(\dt)\times O(\dt)) \, . 
\end{equation}
The group $G=O(\dt,\dt)$ is the split form and the $SO(\dt,\dt; \mathbb{Z})$ part of $\Gamma$ is the Chevalley group based on the vector representation $\mathbf{2\dt}$. Concretely, starting with the $D$-dimensional string frame action 
\begin{equation}
    S= \frac{1}{(2\pi)^7 \alpha'^4} \int \dd^{D}x \sqrt{-g^{s}_D} \, \ee^{-2\phi_D} \left(R^{s}_D+4 \partial_{M}\phi_D \partial^{M}\phi_D - \tfrac{1}{12} H^2 \right) \, ,
\end{equation}
reducing on $T^{\dt}$ and transforming to the Einstein frame gives the $\dext=D-\dt$ dimensional action\footnote{The fields without subscripts refer to quantities of the $\dext$-dimensional theory.} 
\begin{equation}
\label{eq:reduced-action}
     S= \dfrac{M_{P}^{\dext-2}}{2} \int \dd^{\dext}x \sqrt{-g^E} \left(R^E- \partial_{\mu}\lambda_0 \partial^{\mu} \lambda_0 
     + \tfrac{1}{8}\Tr \partial_\mu \mathcal{H}\partial_\mu \mathcal{H}^{-1} \right) \, ,
\end{equation}
where 
\begin{equation}
\label{eq:dilaton}
    \phi = \phi_D - \tfrac{1}{4} \log \det g_{\dt} \equiv - \tfrac12 \sqrt{\dext-2}\,\lambda_0 \, , 
\end{equation}
is the lower-dimensional dilaton, $M_{P}^{\dext-2}=2\text{Vol}(T^{\dt})/(2\pi)^7 \alpha'^4$ and the generalised metric has the familiar form 
\begin{equation} \label{Oddgenmetric}
    \mathcal{H}= \begin{pmatrix}g_{\dt}-Bg_{\dt}^{-1} B &  Bg_{\dt}^{-1}\\
     - g_{\dt}^{-1}B & g_{\dt}^{-1} \\ 
    \end{pmatrix} \, ,
\end{equation}
with $g_{\dt}$ and $B$ the internal metric and $B$-field. 

To connect to our previous group-theoretic discussion of the moduli, note that the $O(\dt,\dt)$ group is defined algebraically as the subgroup of $GL(2\dt,\mathbb{R})$ preserving the metric
\begin{equation}
    \eta = \begin{pmatrix}
        \mathbb{0} & \mathbb{1}_{\dt} \\
        \mathbb{1}_{\dt} & \mathbb{0}
    \end{pmatrix} \, . 
\end{equation}
Thus the vector representation $\mathbf{2\dt}$ is the defining representation, that is $\rho(g)=g$, and the generalised metric~\eqref{eq:gen-metric} is just 
\begin{equation}
    \mathcal{H} = gg^T \, . 
\end{equation}
Decomposing $\mathfrak{o}(\dt,\dt)$, we can choose elements of its Cartan to be diagonal. In block form, 
\begin{align}
    \caa_{\mathfrak{o}(\dt,\dt)} &\ni \begin{pmatrix}
        \Lambda & \mathbb{0} \\
        \mathbb{0} & - \Lambda
        \end{pmatrix} \, , \qquad\text{where} \quad
        \Lambda = \diag ( \lambda_1, \lambda_2, \dots, \lambda_{\dt}) \, , 
\end{align}
and $\cnn^+$ is 
\begin{align}
    \cnn^+ &\ni  \begin{pmatrix}
        U &  B \\ 
        \mathbb{0} & - U^T
        \end{pmatrix} \, ,
\end{align}
where $U$ is upper triangular and $B$ is antisymmetric. Exponentiating and then constructing $\mathcal{H}=gg^T$ one sees that the Cartan elements are related to the radii $R_1, \, ..., \, R_{\dt}$ of the $T^{\dt}$ via $R_i = e^{\lambda_i}$, while $U$ generates off-diagonal elements of $g_{\dt}$ and $B$ gives the $B$-field. 

We can include the dilaton factor as part of the Cartan subalgebra so that, given our choice of positive roots, the closure of the positive Weyl chamber is given by 
\begin{equation}\label{bapl}
    \overline{\caa^+} = \{ (\lambda_0, \lambda_1, \dots \lambda_{\dt}) : \lambda_1\leq \lambda_2 \leq \, ... \, \leq \lambda_{\dt} , \, \lambda_1 + \lambda_2 \geq 0 \} \, .   
\end{equation}
Restricting to $H\in\caa$ we find that the metric~\eqref{eqn:kf} on $\mathcal{M}$ takes the form 
\begin{equation}
    \left<H,H\right> = \dd\lambda_0^2 - \tfrac18 \Tr \dd\mathcal{H}\dd\mathcal{H}^{-1}
        = \dd\lambda_0^2 + \dd\lambda_1^2 + \dots + \dd\lambda_{\dt}^2 \, . 
\end{equation}
In particular, this means the roots of $\mathfrak{o}(\dt,\dt)$ are all length squared two in this metric. 

Let us now turn to the boundary geodesics. Since the $SO(\dt,\dt;\mathbb{Z})$ action is that of a Chevalley group, following~\eqref{eq:Chevalley-boundary}, the boundary is simply the positive Weyl chamber
\begin{equation}
\label{eq:boundaryOdd}
    \partial \mathcal{M} = \overline{\caa^+_1} / \mathbb{Z}_2 \, , 
\end{equation}
where the extra $\mathbb{Z}_2$ factor comes from the fact that we actually quotient by $O(\dt,\dt;\mathbb{Z})$ not $SO(\dt,\dt;\mathbb{Z})$ as we discuss below. Ignoring the $\mathbb{Z}_2$ for now, and taking the canonical geodesic $\gamma_H(t)=\ee^{Ht}$, we get 
\begin{equation}
\label{eqn:oddgm}
\begin{aligned}    
    \mathcal{H}(t) &= \text{diag}(\ee^{2\lambda_1 t}, \dots, \ee^{2\lambda_{\dt} t}, \ee^{-2\lambda_1 t}, \dots, \ee^{-2\lambda_{\dt} t}) \, , \\
    \phi(t) &= - \tfrac12 \sqrt{\dext-2}\,\lambda_0 t \, ,
\end{aligned}
\end{equation}
with 
\begin{equation}
\label{eqn:normonalgebra}
    \sum_{i=0}^{\dt} \lambda_i^2 = 1  \, \qquad
    \text{and  $\lambda_1\leq \lambda_2 \leq \, ... \, \leq \lambda_{\dt} , \, \lambda_1 + \lambda_2 \geq 0$} \, . 
\end{equation}
We see that the boundary points depend only the diagonal radii $R_i$ of the metric on the torus and are independent of the off-diagonal elements and the B-field -- when analysing the boundary geodesics we may effectively take the torus to be rectangular and set $B=0$.

For boundary points corresponding to the minimal parabolic where $H$ lies in the interior of the positive Weyl chamber, there are two possibilities for the metric moduli
\begin{enumerate}
    \item $0<\lambda_1<\lambda_2<...<\lambda_{\dt}$: all the $R_i \to \infty$, corresponding to a decompactification to 26 dimensions (if we use the Type II language, this would be decompactification to Type IIA\footnote{The distinction between IIA and IIB  is arbitrary at this point. We choose this limit to be Type IIA to match with the corresponding limit in M-theory compactifications to be discussed in the next subsection.}  in 10 dimensions choosing appropriately the embedding). 
    \item $\lambda_1 < 0 < \lambda_2 < \, ... \, < \lambda_{\dt}$: $R_1 \to 0, \, R_i \to \infty$ for $i=2,...,\dt$, corresponding to decompactification to 26 dimensions in the T-dual frame (or to 10 dimensional Type IIB in the other language). 
\end{enumerate}
At the same time, the dilaton runs to $\pm\infty$ depending on the sign of $\lambda_0$. From now on we will assume  $\lambda_0  > 0$ \footnote{Actually, since we want to stay in the perturbative regime, where $\phi_D<0$, we restrict to  $\lambda_0 \geq \tfrac{1}{\sqrt{\dext-2}}\sum_{i=1}^{\dt} \lambda_i \ge 0$. }. The remaining cases are infinite distance limits along geodesics generated by elements in $\overline{\caa^{+}_1}$ which saturate some inequalities in equation \eqref{bapl}. As boundary points of the moduli space, they are identified with non-minimal parabolics. They correspond either to partial decompactification limits where some radii do not change size, $\lambda_i = 0$ (at most $\dt-1$ of them), or to decompactification limits for which two or more radii scale the same way.
In the latter case, this occurs when there exist integers $n$ and $i<\dt-n$ such that $\lambda_i=\lambda_{i+1}=...=\lambda_{i+n}$.
If one considers other geodesics in the same class, the metric $g_{\dt}$ can become non-rectangular and the $B$-field can be turned on, but it is still only the radii that grow with $t$, and the growth remains fixed by the choice of $H$.  

Recall that in the case of type II string theory in absence of RR-fluxes, there is no distinction between IIA and IIB. Hence, the two limits described in the bullet points above are equivalent, as in the case of the bosonic string, and correspond to the same point on the moduli space. This can be understood by realising that we have not accounted for the full arithmetic subgroup since we only really quotiented by Chevalley group $SO(\dt,\dt;\mathbb{Z})$ built from the $\mathfrak{o}(\dt,\dt)\cong\mathfrak{so}(\dt,\dt)$ algebra. The T-duality that relates IIA and IIB is a reflection, that is, lives in $O(\dt,\dt;\mathbb{Z})$, and corresponds to an outer automorphisms group of the $\mathfrak{o}(\dt,\dt)$ algebra\footnote{In general, the Chevalley subgroup is the largest group of discrete inner automorphisms, so combining it with discrete outer automorphisms corresponds to quotienting the moduli space by Aut$(D_{\mathfrak{g}})$ where $D_{\mathfrak{g}}$ denotes the Dynkin diagram of $\mathfrak{g}$.}. This outer automorphisms group is a $\mathbb{Z}_2$ corresponding to the exchange of two simple roots shown in the diagram labelled~\eqref{eq:onn-dynk}. The automorphism acts on $\caa$ as $\lambda_1 \leftrightarrow - \lambda_1$ so that $R_1 \leftrightarrow \frac{1}{R_1}$, relating both possibilities given above, and giving the $\mathbb{Z}_2$ quotient in~\eqref{eq:boundaryOdd}. It corresponds, in the $O(2,2)$ example, to restricting to either the left or right half of the purple line in Figure \ref{fig:o22}. 

The states in the theory are charged under the $U(1)^{2\dt}$ symmetries of the metric and B-field. The charges form a lattice $\Pi\cong\mathbb{Z}^{2\dt}$ in $V=\mathbb{R}^{2\dt}$ labelled by momentum and winding numbers on the torus
\begin{equation}
    q = (n_i,w^i) \, .
\end{equation}
If no string oscillator modes are excited, the states satisfy the level matching  condition
\begin{equation}
   q^T \cdot \eta\cdot  q = 0 \, , 
\end{equation}
and the mass of such states, in the $\dext$-dimensional Einstein frame, is given by 
\begin{equation}
    m^2(q)= \ee^{\frac{4}{\dext-2}\phi}
        \Bigl( g^{ij}(n_i-B_{ik}w^{k})(n_j-B_{jl}w^{l})+g_{ij} w^{i}w^{j} \Bigr) \, . 
\end{equation}
It is easy to see that this corresponds to 
\begin{equation}
    m^2(q)=\ee^{\frac{4}{\dext-2}\phi} \, q^T\cdot \mathcal{H}^{-1}\cdot q \, , 
\end{equation}
as in~\eqref{eq:string-mass}. The dilaton factor can be viewed as defining how states in $V$ transform under the action of translations in the $\mathbb{R}$ factor in~\eqref{eq:Mbosonic}. This dilaton weight is not fixed by the coset structure, but is dictated by the string action, unlike when working with the exceptional groups, where the dilaton fits in the relevant $E_{\dt(\dt)}$ representation. If we restrict to the canonical geodesics, the distance-dependent mass is 
\begin{equation}\label{eq:masst}
    m^2(t,H,q) = e^{-\frac{2}{\sqrt{{\dext}-2}}\lambda_0t} \sum_{i=1}^{\dt} \left( e^{-2 \lambda_it} n_i^2 + e^{2\lambda_it} w^{i2} \right) \, .
\end{equation}

As we mentioned above, the charges transform in the vector $\mathbf{2\dt}$ representation of $SO(\dt,\dt)$. The exponential decay rates in~\eqref{eq:masst} have a simple form in terms of the weights of the representation under the Cartan subalgebra $\caa$. That is,
\begin{equation}\label{eq:barlam}
  \tfrac{1}{\sqrt{\dext-2}} \lambda_0 \pm \lambda_i
     = \left< w^\pm_i , H \right> \, ,
\end{equation}
where $H=(\lambda_0,\lambda_1\dots,\lambda_{\dt})\in\overline{\caa_1^+}\slash\mathbb{Z}_2$ and so satisfies $\left<H,H\right>=1$, and
\begin{equation}
  w^\pm_i= \bigl(\tfrac{1}{\sqrt{\dext-2}},0,\dots,\pm1,\dots,0\bigr) \, , 
\end{equation}
are the weights. 

We can then analyse the rate at which the fastest tower in~\eqref{eq:masst} becomes massless. We can just focus on the case $\lambda_i \ge 0$ for all $i=1,...,\dt$ since the other case, $\lambda_1<0,\;\lambda_{i>1}>0$ is related to the former by T-duality as discussed. For a state to become massless the charge vector must have the form $q=(n_i,0)$. Among these vectors, states which become massless the fastest are those with the largest exponent. The hierarchy among $\lambda_i$, given in \eqref{bapl}, is such that $\lambda_{\dt}$ is always at least as large as the other $\lambda_i$ for $i=1,\dots,\dt$ and and so gives the most negative exponent. The tower of momenta with only $n_{\dt}$ non-zero is therefore always among the fastest ones and the fastest decay as a function of $H$ is 
\begin{equation}\label{eq:zeta-bosonic}
    \alpha(H) = \tfrac{1}{\sqrt{\dext-2}} \lambda_0 + \lambda_{\dt}  \, . 
\end{equation}
The minimum $\alpha(H)$ is attained when all the bounds are saturated and $\lambda_i=0$ and $\lambda_0=1$. Such geodesics flow only in the dilaton direction and we find 
\begin{equation}\label{eq:minimum}
    \alpha_{\text{min}} \equiv \min_{H\in\caa^+_1} \alpha(H) 
    = \tfrac{1}{\sqrt{\dext-2}} \equiv \alpha_{\text{St}} \, . 
\end{equation}
This is indeed the lower bound of the Sharpened SDC~\eqref{eqn:sharpeneddc}, corresponding to a string tower. 

On the other hand, saturating  only some of the bounds, leads to local extrema of $\alpha(H)$. In particular, consider the Cartan element, denoted by $H(\cartanelem)$, given by 
\begin{equation}
\label{alphaKKpartial}
\begin{aligned} 
    \lambda_i &= 0 \text{ for all } i=1,\dots,\dt-n \, ,  \\
    \lambda_i &= \cartanelem \text{ for all } i=\dt-n+1,\dots,\dt \, ,  \\
    \lambda_0 &= \sqrt{1 - n \cartanelem^2} \, ,
\end{aligned}    
\end{equation}
together with the charge vector $q$ with momenta $n_i$ non-vanishing for $i=\dt-n+1,\dots,\dt$ and all other momenta and winding set to zero. Extremizing \eqref{eq:barlam} over $p$ gives a local extremum of $\alpha(H)$ and sets $p=p_*$ where  
\begin{align}
     p_*^2 &= \frac{\dext-2}{n\, (\dext+n-2)} \nonumber \, .
\end{align}
The associated exponential rate, common to all the momentum states, is
\begin{equation}\label{alphadn}
    \alpha(H(p_*)) = \sqrt{\frac{1}{\dext-2}+\frac{1}{n}}\equiv \alpha_{KK;n} \, .
\end{equation}
This is precisely the exponential rate for KK towers corresponding to $n$ compact directions decompactifying \cite{Etheredge:2023odp,Castellano:2023jjt}.

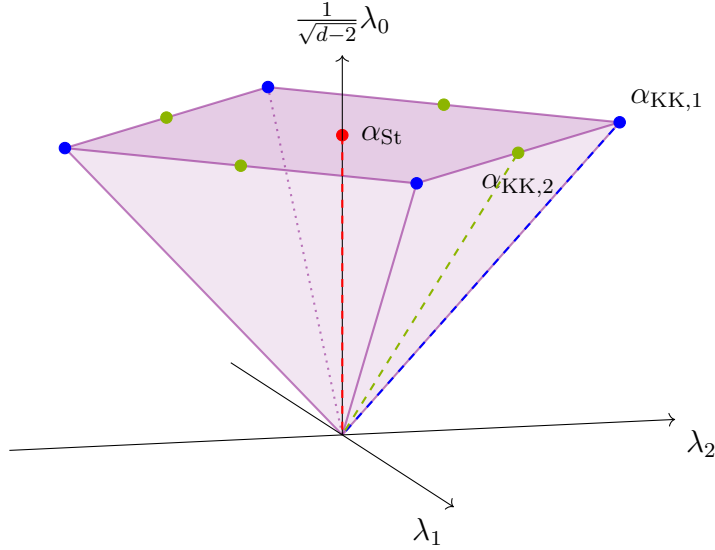
\begin{figure}[t]
\centering
\tdplotsetmaincoords{80}{120}
\begin{tikzpicture}[tdplot_main_coords, scale=2.7]
\def\colone{blue}
\def\coltwo{applegreen}
\def\colthree{cyan}
\def\colinf{red}
\def\colthreeIIB{orange}
\def\colfour{violet}
\coordinate (A) at (1,1,1.5);
\coordinate (B) at (-1,1,1.5);
\coordinate (C) at (-1,-1,1.5);
\coordinate (D) at (1,-1,1.5);
\coordinate (O) at (0,0,0);
\coordinate (U) at (1,0,1.5);
\coordinate (V) at (-1,0,1.5);
\coordinate (W) at (0,-1,1.5);
\coordinate (X) at (0,1,1.5);
\coordinate (Z) at (0,0,1.5);
\fill[violet, opacity=0.1] (A) -- (B) -- (O) -- cycle;
\fill[violet, opacity=0.1] (D) -- (A) -- (O) -- cycle;
\fill[violet, opacity=0.2] (D) -- (A) -- (B) -- (C) -- cycle;
\draw[thick, violet,opacity=0.5] (A) -- (B) -- (C) -- (D) -- cycle;
\draw[thick, violet,opacity=0.5] (O) -- (A);
\draw[thick, violet,opacity=0.5] (O) -- (B);
\draw[dotted, thick, violet,opacity=0.5] (O) -- (C);
\draw[thick, violet,opacity=0.5] (O) -- (D);
\foreach \p in {A,B,C,D}
    \filldraw[\colone] (\p) circle (0.8pt);
\foreach \p in {U,V,W,X}
    \filldraw[\coltwo] (\p) circle (0.8pt);
\filldraw[\colinf] (Z) circle (0.8pt);    
\node at (B) [above right] {$\alpha_{\text{KK},1}$};
\node at (X) [below=4pt] {$\alpha_{\text{KK},2}$};
\node at (Z) [right=3pt] {$\alpha_{\text{St}}$};

\draw[->] (-1.5,-1.5,0) -- (1.5,1.5,0) node[anchor=north east]{$\lambda_1$};
\draw[->] (1.2,-1.2,0) -- (-1.2,1.2,0) node[anchor=north west]{$\lambda_2$};
\draw[->] (0,0,0) -- (0,0,1.9) node[anchor=south]{$\frac{1}{\sqrt{d-2}}\lambda_0$};

\draw[thick, shorten >=2pt,shorten <=2pt,\colone,dashed](0,0,0)--(B);
\draw[thick, shorten >=2pt,shorten <=2pt,\coltwo,dashed](0,0,0)--(X); 
\draw[thick, shorten >=2pt,shorten <=2pt,\colinf,dashed](0,0,0)--(Z);

\end{tikzpicture}
\caption{Plot of weights and extrema of $\alpha(H)$ in $\caa$ for compactification of the bosonic string on $T^2$.}
\label{fig:exampleSO22}
\end{figure}

It is very informative to plot these extrema as points in $\caa$ by defining the vector 
\begin{equation}
\label{eq:zz-def}
    \zz = \alpha(H) H \in \caa \, , 
\end{equation}
so that the length of $\zz$ is the value of $\alpha(H)$ when $H$ points along $\zz$, recalling that $H$ is normalised to length one. In particular, for the case of $\mathbb{R}\times O(2,2)$ one gets Figure~\ref{fig:exampleSO22}. The four weights vectors
\begin{align}
  w^\pm_1 = \bigl(\tfrac{1}{\sqrt{\dext-2}},\pm1,0\bigr) \, , \qquad     
  w^\pm_2 = \bigl(\tfrac{1}{\sqrt{\dext-2}},0,\pm1,\bigr) \, ,
\end{align}
are denoted by the blue nodes. When $H$ points in these directions we get a single radius decompactifying. The green nodes denote the extrema where two radii decompactify. The red node is the extremum corresponding to a massless string tower. (To be slightly more precise, if one restricts to closure of the positive Weyl chamber $\overline{\caa^+}$, the nodes labelled $\alpha_{\text{KK},1}$ and $\alpha_{\text{KK},2}$ are the decompactification limits. The other blue nodes are dual limits where the T-dual radii decompactify). 

Crucially we note that the extrema lie on the boundary of the \emph{convex hull} formed by the set of weights $\{w^\pm_1,w^\pm_2\}$ together with the origin. It is easy to see this remains the case for the reduction on any torus $T^{\dt}$. In fact, as we show in the next section, it is a generic property for any locally symmetric moduli space.

\subsubsection{Toroidal compactifications of the heterotic theory}
\label{sec:exhet} 
The moduli space that arises from compactifications of the heterotic theory on $T^{\dt}$ has the form 
\begin{equation}
\label{eq:Mheterotic}
    \mathcal{M} = \mathbb{R} \times O(\dt,\dt+16;\mathbb{Z}) \backslash O(\dt,\dt+16)/ (O(\dt)\times O(\dt+16)) \, . 
\end{equation}
Concretely, dimensionally reducing the theory one gets a lower-dimensional action of the form~\eqref{eq:reduced-action} but now with 
\begin{equation} \label{Hetgenmetric}
    \mathcal{H} = \begin{pmatrix}g_{\dt} +  C^{T} g^{-1}_{\dt} C+ A_I \kappa^{IJ}A_J & - C^T g^{-1}_{\dt} & C^T g^{-1}_{\dt} A_J +A_J \\
    -g^{-1}_{\dt} C & g^{-1}_{\dt} & - g^{-1}_{\dt} A_J \\
    A_I g^{-1}_{\dt} C +A_I & -A_I g^{-1}_{\dt}  & \kappa_{IJ} + A_I g^{-1}_{\dt} A_J
    \end{pmatrix} \, ,
\end{equation}
with 
\begin{equation}
\label{eq:C-def}
    C_{ij} = B_{ij} + \tfrac{1}{2} A_{iI}\kappa^{IJ}A_{jJ} \, ,
\end{equation}
where $(g_{\dt})_{ij}$ and $B_{ij}$ are the internal metric and $B$-field on $T^{\dt}$ and $A_i{}^I$, $I=1,...,16$ are the 16 Wilson lines . 

The metric $\kappa_{IJ}$ in~\eqref{Hetgenmetric} and~\eqref{eq:C-def} enters into the definition of the algebraic groups $O(\dt,\dt+16)$ and $O(\dt,\dt+16;\mathbb{Z})$. In particular, the latter are the subgroups of $GL(2\dt+16,\mathbb{R})$ and $GL(2\dt+16,\mathbb{Z})$ preserving the metric 
\begin{equation}
\label{invform}
    \eta = \begin{pmatrix}
        \mathbb{0}_{\dt} & \mathbb{1}_{\dt} & \mathbb{0} \\
        \mathbb{1}_{\dt} & \mathbb{0}_{\dt} & \mathbb{0} \\
        \mathbb{0} & \mathbb{0} & \kappa
    \end{pmatrix} \, ,
\end{equation}
where $\kappa$ is the Cartan matrix of $\mathfrak{e}_8\oplus \mathfrak{e}_8$ or $\mathfrak{so}(32)$ depending on whether on has reduced the $E_8\times E_8$ or $Spin(32)/\mathbb{Z}_2$ heterotic theory. Naively one might think these give different reduced theories, but in fact the arithmetic groups defined by the two metrics are the same in that the two corresponding metrics~\eqref{invform} can be related by a $GL(2\dt+16,\mathbb{Z})$ transformation. Furthermore, the lattice $\Pi\cong\mathbb{Z}^{2\dt+16}$ of integer points in the representation space $V\cong\mathbb{R}^{2\dt+16}$ is even self-dual, a condition required by modular invariance of the 10-dimensional theory.  

As mentioned in Section~\ref{sec:funddom}, one can also define the arithmetic group by choosing the canonical metric $\kappa_{IJ}=\delta_{IJ}$ and requiring $\Gamma$ to preserve a particular lattice. The metric~\eqref{invform} with $\kappa=\delta$ can be transformed to the one with $\kappa$ given by the Cartan matrix using a $GL(2\dt+16,\mathbb{Q})$ transformation, and so the two metrics define equivalent $O(\dt,\dt+16,\mathbb{Q})$ groups. However they are not equivalent under $GL(2\dt+16,\mathbb{Z})$ and so the subgroups of $GL(\dt,\dt+16,\mathbb{Z})$ preserving the two metrics are inequivalent. To get the right arithmetic group when $\kappa=\delta$, one instead defines $\Gamma$ as the subgroup of $O(\dt,\dt+16,\mathbb{Q})$ that preserves a particular lattice in $V\cong\mathbb{R}^{2\dt+16}$, namely $\Pi=\Pi_{\dt,\dt}\oplus \Pi_{16}$ where $\Pi_{\dt,\dt}$ is the lattice of integer points in the first $\mathbb{R}^{2\dt}$ component of $V$, and $\Pi_{16}$ is the root lattice of $\mathfrak{e}_8\oplus \mathfrak{e}_8$ or $\mathfrak{so}(32)$ normalised with roots of length-squared two. Although the two lattices $\Pi_{16}$ are not equivalent, the corresponding $\Pi$ are, and hence define the same arithmetic groups. It is this latter formulation, with $\kappa=\delta$ and the arithmetic group fixed by a choice of lattice that is more common in the physics literature and is what we will use in the following.

Unlike the bosonic case, $O(\dt,\dt+16)$ is not the split form and instead we have 
\begin{equation}
    \rank_\mathbb{R} O(\dt,\dt+16) = \rank_\mathbb{Q} O(\dt,\dt+16) = \dt \, . 
\end{equation}
The real Cartan subalgebra of $\mathfrak{o}(\dt,\dt+16)$ can be taken to have the form 
\begin{equation}
    \caa_{\mathfrak{o}(\dt,\dt+16)} \ni \begin{pmatrix}
        \Lambda & \mathbb{0}_{\dt} & \mathbb{0} \\
        \mathbb{0}_{\dt} & -\Lambda & \mathbb{0} \\
        \mathbb{0} & \mathbb{0} & \mathbb{0}
    \end{pmatrix} \, , \qquad\text{where} \quad
        \Lambda = \diag ( \lambda_1, \lambda_2, \dots, \lambda_{\dt}) \, , 
\end{equation}
and we see again that the radii of the internal metric $g_{\dt}$ are given by $R_i=\ee^{\lambda_i}$. Including the dilaton~\eqref{eq:dilaton} in the full Cartan algebra one finds, as for the bosonic case, that the positive Weyl chamber has the form 
\begin{equation}
    \overline{\caa^+} = \{ (\lambda_0, \lambda_1, \dots \lambda_{\dt}) : \lambda_1\leq \lambda_2 \leq \, ... \, \leq \lambda_{\dt} , \, \lambda_1 + \lambda_2 \geq 0 \} \, .   
\end{equation}
Since the arithmetic group $O(\dt,\dt+16;\mathbb{Z})$ cannot be Chevalley (since $O(\dt,\dt+16)$ is not split), we no longer have the result that the boundary $\partial\mathcal{M}$ must be isomorphic to (a quotient of) $\overline{\caa^+_1}$. In fact, from the study of the lattice of states of the theory, there are known results~\cite{Ginsparg:1986bx,Polchinski_1996} implying that the boundary takes the form of a fibration over $\overline{\caa^+_1}$ where each fibre is a discrete set of points. For example, in the case of a compactification on a circle of radius $R$ (i.e.~$d=1$), $\overline{\caa^+_1}=\{pt\}$ is a point but there are two equivalence classes of rational parabolics under $O(1,17; \mathbb{Z})$. These correspond to two decompactification limits corresponding to the inequivalent ten-dimensional $E_8 \times E_8$ and $Spin(32)/\mathbb{Z}_2$ heterotic theories, one of them corresponding to the point $R \to \infty$ and the other to $R \to 0$. 

To see this more explicitly and show how the SDC is satisfied, let us focus on the $O(1,17)$ case since this captures the key points. The geometry has been very nicely analysed in~\cite{Keurentjes:2006cw}, and also more recently in~\cite{Israel:2025ouq}, and we will here  simply show how this fits into the general group-theoretic analysis. For the moment, we will ignore the dilaton factor $\mathbb{R}$ so that $\mathcal{M}=O(\dt,\dt+16;\mathbb{Z}) \backslash O(\dt,\dt+16)/ (O(\dt)\times O(\dt+16))$. Elements of the Lie algebra  take the form
\begin{equation}
\mathfrak{o}(1,17) \ni
\begin{pmatrix}
\lambda & 0 & A^T \\
0 & -\lambda & \alpha^T \\
-\alpha & -A & m
\end{pmatrix}\, ,
\end{equation}
where $\lambda$ is a real number, $m$ is a $16 \times 16$ matrix and $A$ and $\alpha$ are $16$-dimensional vectors, and $m$ satisfies $m^T=-m$. Decomposing we have 
\begin{align}
\caa &\ni 
\begin{pmatrix}
\lambda & 0 & \mathbb{0} \\
0 & -\lambda & \mathbb{0} \\
\mathbb{0} & \mathbb{0} & \mathbb{0}_{16}
\end{pmatrix} \, , &
\cnn^+ &\ni \begin{pmatrix}
0 & 0 & A^T \\
0 & 0 & \mathbb{0} \\
\mathbb{0} &-A & \mathbb{0}_{16}
\end{pmatrix} \, , &
\cmm &\ni \begin{pmatrix}
0 & 0 & \mathbb{0} \\
0 & 0 & \mathbb{0} \\
\mathbb{0} &\mathbb{0} & m
\end{pmatrix} \, . 
\end{align}
The global symmetric space $\mathcal{S}=O(1,17)/O(17)$ is a hyperboloid and can be realised explicitly as a constrained vector $\tau\in\mathbb{R}^{18}$ satisfying $\eta(\tau,\tau)=1$. The vector transforms as $\tau\to g\cdot \tau$ under the action of $g\in O(1,17)$ and, written in terms of horospherical coordinates~\eqref{P0horosperical}, it looks like 
\begin{equation}
\label{eq:hyperboloid}
    \mathcal{S} \ni \tau = \begin{pmatrix}
        R - \frac12 R^{-1} A^TA \\ R^{-1} \\ R^{-1}A 
    \end{pmatrix} 
        = \begin{pmatrix}
            1 & -\frac12 A^TA & A^T \\
            0 & 1 & \mathbb{0} \\
            \mathbb{0} & -A & \mathbb{1}_{16}
            \end{pmatrix}
            \begin{pmatrix}
            R & 0 & \mathbb{0} \\
            0 & R^{-1} & \mathbb{0} \\
            \mathbb{0} & \mathbb{0} & \mathbb{1}_{16}
            \end{pmatrix}\cdot o \, ,
\end{equation}
where $R=\ee^\lambda$ and $o=(1,1,\mathbb{0})$ is stabilised by $O(17)$. 

To describe the boundary of $\cS$ we note that, since $O(1,17)$ is rank one, we only have minimal parabolic groups, with, given the decomposition above, the standard parabolic algebra $\cpp_\emptyset =\caa\oplus\cmm\oplus\cnn^+$. The corresponding canonical geodesic is 
\begin{equation}
    \gamma_\infty(t) = \begin{pmatrix}
        \ee^t \\ \ee^{-t}\\ 0
    \end{pmatrix} \, ,
\end{equation}
which converges to a point $R=\ee^t\to\infty$ on the boundary at $t\to\infty$. The other boundary points of $\mathcal{S}$ come from acting with $g\in O(1,17)$ on $\gamma_{\infty}$
\begin{equation}
    \gamma(t) = g\cdot \gamma_{\infty}(t) 
        = \begin{pmatrix}
            a\ee^t \\ b\ee^t \\ c\ee^t 
    \end{pmatrix} + \cdots \, , 
\end{equation}
where we keep only the leading terms as $t\to\infty$ and $a\in \mathbb{R}$, $b\in\mathbb{R}$ and $c\in\mathbb{R}^{16}$ are elements of the first column of $g$. The requirement that $g\in O(1,17)$ implies $2ab+c^Tc=0$. Comparing with~\eqref{eq:hyperboloid} we see that 
\begin{equation}
\label{eq:genbdry}
    R = b^{-1} \ee^{-t} \, , \qquad A = c/b \, . 
\end{equation}
We see that the rest of the boundary has $R=0$ but a non-zero Wilson line $A\in \mathbb{R}^{16}$ so that
\begin{equation}
    \partial\mathcal{S} = \coprod_{P} \caa^+_{P,1} 
    \cong \{ \infty \}  \cup \{\mathbb{R}^{16} \} \cong S^{16} \, , 
\end{equation}
as expected before restricting to the fundamental domain under T-duality. 

Turning to the local symmetric space $\mathcal{M}=O(1,17;\mathbb{Z})\backslash O(1,17)/O(17)$ we first note that points on the boundary corresponding to rational parabolic subgroups will be of the form~\eqref{eq:genbdry} but with $a,b\in\mathbb{Q}$ and $c\in\mathbb{Q}^{16}$ and hence 
\begin{equation}
    \coprod_{P(\mathbb{Q})} \caa^+_{P(\mathbb{Q}),1} 
        \cong \{\infty\} \cup \mathbb{Q}^{16} \, , 
\end{equation}
and boundary points are associated with rational Wilson lines $A=c/a$ as first noted in~\cite{Keurentjes:2006cw}. To find $\partial\mathcal{M}$ we need to know how many of these points are related by the action of $O(1,17;\mathbb{Z})$. As discussed in~\cite{Keurentjes:2006cw,Israel:2025ouq} these fall into two classes, one set is related to the large $R$ limit and one is represented by a particular Wilson line $\omega_E$ lying at $R\to 0$ on the edge of the fundamental domain. Thus, like the $\Gamma_2\backslash\mathbb{H}^2$ example in Section~\ref{sec:funddom}, the boundary is two disjoint points 
\begin{equation}
    \partial\mathcal{M} = \{ \infty \} \cup \{ \omega_E \} \, . 
\end{equation}
For definiteness we choose to be in the $E_8\times E_8$ frame\footnote{The fixed Wilson line is then given by $\omega_E=(0^7,1,-1,0^7)$ in a standard basis for the $\mathfrak{e}_8\oplus\mathfrak{e}_8$ algebra~\cite{Keurentjes:2006cw}.} so that the $\{\infty\}$ point corresponds to the usual decompactification back to the $E_8\times E_8$ ten-dimensional theory. By making the standard Ginsparg transformation~\cite{Ginsparg:1986bx} between the two equivalent lattices, one can show that the $\{\omega_E\}$ point with $R\to0$ corresponds to the $R\to \infty$ decompactification of the $SO(32)$ theory. Thus in summary, the boundary has two points corresponding to the decompactifications to the ten-dimensional $E_8\times E_8$ and $SO(32)$ theories. 

Turning to the lattice of string states we have 
\begin{equation}
    q = (n, w, \sigma) \in \Pi \, , 
\end{equation}
where $n$ and $w$ label units of momentum and winding and $\sigma\in\Pi_{16}$ is a point in the $\mathfrak{e}_8\oplus \mathfrak{e}_8$ lattice. For BPS states satisfying $q^T\cdot\eta\cdot q=0$, the mass of the state in the nine-dimensional Einstein frame is 
\begin{equation*}
    m^2(q)= \ee^{\frac{4}{7}\phi}
        \Bigl( \left(n+A\cdot\sigma-\tfrac12 w A^2\right)R^{-2} + w^2R^2 + (\sigma-wA)^2
         \Bigr) = \ee^{\frac{4}{7}\phi} q^T\cdot \mathcal{H}^{-1}\cdot q \, , 
\end{equation*}
where $\mathcal{H}$ is given by~\eqref{Hetgenmetric}. Once we include the dilaton, the canonical boundary geodesics come in two classes 
\begin{equation}
    \gamma_H(t) = \Bigg\{ 
    \begin{aligned}
        R &= \ee^{\lambda_1 t}, & A&=0, & \phi &= - \tfrac{\sqrt{7}}{2} \lambda_0 t \, , \\
        R &= \ee^{-\lambda_1 t}, & A&=\omega_E, & \phi &= - \tfrac{\sqrt{7}}{2} \lambda_0 t \, ,
    \end{aligned}
\end{equation}
where $\lambda_1\geq0$ and $\lambda_0^2+\lambda_1^2=1$, with the corresponding masses 
\begin{equation}
    m^2(t,H,q) = \begin{cases}
        \ee^{-\frac{2}{\sqrt{7}}\lambda_0t} \left( e^{-2 \lambda_1t} n^2 + e^{2\lambda_1t} w^{2} + \sigma^2 \right) \, , \\
        \ee^{-\frac{2}{\sqrt{7}}\lambda_0t} \left( e^{2 \lambda_1t} (n + \omega_E\cdot\sigma - w)^2 + e^{-2\lambda_1t} w^{2} + (\sigma-w \, \omega_E)^2 \right) \, ,
    \end{cases}
\end{equation}
where we have used $\omega_E^2=2$. We see that states $q=(n,0,0)$ and $q=(-2m,2m,2m\,\omega_E)$ with $m\in\mathbb{Z}$ become massless in the first and second cases respectively\footnote{Recall that $\omega_E$ is not an element of $\Pi$ but $2\omega_E$ is. Thus we need to take two units of winding $w=2m$ in the second case in order set $\sigma-w\omega_E=0$. In the dual $SO(32)$ theory, these are momentum states that become massless as the dual radius becomes large.}. In both cases the rate of decay is 
\begin{equation}
    \alpha(H) = \tfrac{1}{\sqrt7} \lambda_0 + \lambda_1 \, , 
\end{equation}
just as in the bosonic theory~\eqref{eq:zeta-bosonic}. As there, we can then distinguish different local extrema for $\alpha(H)$ as we vary $H$. These correspond to string and Kaluza--Klein towers
\begin{equation}
\begin{aligned}
    \lambda_0&=1, & \lambda_1&=0 &&:  &&& \alpha(H) &= \sqrt{\tfrac{1}{7}} = \alpha_{\text{St}} \, , \\
    \lambda_0&=\sqrt{\tfrac18}, & \lambda_1&=\sqrt{\tfrac78} &&: &&& \alpha(H) &= \sqrt{\tfrac{8}{7}} = \alpha_{\text{KK},1} \, . 
\end{aligned}
\end{equation}
This generalises straightforwardly so that for the heterotic theory on $T^{\dt}$, the rates match those of bosonic theory on $T^{\dt}$ discussed in the previous section. As there, the information of the rates of decay is completely captured by the weights of the charge-vector representation $\mathbf{2\dt+16}$ under the Cartan subalgebra $\caa$. 

\subsubsection{Toroidal compactifications of M-theory}
\label{sec:Mtori}

Toroidal $T^{\dt}$ compactifications of M-theory have moduli spaces of the form (see Table~\ref{tab:G/K}) 
\begin{equation}
\label{eq:M-Mtheory}
    \mathcal{M} = E_{\dt(\dt)}(\mathbb{Z}) \backslash E_{\dt(\dt)} / K_{\dt} \, . 
\end{equation}
Here $E_{\dt(\dt)}$ is the split form and $E_{\dt(\dt)}(\mathbb{Z})$ is the universal Chevalley group. Elements of $\mathcal{S}=E_{\dt(\dt)}/K_{\dt}$ are parametrised by the choice of internal metric $g_{\dt}$,  3-form $C_3$ and, for $\dt\ge6$, dual 6-form $\tilde{C}_6$. The exceptional group has rank $\dt$ and one can choose the Cartan subalgebra $\caa$ such that, given $H=(\lambda_1\dots,\lambda_{\dt})\in \caa$, we have 
\begin{equation}
    \cS \ni \ee^H\cdot o \qquad \Leftrightarrow \qquad 
    g_{\dt} = \diag (\ee^{2\lambda_1}, \dots, \ee^{2\lambda_{\dt}} ) \, , \quad
    C_3 = \tilde{C}_6 = 0 \, , 
\end{equation}
which parameterises a rectangular torus with radii $R_i=\ee^{\lambda_i}$. One can then choose the subalgebra of positive roots $\cnn^+$ to be the generators of the off-diagonal elements of the metric along with non-trivial $C_3$ and $\tilde{C}_6$ such that the closure of the positive Weyl chamber is given by
\begin{equation}
    \label{eqn:aplusedd}
\overline{\mathfrak{a}^+} = \{ \lambda_1 \leq \, ... \, \leq \lambda_{\dt}, \, \lambda_1 + \lambda_2 + \lambda_3 \geq 0 \} \, .
\end{equation}
Performing the dimensional reduction of eleven-dimensional supergravity on the rectangular torus gives 
\begin{equation}
    S=  \dfrac{M_{P}^{\dext-2}}{2} \int \dd^{\dext}x \sqrt{-g^E} \left(R^E - \sum_{i=1}^{\dt} \partial_{\mu} \lambda_i \partial^{\mu} \lambda_i - \frac{1}{\dext-2} (\partial(\lambda_1+ \dots + \lambda_{\dt}))^2 \right) \, ,
\end{equation}
from which we can read off the metric on $\cS$ restricted to the Cartan subalgebra 
\begin{equation}
\label{eqn:normcart}
    \langle H , H \rangle = \dd\lambda_1^2 + \dots + \dd\lambda_{\dt}^2 + \tfrac{1}{\dext-2} (\dd\lambda_1+\dots+\dd\lambda_{\dt})^2  \, .
\end{equation}

Since $E_{\dt(\dt)}(\mathbb{Z})$ is a Chevalley group, we have a single class of boundary points on $\mathcal{M}$ and  can identify 
\begin{equation}
    \partial\mathcal{M} = \overline{\caa_1^+} \, , 
\end{equation}
along with the canonical geodesics
\begin{equation}
\label{eq:Mth-conanical-geodesic}
    \gamma_H(t) = \left\{ g_{\dt} = \diag (\ee^{2\lambda_1t}, \dots, \ee^{2\lambda_{\dt}t} ), \, C_3=\tilde{C}_6=0 \right\} \, ,
\end{equation}
with $\lambda_1 \leq \, ... \, \leq \lambda_{\dt}$, $\lambda_1 + \lambda_2 + \lambda_3 \geq 0$ and 
\begin{equation}
    \lambda_1^2 + \dots + \lambda_{\dt}^2 + \tfrac{1}{\dext-2} (\lambda_1+\dots+\lambda_{\dt})^2 = 1 \, . 
\end{equation}
Points in $\partial\mathcal{M}=\overline{\caa_1^+}$ corresponding to minimal parabolic subgroups are reached by geodesic trajectories generated by generators of three types:
\begin{enumerate}
    \item all $\lambda_i >0$: decompactification to 11-dimesnional M-theory,
    \item $\lambda_1<0, \, \lambda_i>0 \text{ for } i=2,...,\dt$: decompactification to 10-dimensional Type IIA,
    \item $\lambda_{1,2}<0, \, \lambda_i>0 \text{ for } i=3,...,\dt$: decompactification to 10-dimensional Type IIB.
\end{enumerate}
For parts of the boundary represented by non-minimal parabolics, some inequalities in~\eqref{eqn:aplusedd} are saturated, meaning that some radii stay constant along the geodesic (this corresponds to partial decompactification limits to lower dimensions), or grow at the same rate.

The particle states carry KK momentum, M2-brane charge, M5-brane charge for $\dt\ge5$ and KK-monopole charge for $\dt\ge 7$. The charge lattice $\Pi$ forms a representation of $E_{\dt(\dt)}$ as given in Table~\ref{tab:G/K} and the mass of a state with charges $q\in\Pi$ is given by the standard formula~\eqref{eq:string-mass}. To see this in detail and how the decay rate $\alpha(H)$ of the massless towers depends on $H\in\overline{\caa_1^+}$, we will consider the case of $E_{4(4)}\simeq SL(5)$. The analysis for other $\dt$ follows similarly. 

\subsubsection*{Example: $E_{4(4)} \cong SL(5) $}

Compactifying on $T^4$, the M-theory particle states in the seven-dimensional theory carry internal KK-momentum charge $q_i$ and M2-brane charge $q^{ij}$ where $i=1,\dots,4$ and $q^{ij}=-q^{ji}$. Together, the charges transform in the $\overline{\mathbf{10}}$ representation of $E_{4(4)} \cong SL(5)$ forming an antisymmetric matrix 
\begin{equation}
    q_{AB} = \begin{pmatrix}
        \frac12\varepsilon_{ijkl}q^{kl} & q_i \\ -q_j & 0 
    \end{pmatrix} \, , \quad A,B=1,\dots 5 \, ,
\end{equation}
where $\varepsilon$ is the Levi--Civita symbol. Using the conventions of~\cite{Coimbra:2011ky}, the seven-dimensional Einstein-frame mass is given by 
\begin{equation*}
\begin{aligned}
    m^2(q) &= (\det g)^{-2/5} \Big( g^{ij}(q_i-\tfrac12C_{ikl}q^{kl})(q_j-\tfrac12C_{jmn}q^{mn})+\tfrac{1}{2}g_{ik}g_{jl}q^{ij}q^{kl} \Big) \\
    &= q^T \cdot \mathcal{H}^{-1} \cdot q \, ,   
\end{aligned}
\end{equation*}
where the inverse generalised metric, defined by the $\overline{\mathbf{10}}$ representation acting on $q=(q_i,q^{ij})$ as a vector, has the form
\begin{equation}\label{E4genmetric2}
    \mathcal{H}^{-1} = (\det g)^{-2/5} \begin{pmatrix}
        g^{ik} & \frac12 C_{ijm}g^{mk} \\  \frac12 g^{im}C_{mkl} & \frac12 g_{ik}g_{jl} - \frac14 g^{mn}C_{mij}C_{nkl}
    \end{pmatrix} \, . 
\end{equation}
There is again a level-matching condition which can be written as a ``section condition''
\begin{equation}
    q_{[AB}q_{CD]} = 0 \, . 
\end{equation}

For the canonical geodesic~\eqref{eq:Mth-conanical-geodesic} the mass then runs as 
\begin{equation}\label{SL5mass}
    m^2(t,H,q) = \ee^{-\frac{2}{5} t\Lambda}
        \left(\sum_i \ee^{-2t \lambda_i} q_i^2 
        + \sum_{i < j} \ee^{2t(\lambda_i+\lambda_j)} (q^{ij})^2 \right) \ ,
\end{equation} 
where $\Lambda\equiv \sum_i \lambda_i$. The fastest decay rate for a given choice of $H=(\lambda_1,\dots,\lambda_4)$ is thus  
\begin{equation} \label{eqn:su5z}
\alpha(H) =  \max_{i,j>i} 
\left\{
\lambda_i  ,
-\lambda_i-\lambda_j \right\} + \tfrac15 \Lambda
 = \frac{\lambda_1 + \lambda_2 + \lambda_3 + 6 \lambda_4}{5} 
 \, .
\end{equation}
As in the $O(\dt,\dt)$ case, we can look for the exponential rate associated to decompactification  and emergent string limits as local minima of~\eqref{eqn:su5z}. 
One finds 
\begin{equation}\label{eqn:alphasTowers}
\begin{aligned}
    H &= \sqrt{\tfrac{5}{6}}(0,0,0,1) , 
    &&& \alpha(H) &= \sqrt{\tfrac{6}{5}} = \alpha_{\text{M-KK},1} \, , \\
    H &= \sqrt{\tfrac{5}{14}}(0,0,1,1) , 
    &&& \alpha(H) &= \sqrt{\tfrac{7}{10}} = \alpha_{\text{M-KK},2} \, , \\
    H &= \sqrt{\tfrac{5}{24}}(0,1,1,1) , 
    &&& \alpha(H) &= \sqrt{\tfrac{8}{15}} = \alpha_{\text{M-KK},3} \, , \\
    H &= \sqrt{\tfrac{5}{36}}(1,1,1,1) , 
    &&& \alpha(H) &= \sqrt{\tfrac{9}{20}} = \alpha_{\text{M-KK},4} \, , \\
    H &= \sqrt{\tfrac{5}{36}}(-2,1,1,1) , 
    &&& \alpha(H) &= \sqrt{\tfrac{1}{5}} = \alpha_{\text{IIA-St}} \, , \\
    H &= \sqrt{\tfrac{5}{54}}(-1,-1,2,2) , 
    &&& \alpha(H) &= \sqrt{\tfrac{8}{15}} = \alpha_{\text{IIB-KK,3}} \, ,
\end{aligned}
\end{equation}
corresponding to partial or full decompactifications to M-theory or type IIB, or type IIA  emergent string limits. As expected, we find that the latter is the global minimum
\begin{equation}
    \alpha = \min_{H\in\overline{\caa_+^1}} \alpha(H) = \sqrt{\tfrac{1}{5}} = \alpha_{\text{St}} \, ,
\end{equation}
while the Kaluza--Klein towers have 
\begin{equation}
    \alpha_{\text{KK},n}= \sqrt{\frac{1}{5}+\frac{1}{n}} \ ,
\end{equation}
corresponding to $\dext=7$ in \eqref{eqn:alphakkstring} as expected.

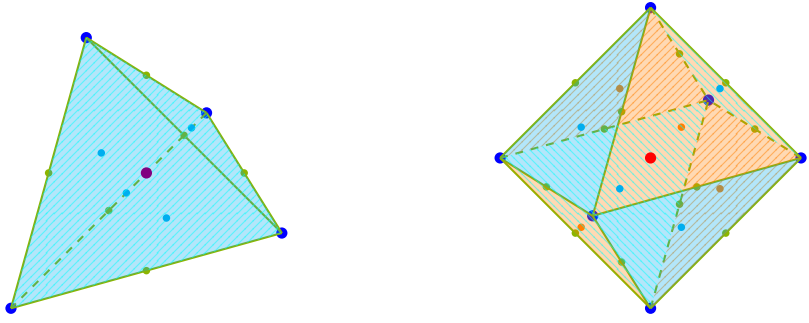
\begin{figure}[t]
\centering
\def\colone{blue}
\def\coltwo{applegreen}
\def\colthree{cyan}
\def\colinf{red}
\def\colthreeIIB{orange}
\def\colfour{violet}
\subfigure{\begin{tikzpicture}[scale=1.3]{
 \protect\coordinate (A) at (1, 1, 1);
  \protect\coordinate (B) at (-1, -1, 1);
  \protect\coordinate (C) at (-1, 1, -1);
  \protect\coordinate (D) at (1, -1, -1);
\def\opaedges{1}
\def\opafill{0.3}
\def\opafilldash{0.7}
\def\sizebig{1.5pt}\def\sizesmall{1.0pt}
\foreach \i in {A,B,C,D}{\path ($(\i)$) node[circle, fill=\colone, inner sep=\sizebig]{};};
\foreach \i/\j in {A/C,B/D,B/C,C/D,A/D}{
\draw[\coltwo, thick] (\i) -- (\j);\path  ($0.5*(\i)+0.5*(\j)$) node[circle, fill=\coltwo, inner sep=\sizesmall]{};};
\foreach \i/\j in {A/B}{
\draw[\coltwo, thick,dashed] (\i) -- (\j);\path  ($0.5*(\i)+0.5*(\j)$) node[circle, fill=\coltwo, inner sep=\sizesmall]{};};
\foreach \i/\j/\k in {A/C/D,B/C/D}{
\fill[\colthree,opacity=\opafill](\i) -- (\j) -- (\k);
\path  ($0.333*(\i)+0.333*(\j)+0.333*(\k)$)  node[circle, fill=\colthree, inner sep=\sizesmall]{};};
\foreach \i/\j/\k in {A/B/C,A/B/D}{
\path[pattern=north east lines, pattern color=\colthree,opacity=\opafilldash](\i) -- (\j) -- (\k);
\path  ($0.333*(\i)+0.333*(\j)+0.333*(\k)$)  node[circle, fill=\colthree, inner sep=\sizesmall]{};
};

\foreach \i in  {O}{\path ($(\i)$) node[circle, fill=\colfour, inner sep=\sizebig]{};};
}\end{tikzpicture}}\hspace{6em}
\subfigure{
\begin{tikzpicture}[scale=2]{
  \protect\coordinate (X)  at (1, 0, 0);
  \protect\coordinate (MX) at (-1, 0, 0);
  \protect\coordinate (Y)  at (0, 1, 0);
  \protect\coordinate (MY) at (0, -1, 0);
  \protect\coordinate (Z)  at (0, 0, 1);
  \protect\coordinate (MZ) at (0, 0, -1);
\def\opaedges{1}
\def\opafill{0.3}
\def\opafilldash{0.7}
\def\sizebig{1.5pt}\def\sizesmall{1.0pt}
\foreach \i in  {X,Y,Z,MX,MY,MZ}{\path ($(\i)$) node[circle, fill=\colone, inner sep=\sizebig]{};};
\foreach \i/\j in {X/Y,X/Z,X/MY,MX/MY,MX/Y,MX/Z,Y/Z,MY/Z}{
\draw[\coltwo, thick] (\i) -- (\j);\path  ($0.5*(\i)+0.5*(\j)$) node[circle, fill=\coltwo, inner sep=\sizesmall]{};};
\foreach \i/\j in {X/MZ,MX/MZ,MY/MZ,Y/MZ}{
\draw[\coltwo, thick,dashed] (\i) -- (\j);\path  ($0.5*(\i)+0.5*(\j)$) node[circle, fill=\coltwo, inner sep=\sizesmall]{};};
\foreach \i/\j/\k in {X/Y/Z,MX/MY/Z}{
\fill[\colthreeIIB,opacity=\opafill](\i) -- (\j) -- (\k);
\path  ($0.333*(\i)+0.333*(\j)+0.333*(\k)$)  node[circle, fill=\colthreeIIB, inner sep=\sizesmall]{};};
\foreach \i/\j/\k in {X/MY/MZ,MX/Y/MZ}{
\path[pattern=north east lines, pattern color=\colthreeIIB,opacity=\opafilldash](\i) -- (\j) -- (\k);
\path  ($0.333*(\i)+0.333*(\j)+0.333*(\k)$)  node[circle, fill=\colthreeIIB, inner sep=\sizesmall]{};};
\foreach \i/\j/\k in {MX/Y/Z,X/MY/Z}{
\fill[\colthree,opacity=\opafill](\i) -- (\j) -- (\k);
\path  ($0.333*(\i)+0.333*(\j)+0.333*(\k)$)  node[circle, fill=\colthree, inner sep=\sizesmall]{};};
\foreach \i/\j/\k in {X/Y/MZ,MX/MY/MZ}{
\path[pattern=north west lines, pattern color=\colthree,opacity=\opafilldash](\i) -- (\j) -- (\k);
\path  ($0.333*(\i)+0.333*(\j)+0.333*(\k)$)  node[circle, fill=\colthree, inner sep=\sizesmall]{};};
\foreach \i in  {O}{\path ($(\i)$) node[circle, fill=\colinf, inner sep=\sizebig]{};};}
\end{tikzpicture}}
\caption{Three-dimensional tetrahedron and octahedron facets of the tetroctahedric convex hull together with the extrema of $\alpha(H)$.}
\label{facets_E4}
\end{figure}

Again there is a simple interpretation of the different decompactifications using the weights of the charge-vector representation and their convex hull. The mass formula~\eqref{SL5mass} reflects the fact that each of the $q_i$ and $q^{ij}$ charges lies in a different weight space. The exponential decay is again given by $\left<w,H\right>$ where $w$ is the weight of the relevant weight space. Concretely we have ten weights
\begin{equation}
    \begin{gathered}
    (1,0,0,0),\, (0,1,0,0),\, (0,0,1,0),\, (0,0,0,1), \\ 
    (\tfrac13,\tfrac13,-\tfrac23,-\tfrac23),\, (\tfrac13,-\tfrac23,\tfrac13,-\tfrac23),\, 
    (\tfrac13,-\tfrac23,-\tfrac23,\tfrac13), \\
    (-\tfrac23,\tfrac13,\tfrac13,-\tfrac23),\, (-\tfrac23,\tfrac13,-\tfrac23,\tfrac13),\, 
    (-\tfrac23,-\tfrac23,\tfrac13,\tfrac13), 
\end{gathered}
\end{equation}
where the first four correspond to momenta $q_i$ and the remaining six to wrapped M2-brane charges $q^{ij}$. The convex hull formed by the weights is a ``tetroctahedric'' (also known as rectified 5-cell), which is one of the three convex semiregular 4-polytopes. It consists of:
\begin{itemize} 
\item $10$ vertices at a distance $\alpha_{\text{M-KK},1}=\sqrt{\frac{6}{5}}$, 
\item $30$ edges all of length $1$ and at a distance $\alpha_{\text{M-KK},2}=\sqrt{\frac{7}{10}}$, 
\item $30$ equilateral triangles at a distance $\sqrt{\frac{8}{15}}$:
\begin{itemize}
\item 10 of them connecting pairs of octahedrons corresponding to $\alpha_{\text{IIB-KK},3}$
\item 20 of them connecting an octahedron with a tetrahedron corresponding to $\alpha_{\text{M-KK},3}$,
\end{itemize}
\item $10$ three-dimensional facets of two types: 
\begin{itemize} \item $5$ tetrahedrons at a distance $\alpha_{\text{M-KK},4}=\sqrt{\frac{9}{20}}$,
\item $5$ octahedrons at a distance $\alpha_{\text{IIA-St}}=\sqrt{\frac{1}{5}}$, 
\end{itemize}
\end{itemize}
where all distances are measured relative to the origin. We see that again all the local extrema of the vector $\zz=\alpha(H)H$ lie on the convex hull of the weights. In particular,  we show the two types of co-dimension-1 facets in Figure~\ref{facets_E4}. The extrema corresponding to $\alpha_{\text{M-KK},1}$, $\alpha_{\text{M-KK},2}$, $\alpha_{\text{M-KK},3}$ and $\alpha_{\text{M-KK},4}$ lie respectively at the centre of vertices, edges, two and three-dimensional facets, and are depicted in blue, green, cyan and violet. The extrema corresponding to $\alpha_{\text{IIB-KK},3}$ are the orange points at the centre of the orange triangles and the one identified with $\alpha_{\text{IIA-St}}$ is the red dot at the centre of the octahedron.

\begin{figure}[t]
\centering
\def\colone{blue}
\def\coltwo{applegreen}
\def\colthree{cyan}
\def\colinf{red}
\def\colthreeIIB{orange}
\def\colfour{RedViolet}
\def\sizesmall{0.2}
\def\colChamber{yellow}
\def\colChamberLines{yellow}
\def\opaChamber{0.6}
\def\radChamber{1}
\def\opaBubbles{0.05}
\def\opaInsideHull{0.05}
\def\radone{0.02}
\def\radtwo{0.02}
\def\fontLabels{\small}

\begin{tikzpicture}[scale=4]\coordinate (A) at (-0.273861, -0.353553); \coordinate (B) at (0.182574, -0.707107); \coordinate (C) at (0.410792, -0.53033); \coordinate (D) at (1.09545, 0.); \coordinate (E) at (0.410792, 0.53033); \coordinate (F) at (0.182574, 0.707107); \coordinate (G) at (-0.273861, 0.353553); \coordinate (H) at (-0.730297, 0.);
\foreach \P/\r/\c in {A/0.223607/\colinf,B/0.365148/\colthree,C/0.33541/\colfour,D/0.547723/\colone,E/0.33541/\colfour,F/0.365148/\colthree,G/0.223607/\colinf,H/0.365148/\colthreeIIB}{\draw[dotted,thick,\c] ($0.5*(\P)$) circle (\r);
\fill[\c,opacity=\opaBubbles] ($0.5*(\P)$) circle (\r);};
\begin{scope}\clip (A)-- (B)-- (C)-- (D)-- (E)-- (F)-- (G)-- (H)-- cycle;\fill[white] (-4,-4) rectangle (4,4);\end{scope};
\path[pattern=north east lines, pattern color=\colChamber,opacity=\opaChamber] 
  (0,0) -- (127.761:\radChamber) arc[start angle=127.761, end angle=52.2388, radius=\radChamber] -- cycle;
\draw[thick,\colChamberLines] (0,0) -- (127.761:\radChamber);
\draw[thick,\colChamberLines] (0,0) -- (52.2388:\radChamber);
\draw[thick,dashed,\colChamberLines] (127.761:\radChamber) arc[start angle=127.761, end angle=52.2388, radius=\radChamber];
\foreach \P/\Q/\R/\cP/\cQ/\cR in {A/H/B/\colinf/\colthreeIIB/\colthree,C/D/B/\colfour/\colone/\colthree,E/F/D/\colfour/\colthree/\colone,G/F/H/\colinf/\colthree/\colthreeIIB}{
\fill[\cQ, opacity=\opaInsideHull](\P) -- (\Q) -- (0,0) -- cycle;
\fill[\cR, opacity=\opaInsideHull](\P) -- (\R) -- (0,0) -- cycle;
};
\foreach \P/\n/\c/\rad/\dist/\name in {
E/$4$/\colfour/\radtwo/\sqrt{\frac{9}{20}}/\alpha_{\text{M-KK},4},
F/$3$/\colthree/\radone/\sqrt{\frac{8}{15}}/\alpha_{\text{M-KK},3},
G/$\infty$/\colinf/\radtwo/\frac{1}{\sqrt{5}}/\alpha_{\text{IIA-St}}}
{
\filldraw[\c] (\P) circle(\rad);
\path ($1.2*(\P)$) node[font=\fontLabels]{$\name$};
\draw[shorten >=2pt,shorten <=2pt,\c,dashed](0,0)--(\P);
};

\foreach \P/\n/\c/\rad/\dist/\name in {
A/$\infty$/\colinf/\radtwo/\frac{1}{\sqrt{5}}/\alpha_{\text{IIA-St}},
B/$3$/\colthree/\radone/\sqrt{\frac{8}{15}}/\alpha_{\text{M-KK},3},
C/$4$/\colfour/\radtwo/\sqrt{\frac{9}{20}}/\alpha_{\text{M-KK},4}}
{\filldraw[\c] (\P) circle(\rad);};

\foreach \P/\n/\c/\rad/\dist/\name in {
D/$1$/\colone/\radone/\sqrt{\frac{6}{5}}/\tilde\alpha_{\text{M-KK},1},
H/$3$/\colthreeIIB/\radone/\sqrt{\frac{8}{15}}/\tilde\alpha_{\text{IIB-KK},3}}
{\filldraw[\c] (\P) circle(\rad);
\path ($1.2*(\P)$) node[font=\fontLabels,yshift=0.25cm]{$\name$};
\draw[shorten >=2pt,shorten <=2pt,\c,dashed](0,0)--(\P);};

\draw[thick, violet,opacity=0.2] (B) -- (D) -- (F) -- (H) -- cycle;

\filldraw[black] (0,0) circle(0.01);
\draw[->,thick] (0,0)-- (0.5, 0.)node[color=black,font=\fontLabels,
xshift=0.2cm, yshift=0.2cm]{$\lambda_1$};
\draw[->,thick] (0,0)-- (0.125, 0.484123)node[color=black,font=\footnotesize,
xshift=0.2cm, yshift=0.2cm]{$\lambda_2$};
\end{tikzpicture}
\caption{Plot of the $\alpha$-hull $\zz=\alpha(H)H$ for the $\lambda_2=\lambda_3=\lambda_4$ slice of the moduli space boundary of M-theory on $T^4$. Blue, cyan and orange regions indicate whether the leading tower is $\tilde\alpha_{\text{M-KK},1}$, $\alpha_{\text{M-KK},3}$ or $\tilde\alpha_{\text{IIB-KK},3}$.
}\label{fig:sliceOfE4}
\end{figure}
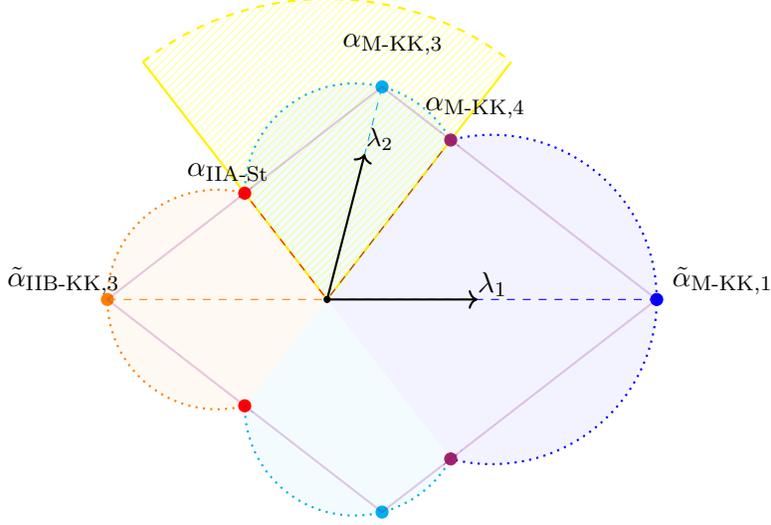

We can also plot the vector $\zz=\alpha(H)H$ for a particular slice of $\overline{\caa}$ given by $\lambda_2=\lambda_3=\lambda_4$. Its intersection with the positive Weyl chamber \eqref{eqn:aplusedd}  is given by:
\begin{equation}
\overline{\mathfrak{a}^+}_{\text{slice}} = \{ -2\lambda_2 \leq \lambda_1 \leq \lambda_2  \} \, .
\end{equation}
The slice is shown in Figure~\ref{fig:sliceOfE4} with $\overline{\mathfrak{a}^+}_{\text{slice}}$ shaded in yellow. We note that only $\alpha_{\text{M-KK},3}$, 
$\alpha_{\text{M-KK},4}$ and $\alpha_{\text{IIA-St}}$ in~\eqref{eqn:alphasTowers} are contained in $\overline{\mathfrak{a}^+}_{\text{slice}}$, shown in cyan, violet and red respectively. This happens because the choice of slice is forcing three of the radii of the $T^4$ to be equal. However, letting $H$ run over the whole of $\mathfrak{a}_{\text{slice}}$ we see that the slice includes points that are dual to decompactifying to M-theory in eight dimensions and IIB in ten dimensions, points we denote by $\tilde\alpha_{\text{IIB-KK},3}$ and $\tilde\alpha_{\text{M-KK},1}$. The diamond is the projection of the convex hull of the weights onto the slice showing again that the extrema lie on the convex hull.

Note that we can repeat this calculation for objects that live in any representation space of $SL(5)$. In particular, we could take the ${\bf 5}$ representation. From Table~\ref{tab:G/K} we see that this space encodes the charges of objects that are string-like in the external space, which arise from  branes wrapping all but one dimension of the internal space. They are labelled by $\hat{q}^i$ for membranes wrapping a one-cycle in $T^4$, and $\hat{q}^{ijkl}=\hat{q}^5\varepsilon^{ijkl}$ for five-branes wrapping the whole of $T^4$. Together they build a five-vector $\hat{q}^A=(\hat{q}^i,\hat{q}^5)$. Their tensions are given by 
\begin{equation} \label{SL5mass2}
    T(\hat{q}) = (\det g)^{-4/5} \left( g_{ij}(\hat{q}^i+\hat{C}^i\hat{q}^5)(\hat{q}^j+\hat{C}^j\hat{q}^5) + (\det g)(\hat{q}^5)^2 \right) 
    = \hat{q}^T \cdot \hat{\mathcal{H}}^{-1} \cdot \hat{q} \, ,
\end{equation}
where $\hat{C}^i=\frac16\varepsilon^{ijkl}C_{jkl}$ and again $\varepsilon$ is the numerical Levi--Civita symbol, so that $\hat{C}^i$ is a vector-density. The inverse generalised metric for the $\mathbf{5}$ representation has the form\footnote{The generalised metrics for the $\mathbf{5}$ and $\overline{\mathbf{10}}$ representations are related by $\mathcal{H}^{-1\, AB,CD}=\frac12(\hat{\mathcal{H}}^{AC}\hat{\mathcal{H}}^{BD}-\hat{\mathcal{H}}^{AD}\hat{\mathcal{H}}^{BC})$.}
\begin{equation}
    \hat{\mathcal{H}}^{-1} = (\det g)^{-4/5} \begin{pmatrix}
        g_{ij} & \hat{C}^ig_{ij} \\
        g_{ij}\hat{C}^j & (\det g) + g_{ij}\hat{C}^i\hat{C}^j
    \end{pmatrix} \, . 
\end{equation}
Taking the canonical geodesics we get 
\begin{equation}
    T(\hat{q},H,t) = \ee^{-\frac{4}{5} t\Lambda}
    \left(  \sum_{i=1}^4 e^{2t \lambda_i} (\hat{q}^i)^2 + 
        \ee^{2t \Lambda} (\hat{q}^5)^2 \right) \ ,
\end{equation}
so that the leading decay rate is 
\begin{equation}
\alpha(H) =  \max_i \left\{ -\lambda_i, -\Lambda \right\} + \tfrac{2}{5}\Lambda \ .
\end{equation}
We find the local minima 
\begin{equation}
\begin{aligned}
    H &= \sqrt{\tfrac{5}{36}}(1,1,1,1) , 
    &&& \alpha(H) &= \sqrt{\tfrac{1}{20}} , &&&& \text{M2} \\
    H &= \sqrt{\tfrac{5}{6}}(0,0,0,1) , 
    &&& \alpha(H) &= \sqrt{\tfrac{2}{15}}  , &&&& \text{M2}\\
    H &= \sqrt{\tfrac{5}{54}}(-1,-1,2,2) , 
    &&& \alpha(H) &= \sqrt{\tfrac{3}{10}} , &&&& \text{IIB F1/D1}\\
    H &= \sqrt{\tfrac{5}{36}}(-2,1,1,1) , 
    &&& \alpha(H) &= \sqrt{\tfrac{4}{5}} ,  &&&& \text{IIA F1}\\
\end{aligned}\label{minima5irrep}
\end{equation} 
corresponding to M2, F1 and D1 states becoming tensionless at the boundary of the moduli space. We note that, in the first line, the M2 can wrap any circle in $T^4$, while in the second it wraps the only circle whose radius shrinks to zero. The first decay rate gives the global minimum
\begin{equation}
\alpha = 
\min_{H\in\overline{\caa_+^1}} \alpha(H) = \tfrac{1}{2\sqrt{5}} = \tfrac12\alpha_{\text{St}} \, .
\end{equation}
Note this rate is lower than the lower bound $\alpha_{\text{St}}$, but this is not surprising since the latter applies to point-like states, which are in the ${\bf 10}$ representation. 

The convex hull of the weights in the $\mathbf{5}$ representation is a ``hypertetrahedron'' (also known as 5-cell or 4-simplex), one of the six convex regular 4-polytopes. It consists of 5 vertices joined by 10 edges, 10 triangles and 5 tetrahedra. The distances from the origin of each of these pieces are given by the corresponding $\alpha(H)$ in \eqref{minima5irrep}.

As explained in \cite{ourpaper}, we could consider the group $SL(5,\mathbb{R})\times \mathbb{R}^+$ and set the charge under the $\mathbb{R}^+$ for the particles to be proportional to $\sqrt{\frac{1}{5}+\frac{1}{\dext-2}}$ for some integer $\dext \geq 3$. The weights form the base of a 4-simplicial cone, and  the distance to each face of the 4-simplex to the origin in that case is given by
\begin{equation}
\alpha_{n} =  \sqrt{\frac{1}{\dext-2} + \frac{1}{n}}  \ ,
\end{equation}
matching the pattern~\eqref{eqn:alphakkstring} for the chosen value of $\dext$.

\subsubsection{Universal vector multiplet of IIA on Calabi--Yau}
\label{sec:UniV}

As a final example not based on toroidal compactifications, we consider the universal vector multiplet moduli space that arises from reductions of type IIA on any Calabi--Yau manifold. Dimensional reduction of the ten-dimensional supergravity theory gives a sector of the form 
\begin{equation}
    S=  \dfrac{M_{P}^{2}}{2} \int \dd^{4}x \sqrt{-g^E} \left(R^E - \tfrac{6}{\tau_2^2}\left((\partial\tau_1)^2 + (\partial\tau_2)^2 \right) \right) \, ,
\end{equation}
with $\tau=b+\ii\,\ee^{2\lambda}$, where $\ee^{6\lambda}$ is the volume modulus of the Calabi--Yau and $b$ is the modulus for taking a B-field $B\sim b\,\omega$ where $\omega$ is the Kähler form on the Calabi--Yau manifold. One can view this as a slice of the action~\eqref{eq:reduced-action} for the bosonic theory on $T^6$, viewed as a Calabi--Yau manifold, setting the four-dimensional dilaton $\lambda_0$ to zero and the six radii equal with $\lambda_i=\lambda$. Locally they give coordinates on a patch of $SL(2,\mathbb{R}) / SO(2)$. The normalisation is such that the root has length $2/3$, matching the third line of table~\ref{tab:sk} in the appendix, and so is consistent with the moduli-space metric being special Kähler as required for an $\mathcal{N}=2$ four-dimensional vector multiplet. 

Of course, the moduli space metric gets $\alpha'$ string corrections (but no loop corrections, since the dilaton is in a separate hypermulitplet) which will stop it being a symmetric space. However, these will be suppressed at large volume and so we expect as $\lambda\to\infty$ the moduli space is approximately locally symmetric, that is
\begin{equation}
\label{eq:M-UV}
    \left.\mathcal{M}\right|_{\text{large $\lambda$}} \approx \left.SL(2,\mathbb{Z}) \backslash SL(2,\mathbb{R}) / SO(2) \right|_{\text{large $\lambda$}}\, . 
\end{equation}
Note that in this limit, the relevant part of the duality group is just shifts of $B$ given by $b\to b+n$ for some $n\in\mathbb{Z}$ since T-duality-like transformations take us out of the large $\lambda$ region. Nonetheless we can still use our formalism to describe what happens at the large $\lambda$ cusp. (Put another way we are working in the Siegel set $S_{P_1,T_1}$ in Figure~\ref{fig:Siegel}.)

There are four-dimensional particle states coming from D0-branes, D2-branes wrapped on 2-cycles Poincaré dual to the cohomology class of $\frac12\, \omega\wedge\omega$, D4-branes wrapped on 4-cycles Poincaré dual to the cohomology class of $\omega$ and D6-branes wrapping the whole Calabi--Yau space. Along the canonical geodesic that goes to the large-volume boundary point
\begin{equation}
    \tau(t) = \ii\, \ee^{2t}
\end{equation}
the mass-squared of a particle of D-brane charge $q=(q_0,q_2,q_4,q_6)$  is given (in the four-dimensional Einstein frame) by 
\begin{equation}
    m^2(t,q) = \ee^{-6t}q_0^2 + \ee^{-2t}q_2^2 + \ee^{2t}q_4^2 + \ee^{6t}q_6^2 \, .  
\end{equation}
We see that the states transform in the four-dimensional representation of $SL(2,\mathbb{R})$ with weights $w=\pm1,\pm3$. As $t\to\infty$ the leading massless tower comes from the D0-branes. One can make a similar analysis of the universal hypermultiplet space based on $SU(2,1)/U(2)$ although in this case the relevant states are external strings that come from fundamental strings, D4-branes on three-cycles and NS5-branes on four-cycles, all in the eight-dimensional adjoint representation of $SU(2,1)$. 

\section{The Swampland Distance Conjecture for symmetric spaces}
\label{sec:weight-polytopes}
In the previous section, we saw how the familiar string and M-theory examples of locally symmetric moduli spaces satisfying the SDC can be understood using a group-theoretic language. 

We now use this formalism to give a ``bottom-up'' proof of the SDC for generic locally symmetric spaces, independent of the high-energy theory from which the moduli space arises. We will assume only the following:
\begin{enumerate}
    \item[\textbf{A1:}] $\mathcal{M}=\Gamma\backslash\cS$ is any locally symmetric space, provided it contains no $\mathfrak{so}(1,k)$  or $\mathfrak{su}(1,k)$ factors in the decomposition~\eqref{eq:simple-decomp} and $\Gamma=\Gamma_1\times\dots\times \Gamma_s$ with $\Gamma_i\subset G_i$;
    \item[\textbf{A2:}] the asymptotic volume growth of any geodesic ball of radius $R$ in $\mathcal{M}$ is no faster than $R^{\tilde{d}}$ where $\tilde{d}=\dim \mathcal{M}$;
    \item[\textbf{A3:}] there is a ``semi-complete'' set of states in the theory with a non-trivial monodromy group and for which at least one state becomes massless somewhere on the boundary of the moduli space. 
\end{enumerate}
(We will define the notion of ``semi-complete'' below.) A2 is the ``compactifiability'' condition of~\cite{Delgado:2024skw}, argued to be a consequence of the finiteness of quantum gravity amplitudes. The condition on $\Gamma$ and the exclusion of the rank-one factors $\mathfrak{so}(k,1)$ or $\mathfrak{su}(k,1)$ in A1 are technical, used to show that the duality group $\Gamma$ is arithmetic. One can also simply assume $\Gamma$ is arithmetic (as is for example the case for $O(1,17)$ in the heterotic supersymmetric and non-supersymmetric theories) and the proof again goes through. Explicitly we will prove the following: 
\begin{thm*}
Under assumptions A1-A3 the SDC holds. The exponential decay rate $\alpha$ of the leading tower is a continuous but non-smooth function of the asymptotic point on the boundary. Each of its extrema lies at an extremal distance point to a face\footnote{Here we include $C$ itself as one of the faces of $C$.} of the convex hull $C$ of the weights of a representation\footnote{In comparing with~\cite{Delgado:2024skw}, we note that since $G$ is reducible every finite-dimensional representation is a sum of irreducible representations.} of $G$. 
\end{thm*}
\noindent
We will do the proof in  parts, deriving two related theorems: the first on the SDC and the second on the convex hull. As we will see, the analysis also proves a number of other conjectures closely related to the SDC, including some of those in~\cite{Etheredge:2023usk}.

\subsection{Proof of the SDC}

The first part of the proof relies on a number of theorems from the literature on locally symmetric spaces. We use these and the first two assumptions to show that the volume of $\mathcal{M}$ is finite and that the duality group $\Gamma$ is arithmetic. 

The volume of the geodesic ball $B(R)$ of radius $R$ in a globally symmetric space $\mathcal{S}=G/K$ scales exponentially as (see for example~\cite{Kneiper1997})
\begin{equation}
\label{eq:ball-volume}
    \vol B(R) \approx R^{(r-1)/2} \ee^{2\lVert\rho\rVert R} \, , 
\end{equation}
where $r=\rank G$, $\rho=\frac12\sum_{\alpha\in\Phi^+}n_\alpha \alpha$, with $n_\alpha=\dim \cgg_\alpha$, is the Weyl vector, and $\lVert\rho\rVert$ is the length of $\rho$ calculated using the Killing form. This suggests that finding non-compact quotients $\mathcal{M}=\Gamma\backslash\cS$ that scale no faster than $R^{\tilde{d}}$ is hard. This is indeed the case. Fraczyk and Gelander~\cite{Fraczyk23} have recently proven a conjecture of Margulis, which implies the following\footnote{Theorem~1.1~\cite{Fraczyk23} assumes rank greater than one.  However, as noted there, Theorem 9.13 extends the result to include rank-one spaces satisfying ``Kazhdan's property (T)'' which means it also covers the case of the rank-one spaces based on $\mathfrak{sp}(1,k)$ and $\mathfrak{f}_{4(-20)}$, leaving only those based on $\mathfrak{so}(1,k)$ and $\mathfrak{su}(1,k)$.}:  
\begin{proposition}[Corollary of Theorems 1.1 and 9.13 of~\cite{Fraczyk23}]
  Let $G'$ be a connected centre\mbox{-}free simple Lie group with Lie algebra not equal to $\mathfrak{so}(1,k)$ or $\mathfrak{su}(1,k)$ and let $\mathcal{S}'=G'/K'$ be the associated symmetric space. Let $\Gamma'\subset G'$ be a discrete group. The injectivity radius of $\mathcal{M}'=\Gamma'\backslash\mathcal{S}'$ is unbounded unless $\mathcal{M}'$ has finite volume. By A1 and A2 this implies that the semi-simple part of $\mathcal{M}$ has finite volume.
\end{proposition}
\noindent
By definition, the injectivity radius $R$ of a locally symmetric space $\mathcal{M}'$ is the radius $R$ of the largest geodesic ball in $\mathcal{M}'$ that fits inside a fundamental domain. If $R$ is large, the volume of this ball will be approximately given by~\eqref{eq:ball-volume}, and hence if the injectivity radius is unbounded then $\mathcal{M}'$ violates our compactifiability assumption A2. Using A1 one can apply this result to each $G_i/K_i$ factor in the decomposition~\eqref{eq:Sfactors} separately and hence deduce that the semi-simple part of $\mathcal{M}$ has finite volume.

We can then use the famous theorem of Margulis and its extension by Corlette to say that the finiteness of the volume implies 
\begin{proposition}[from~\cite{Margulis-arithmetic} and~\cite{Corlette}]
  The duality group $\Gamma$ is arithmetic. 
\end{proposition}
\noindent This is the key condition for deriving the SDC. As noted above, one can alternatively simply assume $\Gamma$ is arithmetic, and then the proof goes through for any locally symmetric space, including all rank-one cases. 

We now turn to the space of states $\Pi$. By definition $\mathcal{S}$ is simply connected and the fundamental group of $\mathcal{M}$ is therefore the duality group $\Gamma$. In general, as we move around a loop in $\mathcal{M}$ we expect the states to come back to themselves up to some monodromy. That is to say, the states should transform under some real representation $\rho:\Gamma\to GL(\dimV,\mathbb{R})$. Thus if $V\simeq\mathbb{R}^{\dimV}$ is the vector space on which the representation acts, we have a space of states $\Pi\subset V$ that is invariant under the action of $\rho$. We can then use Margulis's famous ``superrigidity'' result (which actually is the origin of the arithmeticity theorem just mentioned) along with Corlette's extension to give\footnote{The theorems of~\cite{Margulis-rigidity} and~\cite{Corlette} strictly also require topological conditions involving, for example, the Zariski closure of $\rho(\Gamma)$ which we will assume hold. Also we assume there is a countably infinite number of states in the theory so that $\Pi\subseteq L$.}
\begin{proposition}[from~\cite{Margulis-rigidity} and~\cite{Corlette}]
    The representation $\rho:\Gamma\to GL(\dimV,\mathbb{R})$ extends to a representation $\rho: G\to GL(\dimV,\mathbb{R})$, and there is a lattice $L\simeq\mathbb{Z}^{\dimV}\subset V$ invariant under the action of $\Gamma$ such that $\Pi\subseteq L$. 
\end{proposition}
\noindent This result is particularly useful because, since $G$ is reductive, we can analyse representations of $G$, and hence the space of states $\Pi$, using weight spaces. 

As in the string theory examples, the spectrum of states $q\in\Pi$ can depend on the moduli $\phi$, but should be invariant under the action of $\Gamma$. In other words, if $g_\phi\in G$ is a representative of a point in $G/K$, the mass of the state $q\in\Pi$ should be a function invariant under the action of $\Gamma$ and $K$ on $g_\phi$ and $q$. Since $q$ transforms linearly under $\Gamma$ this implies that the moduli dependence of the mass is through the dressed charge $v(q,\phi)$ given in~\eqref{eq:dressed-q}, 
\begin{equation}
    m^2(q,\phi) = f( v(q,\phi)) \, , 
\end{equation}
where $f$ is some $K$-invariant function.\footnote{It can of course also depend on $G$-invariant combinations of $q$, such as $\eta(q,q)$ in the string $O(\dt,\dt)$ case, but these introduce no moduli space dependence.} 

To understand how the SDC is satisfied, we start by observing how $v(q,\phi)$ changes as one goes to the boundary. Since $\Gamma$ is arithmetic, for each point on the boundary $\partial\mathcal{M}$ there is a canonical infinite distance geodesic of the form~\eqref{eq:rational-geodesic}
\begin{align}
    \gamma_H(t)=\ee^{tH}\cdot o \, , 
\end{align}
where $H\in\caa_{P(\mathbb{Q}),1}^+$ for some rational parabolic subgroup $P(\mathbb{Q})$. Viewed as a representation space of $G$ (or $G(\mathbb{Q})$), we can decompose $V$ into weight spaces $V_w$ of fixed weight $w$ with respect to the action of the Cartan subalgebra $\caa$. We have
\begin{align}
    V = \bigoplus_{w\in W} V_w \, ,
\end{align}
where $W$ is the set of weights. Furthermore, since all Cartan subalgebras are conjugate under the action of $G(\mathbb{Q})$, we can always choose $\caa$ to be aligned with $P(\mathbb{Q})$ so that $\caa_{P(\mathbb{Q})}\subset \caa$. 

The dressed charged vector as one moves along the geodesic is then given by\footnote{Here we are viewing the weight $w$ as an element of $\caa$ rather that $\caa^*$.}
\begin{align}
\label{eq:dressed-canon}
    v(t,H,q) = \rho(\ee^{-tH})\cdot q
       = \sum_{w\in W} \ee^{-t\langle w,H\rangle} q_w \, , 
\end{align}
where $q_w$ is the component of the charge vector $q$ in the weight space $V_w$ and the sum is over the weights appearing in the representation space. Thus we see it is the inner products $\langle w,H\rangle$ that control the exponential decay (or growth) of the dressed charge vector as one approaches the boundary of $\mathcal{M}$. Any representation will have at least one weight in the closure of the positive Weyl chamber $\overline{\caa_{P(\mathbb{Q})}^+}$ and so at least one $\langle w,H\rangle$ will be positive. Picking out the fastest decay we have 
\begin{align} \label{eq:alpha-w*}
    \alpha(H) = \max_{w\in W} \langle w,H\rangle > 0 \, , 
\end{align}
and can define the subspace of fastest decaying states\footnote{Typically there is only a single weight space in $V_*$. However, for certain $H$ there can be multiple spaces with the same $\langle w,H\rangle$.}
\begin{align}
    V_* = \sum_{\substack{w\in W \\ \langle w,H\rangle =\alpha(H)}} V_w \, . 
\end{align}
Physically speaking, any state $q_*\in V_*$ has the fastest decaying dressed charge vector
\begin{equation}
    v(t,H,q_*) = \ee^{-t\alpha(H)} q_* \, , 
\end{equation}
for a given fixed $H$. These are the states that we will argue become massless. 

\begin{figure}[t]
	\begin{center}
\begin{tikzpicture}[scale=1.5]
\begin{scope}
\def\pointsize{1.5pt}
\pgftransformcm{0.4}{0}{0.08}{0.4}{\pgfpoint{1cm}{1.5cm}} 
\foreach \x in {-6,-5,...,3}{\foreach \y in {-4,-3,...,2}{we have drawn
\node[draw,circle,inner sep=\pointsize,fill] at (\x,\y) {}; }}
\draw [ultra thick,-latex,NavyBlue] (-4,-3) -- (-3,-3) node [below right] {$e_1$};
\draw [ultra thick,-latex,NavyBlue] (-4,-3) -- (-4,-2) node [above left] {$e_2$};
\draw [ultra thick,RedViolet] (-5.8,-4.2) -- (3.5,2) node [above right] {$V_{*}$};
\node[draw,circle,inner sep=\pointsize,fill=RedViolet] at (-4,-3) {};
\node[draw,circle,inner sep=\pointsize,fill=RedViolet] at (-1,-1) {};
\node[draw,circle,inner sep=\pointsize,fill=RedViolet] at (2,1) {};
\end{scope}
\end{tikzpicture}
    \caption{Lattice points in $V_{*}\cap \Pi$ for a two-dimensional representation.}
	\label{fig:Vw*}
	\end{center}
\end{figure}
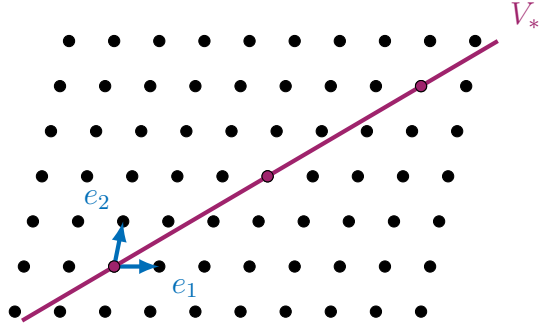

First the question remains however as to whether there are any such states, or better still if there is an infinite tower of such states, lying in $V_*\cap \Pi$. Recall that $\Pi\subset L$ where $L$ is a lattice in $V$. We then have 
\begin{proposition}
\label{prop:V*}
    $V_*\cap L$ contains an infinite number of points. 
\end{proposition}
The fact that boundary points are defined by rational parabolic subgroups is key to this result. By definition, the Cartan subalgebra relevant to $P(\mathbb{Q})$ is defined over the rationals, that is, the adjoint action of elements in $\caa$ is diagonalisable over $\mathbb{Q}$, which in turn means the action of $\caa$ on $V$ is also diagonalisable over $\mathbb{Q}$. Let $\{e_I\}$ be a basis for the lattice $\Pi$, so that $q=q^I e_I$. By definition, the weight spaces each have a basis $\{f_{w,\hat{a}}\}$ that is diagonal with respect to the action of $\caa$. (The extra index $\hat{a}$ runs over the dimension of $V_w$, given that the weight spaces are not necessarily one-dimensional.) Since the action of $\caa$ can be diagonalised over the rationals, the two bases can be related by a \emph{rational matrix} $R\in GL(\dimV,\mathbb{Q})$. 
In particular, we can write basis vectors for $V_*$ as
\begin{align}
\begin{aligned}
    f_{*,1} &= r_1^1  e_1 + \dots + r_1^{\dimV} e_{\dimV},  \\
    f_{*,2} &= r_2^1  e_1 + \dots + r_2^{\dimV} e_{\dimV},  \\
    & \dots  \\
    f_{*,\dimVstar} &= r_{\dimVstar}^1  e_1 + \dots + r_{\dimVstar}^{\dimV} e_{\dimV},  
\end{aligned}   & \qquad \qquad 
    \qquad r_{\hat{a}}^I \in \mathbb{Q} \, , 
\end{align}
where $\dimVstar$ is the dimension of $V_*$. Focussing on $f_{*,1}$, if $c$ is the lowest common denominator of $\{r_1^1,\dots,r_1^{\dimV}\}$, we immediately see that $V_{*}$ contains an infinite number of lattice points given by 
\begin{align}
    q_{\dimV} =  k c (r_1^1  e_1 + \dots + r_1^{\dimV} e_{\dimV}) \in V_{*}\cap L \, , 
    \qquad k\in \mathbb{Z} \, . 
\end{align}
By considering multiple basis vectors $f_{*,\hat{a}}$, we get a tower with $\mathbb{Z}^{\dimVstar}$ states. For the simple case of a two-dimensional representation space $V$ with a one-dimensional $V_{*}$ space, this is depicted in Figure~\ref{fig:Vw*}. 

As we move around the boundary of $\mathcal{M}$ by varying $H\in\caa_{P(\mathbb{Q}),1}^+$, the rate $\alpha$ in~\eqref{eq:alpha-w*} will change, as will the weight space $V_{*}$ with the fastest decay. Moving to a point on the boundary with a different rational parabolic group $P(\mathbb{Q})$, the relevant Cartan $\caa$ subalgebra may change but it will still be diagonalisable over the rationals. Thus the argument for the infinite tower of massless states applies to all points on the boundary. 

Finally, note that we used the canonical geodesic~\eqref{eq:rational-geodesic} in making the argument. One would expect that any other geodesic in the same class would have the same set of states appearing. This is indeed that case as we will now show, namely that changing the representatives does not change the rate of exponential decay. To see this, recall that a generic geodesic has the form~\eqref{eq:gen-geo-class} where $P$ is the real locus of the rational parabolic $P(\mathbb{Q})$. Thus for $q$ lying only along $q_{*}$, the corresponding dressed vector is 
\begin{align}
    v(t,q_*) = \rho(\gamma(t)^{-1}) \cdot q_{*}
       = \rho(m^{-1}\, \ee^{-tH} a^{-1} n^{-1}) \cdot q_{*} \, , 
\end{align}
where $m\in M_{P(\mathbb{Q})}$ is a representative of the point in $M_{P(\mathbb{Q})}/K_{P(\mathbb{Q})}$. Next we argue that $\rho(n^{-1}) \cdot q_*\in V_*$. Recall that $\cnn_{P(\mathbb{Q})}$ is a sum of positive root spaces $\cgg_\beta$ and consider $q'=\tilde{\rho}(X)\cdot q_{*}$ for some $X\in \cgg_\beta$ where $\tilde{\rho}$ is the Lie algebra representation induced by $\rho$. By definition, $\tilde{\rho}(H)\cdot q'= (\alpha(H)+\langle H, \beta\rangle)q'$ and, since $\beta$ is a positive root and $H\in\caa_{P(\mathbb{Q}),1}^+$, we have also $\langle H, \beta\rangle\geq 0$. But by definition $\alpha(H)$ is the maximum possible eigenvalue  for the action of $H$ on an element of $V$, so we must have either $q'=0$ or $\langle H, \beta\rangle =0$. Either way, when we exponentiate, we get $\rho(n^{-1})\cdot q_*\in V_*$ as required, and hence it is still an eigenstate of the action of $H$ with eigenvalue $\alpha(H)$. Since $a^{-1}$ and $\ee^{-tH}$ commute we then have 
\begin{align}
    v(t,q_*) = \rho(m^{-1}a^{-1})\rho(\ee^{-tH})\rho(n^{-1})\cdot q_{*} 
       = \ee^{-t\alpha(H)} \rho(m^{-1}a^{-1}n^{-1}) \cdot q_{*} \, . 
\end{align}
We have therefore shown that this dressed vector has the same rate of decay along any geodesic in the class.  In the context of string compactifications, this means that geodesics that go to the boundary in directions that combine radii and $B$-field have the same exponential decay rate as the the ones moving purely along radii. 

Finally we need to argue that the states in $V_*$ become massless. For this we first invoke assumption A3 to argue
\begin{proposition}
\label{prop:mass}
    Generically there exists a representation space $V$ such that, for all $q\in\Pi\subset V$, as one approaches the boundary 
    \begin{equation}
    \label{eq:asymp-mass}
        m^2(q,\phi) \approx \left|v(q,\phi)\right|^2 + \dots \, , 
    \end{equation}
    for some $K$-invariant positive-definite metric, as in~\eqref{eq:string-mass}, and where we are dropping terms of higher-order in $v(q,\phi)$. 
\end{proposition}
\noindent
To see this, recall that A3 assumed that there is at least one state $q\in\Pi$ that becomes massless at some point $p\in\partial\mathcal{M}$ on the boundary and that it transforms non-trivially under $\Gamma$ (and hence $G$). We then define $V$ to the be smallest representation of $G$ that contains the massless state (and any others that may exist at other points). Recall that $m^2(q,\phi)$ is a $K$-invariant function of the dressed vector $v(q,\phi)$ and so can be expanded in terms of invariant symmetric polynomials. If $V$ is irreducible as a $K$ representation and non-trivial, the lowest degree polynomial (other than a constant) is given by the quadratic, $K$-invariant positive-definite metric, unique up to scale. There could be higher-order invariants but these will be sub-leading in $v(q,\phi)$. We can thus expand the mass-formula in small $v(q,\phi)$ as 
\begin{equation}
    m^2(q,\phi) \approx \text{const.} + \left|v(q,\phi)\right|^2 + \dots \, . 
\end{equation}
The only way to get a massless state at the point $p\in\partial\mathcal{M}$ is thus for $v(q,\phi)\to 0$ and for the constant term to vanish, at least near $p\in\partial\mathcal{M}$. But since the expressions are $G$-invariant, the constant must vanish for all $q\in\Pi\subset V$ near $p\in\partial\mathcal{M}$ and we have~\eqref{eq:asymp-mass}. If $V$ is not irreducible then we can have a different quadratic norm for each irreducible component. However, since there must be a massless state in each component (by our definition of $V$ as the smallest representation) there will again be some positive-definite $K$-invariant metric on $V$ such that~\eqref{eq:asymp-mass} holds. 

Putting everything together we then have the first part of the theorem
\begin{thm}[SDC]
\label{thm:SDC}
    Under assumptions A1--A3, and further assuming the space of states is complete, that is $\Pi=L$, then the SDC holds with the massless states lying in $V_*$ and decaying with rate $\alpha(H)$.  
\end{thm}
\noindent
For simplicity we have slightly strengthened A3 to assume a complete set of states. Note that this is actually not the case in several of our examples because of the BPS condition, but we will see shortly how to relax this condition to a notion of ``semi-complete''. The final steps here are that, since $\Pi=L$, we have $V_*\cap\Pi$ is infinite as required from Prop.~\ref{prop:V*}. Furthermore from Prop.~\ref{prop:mass}, their masses scale as $m^2 \sim \ee^{-2t\alpha(H)}$.  

\subsection{The leading towers as the convex hull of weight spaces}

We now turn to the second part of the proof, showing that the convex hull of the leading towers of states is just the convex hull of weight spaces. This is most easily seen graphically, although we will then also give a proof. 

\begin{figure}[t]
\centering
\begin{tikzpicture}[scale=0.8]
\draw[thick, purple] (0,0) circle [radius=2.5];
\draw[->, brown, thick] (-1.75,-1.785) -- (1.75,1.785) node[midway, above left] {$\omega$};
\draw[->, NavyBlue, thick] (-1.75,-1.785) -- (0.7,-0.5) node[midway, below right] {$H$};
\draw[gray, dashed] (-2,-1.91612) -- (2.65,0.522755);
\fill[gray] (2.46361,0.424996) circle (2pt) node[right] {$\langle \omega, H \rangle$};
\end{tikzpicture}
\qquad 
\begin{tikzpicture}[scale=0.8]
\draw[thick, purple] (0,0) circle [radius=2.5];
\draw[thick, purple] (0.675,-1.17724) circle [radius=2.5];

\draw[thick, NavyBlue, decorate, decoration={snake, segment length=6pt, amplitude=1.6pt}] 
    (2.42488,0.608236) arc[start angle=15, end angle=225, radius=2.5];

\draw[thick, NavyBlue, decorate, decoration={snake, segment length=6pt, amplitude=1.6pt}]  (2.42488,0.608236) arc[start angle=45, end angle=-166, radius=2.5];

\draw[->, brown, thick] (-1.75,-1.785) -- (1.75,1.785) node[midway, above left] {$\omega$};
\draw[->, brown, thick] (-1.75,-1.785) -- (3.1,-0.569475) node[midway, above left] {$\omega'$};
\draw[applegreen, thick, dashed](-1.75,-1.785)--(2.42488,0.608236)
node[above right, applegreen] {$P$};
\fill[applegreen] (2.42488,0.608236) circle (2.pt);

\draw[->, NavyBlue, thick] (-1.75,-1.785) -- (0.234141, -2.03637) node[midway, below right,yshift=-1pt] {$H$};
\draw[gray, dashed] (-1.75,-1.785) -- (2.87817, -2.37133);
\fill[gray] (2.87817, -2.37133) circle (2pt) node[right] {$\langle \omega', H \rangle$};
\fill[gray] (1.25, -2.16506) circle (2pt) node[below] {$\langle \omega, H \rangle$};

\end{tikzpicture}
\caption{Graphic representation of $\alpha(H)$ as a function of $H$ for a single weight space and a pair of weight spaces.}
\label{fig:weight}
\end{figure}
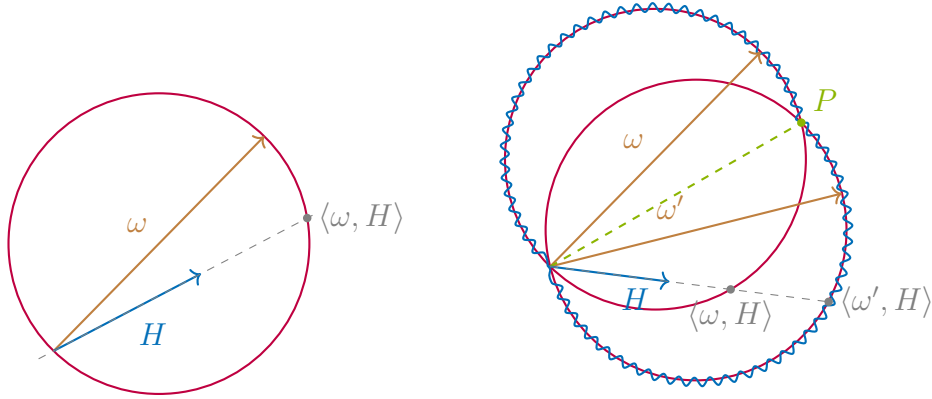

Recall that points on the boundary are labelled by a choice of rational parabolic subgroup $P(\mathbb{Q})$ and an element of the Cartan $H\in\caa_{P(\mathbb{Q}),1}^+$. The fastest decay of the dressed charge vector as one approaches this point is given by
\begin{align}
    \alpha(H) = \max_{w\in W} \langle w,H\rangle \, . 
\end{align}
To see how $\alpha(H)$ depends on $H$ we can plot the ``$\alpha$-hull'' as in~\cite{Etheredge:2023odp}. This is the locus of the 
\begin{equation}
    \zz = \alpha(H) H\;,
\end{equation}
as one varies the unit-length vector $H\in\caa_{P(\mathbb{Q}),1}$. For a single weight $w$ we get a sphere, centred on $\frac12 w$ as in the first diagram in Figure~\ref{fig:weight}. If we have two weight spaces we get a pair of spheres and the $\alpha$-hull is given by the longest distance from the origin to the union of the spheres shown by the wiggly blue line in the second diagram in Figure~\ref{fig:weight}. We see that the fastest-decay space $V_*$ switches from $V_w$ to $V_{w'}$ as $H$ changes. Furthermore there is a local minimum at $H=P$ where $\langle w,H\rangle =\langle w',H\rangle$ and we get $V_*=V_w\oplus V_{w'}$. 

\begin{figure}[t]
	\begin{center}
 \begin{tikzpicture}[scale=1.3]
    \node[above right, xshift=0.1cm, yshift=0.1cm] at (1, 0) {$\rho_{\mathbf{3}}$};
    \filldraw[blue, thick] (1, 0) circle(0.05);
    \filldraw[blue, thick] (-0.5, -0.866025) circle(0.05);   
    \filldraw[blue, thick] (-0.5,0.866025) circle(0.05);
    \draw[blue,thick] (1, 0)--(-0.5, -0.866025)--(-0.5,0.866025)--cycle;
    \draw [thick,dotted,blue] (1,0) arc (0:120:0.5);
    \draw [thick,dotted,blue] (1,0) arc (0:-120:0.5);
       \draw [thick,dotted,blue] (-0.5,0.866025) arc (120:240:0.5);
          \draw [thick,dotted,blue] (-0.5,0.866025) arc (120:0:0.5);
   \draw [thick,dotted,blue] (-0.5,-0.866025) arc (-120:-240:0.5);
      \draw [thick,dotted,blue] (-0.5,-0.866025) arc (-120:0:0.5);
   
    \filldraw[cyan, thick] (2.5, -0.866025) circle(0.0);
    \filldraw[cyan, thick] (2.5,0.866025) circle(0.0);
    \filldraw[cyan, thick] (-0.5,2.598) circle(0.0);
    \filldraw[cyan, thick] (-2,1.73205) circle(0.0);
    \filldraw[cyan, thick] (-0.5, -2.598) circle(0.0); 
    \filldraw[cyan, thick] (-2,-1.73205) circle(0.0);
    \draw[->](0,0)--(1.4,0);
    \draw[->](0,0)--(0.7,1.21244);  
\fill[blue, opacity=0.1](1, 0)--(-0.5, -0.866025)--(-0.5,0.866025)--cycle;
    \end{tikzpicture}
\begin{tikzpicture}[scale=1.3]
    \node[above right, xshift=0.1cm, yshift=0.1cm] at (2.5,0.866025) {$\rho_{\mathbf{15}}$};
    
\coordinate (A) at (2.5, -0.866025);
\coordinate (B) at (2.5, 0.866025);
\coordinate (C) at (-0.5, 2.598);
\coordinate (D) at (-2, 1.73205);
\coordinate (E) at (-2, -1.73205);
\coordinate (F) at (-0.5, -2.598);

\coordinate (G) at (1,0);
\coordinate (H) at (-0.5, -0.866025);
\coordinate (I) at (-0.5,0.866025) ;
\coordinate (J) at (1, -1.73205);
\coordinate (K) at (-0.5, 0.866025);
\coordinate (L) at (1,1.73205);
\coordinate (M) at (-2,0);

\foreach \P in {A,B,C,D,E,F}{\draw[dotted,thick,cyan] ($0.5*(\P)$) circle (1.32288);};

\begin{scope}
    \clip (A) -- (B) -- (C) -- (D) -- (E) -- (F)  -- cycle;
    \fill[white] (-4,-4) rectangle (4,4); 
\end{scope};

\fill[cyan, opacity=0.1] (A) -- (B) -- (C) -- (D) -- (E) -- (F) -- cycle;
\draw[cyan, thick] (A) -- (B) -- (C) -- (D) -- (E) -- (F) -- cycle;

\foreach \P in {A,B,C,D,E,F} {\filldraw[cyan] (\P) circle(0.07);}
\foreach \P in {G,H,I,J,K,L,M} {\filldraw[cyan] (\P) circle(0.05);}

\filldraw[black] (0,0) circle(0.03);
\draw[->] (0,0) -- (1.4,0);
\draw[->] (0,0) -- (0.7,1.21244);

\end{tikzpicture}

\caption{Convex hulls of the fundamental weight bases for the $\mathbf{3}$ and $\mathbf{15}$ irreducible representations of $\mathfrak{sl}(3,\mathbb{R})$. Weights are shown as dots with the highest weights labelled $\rho_{\mathbf{3}}$ and $\rho_{\mathbf{15}}$. The $\alpha$-hull given by the bubbles in dashed lines. Local minima of $\alpha(H)$ occur at edges of the convex hull \cite{Calderon-Infante:2020dhm}, where pairs of bubbles meet.}
\label{fig:SU3_3_and_SU3_3bar}
\end{center}
\end{figure}
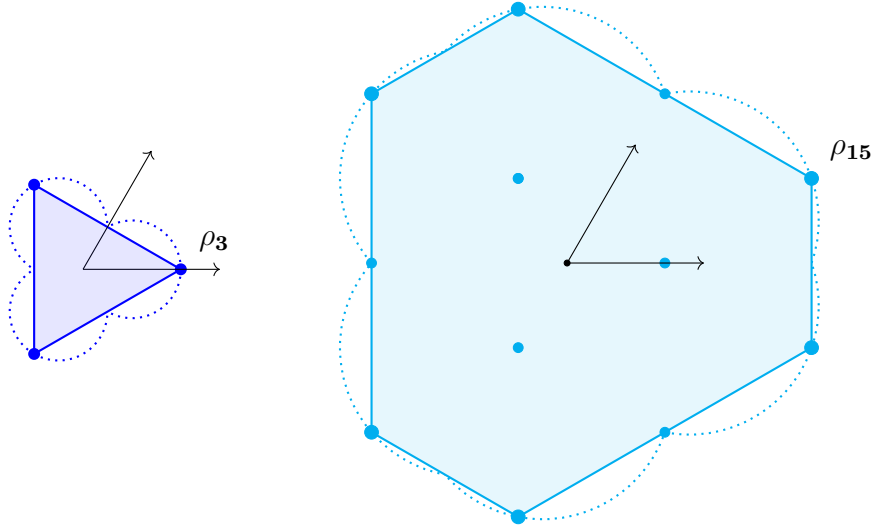

Including all the weights in the representation we get a diagram like Figure~\ref{fig:SU3_3_and_SU3_3bar}. Each weight vector defines a sphere and the $\alpha$-hull is 
shown by the dashed lines in the figure. The local extrema occur at the vertices or on edges of the convex hull of the weights, at points where bubbles meet. Recall that for a given boundary parabolic $P(\mathbb{Q})$, we should only really take $H$ in the positive Weyl chamber $\caa_{P(\mathbb{Q}),1}^+$ and so in these diagrams we should restrict to points within the sector enclosed by the two arrows. The other parts of the diagram are either related by duality transformations by elements of $\Gamma$, as in the bosonic string and M-theory on a torus examples, or alternatively correspond to positive Weyl chambers for different boundary cusps (that is, different equivalence classes of $P(\mathbb{Q})$ under $\Gamma$), as in the heterotic string example. 

For any convex hull $C$ there is a set of points called ``extremal'' which are the minimum set such that the convex hull is equal to the convex hull of the extremal set. Equivalently they are the vertices of the convex hull polytope and, to avoid confusion, we will refer to them as simply the vertex set of $C$. With this in mind, one can show the second part of our theorem
\begin{thm}[Convex hull]
\label{thm:convexhull}
    Let $C$ be the convex hull of the weights of the representation space $V$. For any $H\in\caa_{P(\mathbb{Q}),1}$ the fastest decay rate $\alpha(H)$ is equal to $\langle w,H\rangle$ for one (or more) $w$ in the vertex set of $C$. The rate $\alpha(H)$ is always greater than or equal to the distance along $H$ to any point in $C$ and, furthermore, at the extrema of $\alpha(H)$ the vector $\zz$ lies at an extremal distance point to some face of $C$. 
\end{thm}
\noindent
The proof is relatively straightforward. Let $E\subset W$ be the vertex set of weights defining the convex hull $C$. By definition $C$ is the set of points 
\begin{equation}
    C = \Bigl\{ {\textstyle \sum_{w\in E} \mu_w w : \mu_w \geq 0 , \sum_{w\in E}\mu_w=1 } \Bigr\} \, . 
\end{equation}
Given $p\in C$, because of the conditions on $\mu_w$ we have 
\begin{equation}
    \langle p,H\rangle = \sum_{w\in E} \mu_w \langle w,H\rangle
        \leq \max_{w\in E} \langle w,H\rangle \, . 
\end{equation}
Since every weight $w$ lies in $C$, this proves that for any $H$, 
\begin{equation}
    \alpha(H) = \max_{w\in E} \langle w,H\rangle \, , 
\end{equation}
as required. If $p$ lies along $H$, that is $p=lH$, the inequality then reads $l\leq\alpha(H)$. In other words, $\alpha(H)$ is greater than or equal to the distance along $H$ to any point in $C$ (corresponding to the bubbles we saw in Figure~\ref{fig:SU3_3_and_SU3_3bar}). 

To derive the condition on the extrema of $\alpha(H)$, we start by deriving the following:
\begin{proposition}
    $H$ cannot be an extremum of $\alpha(H)$ unless $\zz=\alpha(H)H$ lies inside the convex hull $C$ formed by the vertices in $E$. Furthermore, either $\zz$ lies on the boundary of $C$ or $\zz$ is the point on $C$ closest to the origin. 
\end{proposition}
\noindent
To see this, recall that it is a classic result for convex sets (known as the ``hyperplane separation theorem'') that if $\zz$ lies outside $C$, and $p_\text{min}$ is the point in $C$ closest to $\zz$, then $\langle p-\zz, p_\text{min}-\zz\rangle>0$ for all $p\in C$. (Put another way, the convex hull $C$ and $\zz$ lie either side of a plane perpendicular to $\Delta=p_\text{min}-\zz$.) Consider varying $H$ along the vector $(\Delta - \langle\Delta,H\rangle H)$ so that 
\begin{equation}
    \dd H = (\Delta - \langle\Delta,H\rangle H) \dd\gamma \, , 
\end{equation}
for some $\dd \gamma$ (note that this has $\langle H,\dd H\rangle=0$ as required). We then have 
\begin{equation}
\label{eq:slope}
    \frac{\dd\langle w,H\rangle}{\dd\gamma} = \langle w, \Delta \rangle - \langle\Delta,H\rangle \langle w, H \rangle
    = \langle w-\zz,\Delta\rangle > 0 \, , \quad\forall w\in E\, ,
\end{equation}
by the separation theorem. In general $\dd\alpha(H)/\dd\gamma$ may be discontinuous at $H$, since it might be a point where the weight giving the maximum of $\langle w, H\rangle$ changes. However, even if this is the case, the inequality~\eqref{eq:slope} implies that $\dd\alpha(H)/\dd \gamma$ cannot change sign as we pass through $H$ in the direction of $\Delta - \langle\Delta,H\rangle H$. Thus we cannot have an extremum, and instead $\zz$ must lie in $C$. 

Now consider the hyperplane $P$ through a putative extremum point $\zz=l H\in C$, perpendicular to $\zz$, defined by $P=\{ x : \langle x, H \rangle = l\}$. Again by the standard ``half-plane theorem'' this will divide $E$ into three disjoint sets, depending whether they lie above, below or on $P$. 
\begin{equation}
    E_\pm = \{ w\in E : \langle w, H \rangle \gtrless l \} \, , \qquad 
    E_0 = \{ w\in E : \langle w, H \rangle = l \} \, .     
\end{equation}
Clearly if $E_+$ is non-empty then $l$ cannot be equal to $\alpha(H)$ since $\langle w, H\rangle > l$ for any $w\in E_+$ and hence the point is excluded. Since $\zz\in C$, $E_0$ cannot be empty. This means either $E_0=E$ and $\zz$ is the minimum distance\footnote{Note that this can only occur if $\dim C< \dim \caa_{P(\mathbb{Q})}$, that is if $G$ has some abelian $\mathbb{R}^p$ factor.} to the plane through $E$, or $E_0=F\subset E$ with $E_-$ non-empty and $\zz$ lies on the boundary of $C$ and on the face  defined by $F$. 

Repeating this argument for each face subset $F$, and iterating as necessary, one gets the second part of theorem~\ref{thm:convexhull}. It also implies that an extremum on a given face $F\subset E$ is a minimum for variations within the face, and a maximum for variations away from the face. 

\begin{figure}[t]
\centering
\def\colA{orange}
\def\colB{cyan}
\def\colone{blue}
\def\coltwo{RedViolet}
\def\colthree{cyan}
\def\colinf{red}
\def\colthreeIIB{orange}
\def\colfour{RedViolet}
\def\sizesmall{0.4}
\def\opaBubbles{0.05}
\def\opaInsideHull{0.0}
\def\radone{0.05}
\def\radtwo{0.02}
\def\fontLabels{\small}
\def\scale{2}
\def\radInternal{0.035}

\begin{tikzpicture}[scale=\scale]\coordinate (A) at (-2.12132, -0.353553); \coordinate (B) at (-1.41421, -0.707107); \coordinate (C) at (0., -0.707107); \coordinate (D) at (1.41421, -0.707107); \coordinate (E) at (2.12132, -0.353553); \coordinate (F) at (2.82843, 0.); \coordinate (G) at (2.12132, 0.353553); \coordinate (H) at (1.41421, 0.707107); \coordinate (I) at (0., 0.707107); \coordinate (J) at (-1.41421, 0.707107); \coordinate (K) at (-2.12132, 0.353553); \coordinate (L) at (-2.82843, 0.);
\foreach \P/\r/\c in {B/0.790569/\colA,D/0.790569/\colA,F/1.41421/\colB,H/0.790569/\colA,J/0.790569/\colA,L/1.41421/\colB}{
\draw[dotted,ultra thick,\c] ($0.5*(\P)$) circle (\r);};
\foreach \P/\r/\c in {B/0.790569/\colA,D/0.790569/\colA,F/1.41421/\colB,H/0.790569/\colA,J/0.790569/\colA,L/1.41421/\colB}{\fill[white] ($0.5*(\P)$) circle (\r);};
\foreach \P/\r/\c in {B/0.790569/\colA,D/0.790569/\colA,F/1.41421/\colB,H/0.790569/\colA,J/0.790569/\colA,L/1.41421/\colB}{\fill[\c,opacity=\opaBubbles] ($0.5*(\P)$) circle (\r);\filldraw[\c] (\P) circle(\radone);};
\foreach \P/\Q/\R/\cP/\cQ/\cR in {A/B/L/\coltwo/\colone/\colone,C/D/B/\coltwo/\colone/\colone,E/F/D/\coltwo/\colone/\colone,G/F/H/\coltwo/\colone/\colone,I/H/J/\coltwo/\colone/\colone,K/J/L/\coltwo/\colone/\colone}{
\fill[\cQ, opacity=\opaInsideHull](\P) -- (\Q) -- (0,0) -- cycle;
\fill[\cR, opacity=\opaInsideHull](\P) -- (\R) -- (0,0) -- cycle;
\draw[\cP, thick] (\Q) -- (\R);};
\foreach \P/\n/\c/\rad/\dist/\name in {F/$1$/\colone/\radone/2 \sqrt{2}/\rho_{(\mathbf{5},\mathbf{1})},H/$1$/\colone/\radone/\sqrt{\frac{5}{2}}/\rho_{(\mathbf{3},\mathbf{2})}}{
\path ($1.1*(\P)$) node[font=\fontLabels]{$\name$};};
\foreach \P/\c in {(-1.41421, -0.707107)/\colA,(0., -0.707107)/\colA,(-1.41421, 0.707107)/\colA,(1.41421, -0.707107)/\colA,(0., 0.707107)/\colA,(1.41421, 0.707107)/\colA,(-2.82843, 0.)/\colB,(-1.41421, 0.)/\colB,(1.41421, 0.)/\colB,(2.82843, 0.)/\colB,(0., 0.)/\colB}
{\filldraw[\c] \P circle(\radInternal);};
\filldraw[black] (0,0) circle(0.01);
\draw[->,thick] (0,0)-- (0.707107,0.)node[color=black,font=\fontLabels,
xshift=0.2cm, yshift=0.2cm]{};
\draw[->,thick] (0,0)-- (0.,0.707107)node[color=black,font=\footnotesize,
xshift=0.2cm, yshift=0.2cm]{};
\end{tikzpicture}

\caption{
Convex hull (in purple) of the weights of the reducible $(\mathbf{5},\mathbf{1})\oplus (\mathbf{3},\mathbf{2})$ representation of $\mathfrak{so}(2,2)\simeq \mathfrak{sl}(2,\mathbb{R})\oplus\mathfrak{sl}(2,\mathbb{R})$. The $(\mathbf{5},\mathbf{1})$ and $(\mathbf{3},\mathbf{2})$ weights are shown as cyan and orange dots respectively. Bubbles indicate the value of $\langle \omega, H \rangle$ for each weight, while the dashed line indicates the value of $\alpha(H)$ for the leading tower in that direction, that is the ``$\alpha$-hull''. Local extrema of $\alpha(H)$ occur at faces of the convex hull, either as maxima at vertices, or minima where pairs of bubbles meet. Note that at the points where bubbles from different irreps meet, there is a discontinuity in $\dd\alpha(H)/\dd H$ but $\alpha(H)$ is not extremized.}
\label{fig:SO4reducible}
\end{figure}
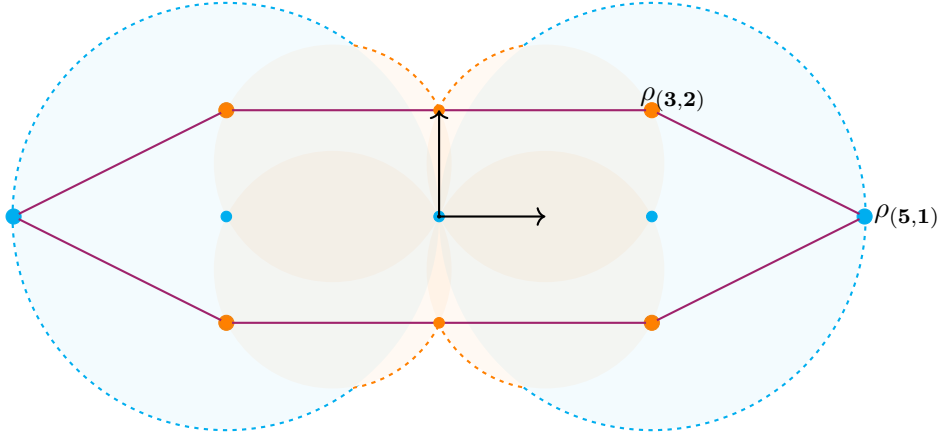

It is worth stressing that not all minimal distance points to a face give an extremum, as we see from the example of Figure~\ref{fig:SO4reducible} for which the vertex $\rho_{(\mathbf{3},\mathbf{2})}$ is not an extremum. Note further, that the face $F=\{\rho_{(\mathbf{3},\mathbf{2})},\rho_{(\mathbf{5},\mathbf{1})}\}$ is an example of where the point closest to the plane through $F$ lies outside the convex hull of $F$. Namely, where the orange and cyan bubbles meet. This is a non-smooth point of $\alpha(H)$ but not an extremum. 

\bigskip

\noindent
Finally we note that the first part of theorem~\ref{thm:convexhull} implies that for any $H$, there is some $w$ in the vertex set such that $V_w\subseteq V_*$, where $V_*$ is the subspace of fastest decaying states. This naturally allows us to relax the complete set of states condition which demanded that $\Pi$ was equal to the full lattice $L$, used in the derivation of theorem~\ref{thm:SDC}. Instead we can define 
\begin{def*}[Semi-complete spectrum]
    We say the set of states $\Pi\subset L\subset V$ is semi-complete if $\Pi\cap V_w=L\cap V_w$ for all $w$ in the vertex set of $C$. 
\end{def*}
\noindent
In other words if $w$ is in the vertex set of $C$ then every lattice point in $V_w$ is also in $\Pi$.\footnote{Since $\Pi$ and $L$ are both by definition invariant under the action of $\Gamma$, the definition of semi-complete is independent of the $P(\mathbb{Q})$ used to fix the Cartan subgroup and hence the weight spaces.} In particular, there will be an infinite number of such points. (Note in particular that the BPS-like conditions in our examples give instances of semi-complete spectra.) Since $V_*$ always contains at least one of these spaces, there is thus always an infinite set of states going massless. 

Putting everything together we finally reach the statement of the general theorem given at the beginning of this section.

\section{Summary and conclusions}
\label{sec:conclusions}
In this paper we have given a unified description of the boundary and geodesics of locally symmetric moduli spaces $\mathcal{M}$, and proved the Swampland Distance Conjecture for such spaces, under a fairly general set of assumptions. We also saw how the decay rate of the leading tower is related to the convex hull of the weights of a representation of the reductive group $G$ defining $\mathcal{M}$. Our analysis was purely group-theoretic and hence generic, not relying on any specific compactification. It is perhaps helpful to summarise the key points: 

\paragraph{Locally symmetric spaces and their geodesics:}
Non-compact locally symmetric spaces have non-positive sectional curvature and take the form $\mathcal{M}=\Gamma\backslash G/K$ where $G$ is a reductive Lie group, $K$ its maximally compact subgroup and $\Gamma$ is a discrete subgroup of $G$, equal to the fundamental group of $\mathcal{M}$. If $\Gamma$ is arithmetic then we have the following:
\begin{itemize}
\renewcommand\labelitemi{--}
    \item Points on the boundary $\partial\mathcal{M}$ are labelled by a pair $(P(\mathbb{Q}),H)$ of a \emph{rational parabolic subgroup} $P(\mathbb{Q})\subset G$ and a unit-length element of the positive Weyl chamber $H\in\caa_{P(\mathbb{Q}),1}^+$. 
    \item Two rational parabolic subgroups $P(\mathbb{Q})$ describe the same boundary point if they are related by an element of $\Gamma$. There is only a finite number of such equivalence classes each defining a different ``cusp''. 
    \item Generic geodesics in $\mathcal{M}$ are ergodic, failing to reach the boundary. For these, the distance away from the initial point scales as $\log t$ where $t$ is the distance along the geodesic.
    \item Geodesics that reach the boundary are dense within the set of all geodesics. For each boundary point $(P(\mathbb{Q}),H)$ there is a canonical geodesic of the form $\gamma(t)=\ee^{Ht}\cdot o$ among the equivalence class of geodesics that reach the same boundary point. 
\end{itemize}

\paragraph{Locally symmetric spaces and string compactifications:} 
Using this framework to reanalyse familiar classes string compactifications (notably on tori) one recalls first that the string states generically transform in some representation $\rho$ of $G$ (and hence $\Gamma$) with a simple quadratic mass formula. One then has:
\begin{itemize}
\renewcommand\labelitemi{--}
    \item The Cartan subalgebra $\caa_{P(\mathbb{Q})}$ can be chosen to align with the radii of the torus, together with the dilaton, showing that one can effectively ignore all other moduli in analysing the SDC. Different Weyl chambers of $\caa_{P(\mathbb{Q})}$ correspond to different (equivalent) duality frames.
    \item For the bosonic, type II and M theories on tori, the duality group $\Gamma$ is a Chevalley subgroup of $G$ and as a result $\partial\mathcal{M}$ is a point and as a result $\mathcal{M}$ has only one cusp. For the heterotic theory, it has two cusps corresponding to $E_8\times E_8$ and $SO(32)$ limits. 
    \item The ``tower-vectors'' $\zeta$~\cite{Calderon-Infante:2020dhm,Etheredge:2024tok} of different towers of states are simply the weights $w$ of the representation $\rho$ when decomposed using $\caa_{P(\mathbb{Q})}$ at that cusp, and the corresponding convex hull is just the convex hull of the weights of $\rho$. 
\end{itemize}

\paragraph{Proof of the SDC and convex hulls:} We were then able to prove the following, which applies to (almost) any locally symmetric moduli space, irrespective of how it arises: 
\begin{thm*}
Under assumptions A1-A3 the SDC holds. The exponential decay rate $\alpha$ of the leading tower is a continuous but non-smooth function of the asymptotic point on the boundary. Each of its extrema lies at an extremal distance point to a face of the convex hull $C$ of the weights of a representation of $G$. 
\end{thm*}
\noindent
The assumption A1 was essentially technical, slightly constraining the allowed class of symmetric spaces. Assumption A2 was the ``compactifiability'' assumption of~\cite{Delgado:2024skw}. Using a recent theorem of~\cite{Fraczyk23} these two are enough to prove that ``compactifiable'' is equivalent to finite volume (or more precisely that $\mathcal{M}$ is the product of a finite volume space and flat space) for locally symmetric spaces. A classic result of Margulis~\cite{Margulis-arithmetic} and its extension by Corlette~\cite{Corlette} then imply that $\Gamma$ is arithmetic and so we can use all the analysis of geodesics and characterisation of the boundary of $\mathcal{M}$ discussed above. 

To get a leading tower going massless and the relation to the weights we needed to make a fairly weak set of assumptions (A3) about the states. The first part was that they have non-trivial monodromy and so form a representation of $\Gamma$. By another rigidity theorem of Margulis~\cite{Margulis-rigidity} this implies the representation extends to a representation of $G$ and so can be analysed using roots and weights, since $G$ is reductive. The states then lie on some lattice $L$ inside the representation space $V$. The second part was that at least one state becomes massless somewhere on the boundary. Invariance under $K$ and $\Gamma$ then implies that as one approaches $\partial\mathcal{M}$, the mass formula generically takes a simple quadratic form.\footnote{Note that even for special cases where the leading term is higher-order, the proof of the SDC conjecture still holds, although the $\alpha$-hull analysis will change.} Finally, the third part was that the states are ``semi-complete'', meaning, not necessarily that there is a state for every point in the lattice, but there is for every lattice point that lies in a ``vertex'' weight space, that is, with a weight that lies at a vertex of the convex hull $C$. It is satisfied by the BPS-like states that appear in the string theory examples, but it would be interesting to understand this semi-complete condition, or some variant of it, in more detail. In particular, a likely alternative is that (1)~the orbits of allowed states are characterised by homogeneous conditions on the charge $q$ and (2)~if $q$ is allowed then so is $nq$ for any $n\in\mathbb{Z}$. 

\bigskip

It is of course very natural to ask how this formalism might help analyse other distance conjectures. In the companion paper~\cite{ourpaper}, we focus on the implication for the Emergent String Conjecture (EMS)~\cite{Lee:2019wij} which restricts $\alpha(H)$ to correspond to those coming from the exponential rates for string oscillators and (BPS) Kaluza--Klein towers. Assuming the ESC and that the particles states transform in an irreducible representation, we show how to classify all the possible moduli spaces, particle content and spacetime dimension that are allowed.\footnote{Technically, we restrict the analysis to the split forms of the groups $G_i$ in the decomposition~\eqref{eq:simple-decomp}. The non-split forms can be analysed analogously, but for simplicity we leave this for future work.} 

The ESC assumption turns out to be quite constraining. In particular, for symmetric moduli spaces up to normalization of the root system, all the consistent ``global'' polytopes, for which the dualities are associated to the full Weyl group of an algebra, are either the exceptional ones of M-theory on tori or associated to (regular or special) subgroups of these, and the particle content follows from branching rules of the allowed exceptional representation. Other possible theories, with smaller duality groups, would be associated to slices of the exceptional Weyl polytopes that break the Weyl group to a smaller subgroup. Moreover, for each one of the allowed theories featuring an emergent string limit, there exists (at least one) consistent representation for string-like objects in the low energy theory.

Connections to the Scalar Weak Gravity Conjecture~\cite{Arkani-Hamed:2006emk,Palti:2017elp} (and the related Convex Hull SWGC~\cite{Etheredge:2023odp}) are left to future work. However, with respect to the conjectures in~\cite{Etheredge:2023usk}, it is straightforward to see that under our assumptions A1--A3:
\begin{itemize}
\renewcommand\labelitemi{--}
    \item The ``dense-direction conjecture'' states that geodesics through a point $x\in \mathcal{M}$ that go to infinite distance are dense in the set of all geodesics through $x$. This follows directly from the fact that the set of rational parabolic subgroups $P(\mathbb{Q})$ is dense in the set of real parabolic subgroups. 
    \item The ``heavy-tower conjecture'' states that there is always an infinite tower of states with exponentially \emph{divergent} masses as one goes to the boundary. This follows, in the case with no $\mathbb{R}^p$ factor, directly from the symmetry of the convex hull under the Weyl group of $G$, implying that there will always be weights in the ``opposite Weyl chamber'' $-\overline{\caa^+_{P(\mathbb{Q})}}$, which will necessarily have $\langle w,H\rangle<0$. When there is an $\mathbb{R}^p$ factor it does not necessarily hold unless the representation has states with both positive and negative  $\mathbb{R}^p$ charge. 
\end{itemize}

A second very natural question is how one might relax the assumptions that led to the proof of the SDC, most notably the very strong one that $\mathcal{M}$ is locally symmetric. It is perhaps first worth noting that locally symmetric spaces are part of a larger class of metric spaces where one simply constrains the sectional curvature to be non-positive. (When simply-connected and complete these are known as ``Hadamard manifolds''. Thus generically a space with non-negative sectional curvature has the form $\mathcal{M}=\Gamma\backslash\cS$ where $\cS$ is Hadamard.).  Such spaces still admit a notion of rank, and strikingly ``rank rigidity''~\cite{Ballman,BurnsSpatzier} implies that if the rank is greater than one, and the space has finite volume, then it must be locally symmetric. However, as is well known, there are string examples of moduli spaces with infinite distance limits where even this bound on the sectional curvature fails to be true~\cite{trenner2011asymptotic} (although see also the discussions in~\cite{Cecotti:2021cvv,Lanza:2021udy,Raman:2024fcv,Marchesano:2023thx,Castellano:2024gwi}). The most straightforward possibility (as in our trivial IIA universal vector example) is that the moduli space near the boundary contains a factor that looks like a symmetric space cusp, or more precisely is isometric to a Siegel set $S_{P_i,T_i}$ of a symmetric space. Even then the duality group on $S_{P_i,T_i}$ is only $N^+(\mathbb{Z})$ and one may need to also make assumptions about the reducibility of the representation as in~\cite{Delgado:2024skw}.

\vspace{0.5cm}

\noindent {\bf \large Acknowledgments\vspace{0.1in}}\\
We would like to thank Jos\'e Calder\'on Infante, \'Alvaro Herráez, H\'ector Parra de Freitas, Ignacio Ruiz Garc\'ia and Irene Valenzuela for useful discussions and specially to Muldrow Etheredge and Carmine Montella for valuable comments on the manuscript. This work was supported in part by CNRS--Imperial College PhD fellowship scheme, by an STFC Doctoral Training Grant strudentship, by the ERC Consolidator Grant 772408-Stringlandscape, by the ERC Starting Grant QGuide101042568 - StG 2021 and by the STFC Consolidated Grant ST/X000575/1. V.C would like to thank the Institut de Physique Theorique at Saclay, and B.F. the Theoretical Physics Department at CERN and the Instituto de F\'isica Te\'orica in Madrid for hospitality and support during the early stages of this work.

\appendix

\section*{Appendix: Symmetric spaces and supersymmetry} \label{sec:normalisation}

In this appendix, we comment on the possible symmetric spaces compatible with supersymmetry and the corresponding normalisation of the symmetric space metric. 

All maximal and half-maximal supergravities have symmetric space moduli spaces, given in table~\ref{tab:max-halfmax} (see for example~\cite{Samtleben:2008pe}). They are all symmetric spaces and supersymmetry further constrains the metric on the moduli space is normalised so that the roots of $\cgg$ all have length two $\langle \beta,\beta\rangle = 2$. 
\begin{table}
\centering
\begin{tabular}{|c|c|c|} \hline
    $\dext$ & maximal $G/K$ & half-maximal $G/K$\\
    \hline
    $9$ & $GL(2)/SO(2)$ & $\mathbb{R}\stimes SO(1,1+\rt)/SO(1+\rt)$ \\
    $8$ & $SL(2)\stimes SL(3)/SO(3)\stimes SO(3)$ & 
        $\mathbb{R}\stimes SO(2,2+\rt)/SO(2)\stimes SO(2+\rt)$ \\
    $7$ & $SL(5)/SO(5)$ & $\mathbb{R}\stimes SO(3,3+\rt)/SO(3)\stimes SO(3+\rt)$ \\
    $6$ & $SO(5,5)/SO(5)\stimes SO(5)$ & $\mathbb{R}\stimes SO(4,4+\rt)/SO(4)\stimes SO(4+\rt)$ \\
    $5$ & $E_{6(6)}/Sp(4)$ & $\mathbb{R}\stimes SO(5,5+\rt)/SO(5)\stimes SO(5+\rt)$ \\
    $4$ & $E_{7(7)}/SU(8)$ & $SL(2)\stimes SO(6,6+\rt)/SO(2)\stimes SO(6)\stimes SO(6+\rt)$ \\
    $3$ & $E_{8(8)}/SO(16)$ & $\mathbb{R}\stimes SO(8,8+\rt)/SO(8)\stimes SO(8+\rt)$ \\
    \hline
\end{tabular}
\caption{Symmetric spaces $G/K$ for maximal and half-maximal supergravities.}\label{tab:max-halfmax}
\end{table} 

For eight-supercharge theories (for which $\dext\leq 6$) one has moduli spaces for scalars in  vector or hypermultiplets. (Or more generally the vector multiplet moduli space becomes the moduli space for vector-tensor multiplets.) Supersymmetry implies that the metric $g_{ij}$ on the moduli spaces of hypermultiplets in any dimension is quaternionic--Kähler and is Einstein 
\begin{equation}
\label{eq:QH-Einstein}
    R_{ij}=-kg_{ij} \, , 
\end{equation}
with $k=n_H+2$ whiere $n_H$ is the number of hypermultiplets. The irreducible non-compact quaternionic--Kähler manifolds that are also locally symmetric were first classified by~\cite{Alekseevsky:2005} and are listed in table~\ref{tab:qk}. In all cases we see that $2+n_H=h^\vee$ and so again we find that the metric is such that longest roots of $\cgg$ are normalised to $\langle \beta_{\text{long}},\beta_{\text{long}}\rangle=2$. 
\begin{table}
\centering
\begin{tabular}
			{|>{$}c<{$}|
				>{$}c<{$}|} \hline
G/K & 2+n_{H}=h^{\vee}  \\ \hline
 SU(p,2) / S(U(2) \times U(p)) & p+2   \\
 SO_e(p,4)/SO(p)\times SO(4) & p+2  \\
 Sp(p,1)/Sp(1)\times Sp(p) & p+2  \\
 E_{6(2)}/SU(2) \cdot SU(6) & 12  \\
 E_{7(-5)}/SU(2) \cdot Spin(12) & 18  \\
 E_{8(-24)}/SU(2) \cdot E_7 & 30 \\
 F_{4(4)}/Sp(1) \cdot Sp(3) & 9 \\
 G_{2(2)}/SO(4) & 4 \\  \hline
\end{tabular}
\caption{Non-compact symmetric spaces of quaternionic K{\"a}hler type, with number of hypermultiplets and dual Coxeter number $h^{\vee}$ of $G$. We use the ``central product'' notion $H\cdot G=H\times G/Z(H)$ where $Z(H)$ is the centre of $H$ and $G_e$ denotes the component of $G$ connected to the identity.}
\label{tab:qk}
\end{table} 

Turning to vector multiplet moduli spaces and restricting to $\dext=4$ means that $\mathcal{M}$ is special Kähler and so is again Einstein. The irreducible non-compact locally symmetric special Kähler manifolds were classified in~\cite{Cremmer:1984hc} (see also~\cite{Cecotti:2020rjq}) and are listed together with the value of the cosmological constant $k$ and the dual Coxeter number in table~\ref{tab:sk}. We see that here the longest root can be shorter, namely the normalisation is such that $\langle\beta_{\text{long}},\beta_{\text{long}}\rangle\leq 2$.
\begin{table}
\centering
\begin{tabular}
{|>{$}c<{$}|>{$}c<{$}|>{$}c<{$}|>{$}c<{$}|} \hline
G/K & k & h^{\vee} & \langle\beta_{\text{long}},\beta_{\text{long}}\rangle \\ \hline
 SU(p,1) / S(U(1) \times U(p)) & p+1 & p+1 & 2  \\
 SO(p,2) \times SL(2)/SO(p)\times SO(2)^2 & p \text{ or } 1 & p \text{ or } 1 & 2  \\
 SL(2)/SO(2) & \frac13 & 1 & \frac{2}{3}\\
 (SL(2)/SO(2))^3 & 1 & 1 & 2 \\
 Sp(6)/U(3) & 2 & 4 & 1 \\
 U(3,3)/U(3)^3 & 3 & 6 & 1\\
 SO^*(12)/U(6) & 5 & 5 & 2\\
 E_{7(-25)}/(E_6 \times SO(2) & 9 & 18 & 1\\  \hline
\end{tabular}
\caption{Non-compact symmetric spaces of special K{\"a}hler type, with cosmological constant $k$ and $h^{\vee}$ the dual Coxeter number of $G$.}\label{tab:sk}
\end{table}

\bibliographystyle{JHEP}
\bibliography{refs}

\providecommand{\href}[2]{#2}\begingroup\raggedright\begin{thebibliography}{100}

\bibitem{Delgado:2024skw}
M.~Delgado, D.~van~de Heisteeg, S.~Raman, E.~Torres, C.~Vafa, and K.~Xu, {\it {Finiteness and the Emergence of Dualities}},  \href{http://arxiv.org/abs/2412.03640}{{\tt arXiv:2412.03640}}.

\bibitem{ourpaper}
S.~Baines, V.~Collazuol, B.~Fraiman, M.~Graña, and D.~Waldram, ``Locally symmetric spaces and the sharpened distance conjecture (to appear).''

\bibitem{Vafa:2005ui}
C.~Vafa, {\it {The String landscape and the swampland}},  \href{http://arxiv.org/abs/hep-th/0509212}{{\tt hep-th/0509212}}.

\bibitem{Brennan:2017rbf}
T.~D. Brennan, F.~Carta, and C.~Vafa, {\it {The String Landscape, the Swampland, and the Missing Corner}},  {\em PoS} {\bf TASI2017} (2017) 015, [\href{http://arxiv.org/abs/1711.00864}{{\tt arXiv:1711.00864}}].

\bibitem{Palti:2019pca}
E.~Palti, {\it {The Swampland: Introduction and Review}},  {\em Fortsch. Phys.} {\bf 67} (2019), no.~6 1900037, [\href{http://arxiv.org/abs/1903.06239}{{\tt arXiv:1903.06239}}].

\bibitem{Agmon:2022thq}
N.~B. Agmon, A.~Bedroya, M.~J. Kang, and C.~Vafa, {\it {Lectures on the string landscape and the Swampland}},  \href{http://arxiv.org/abs/2212.06187}{{\tt arXiv:2212.06187}}.

\bibitem{vanBeest:2021lhn}
M.~van Beest, J.~Calder\'on-Infante, D.~Mirfendereski, and I.~Valenzuela, {\it {Lectures on the Swampland Program in String Compactifications}},  \href{http://arxiv.org/abs/2102.01111}{{\tt arXiv:2102.01111}}.

\bibitem{Grana:2021zvf}
M.~Gra\~na and A.~Herr\'aez, {\it {The Swampland Conjectures: A Bridge from Quantum Gravity to Particle Physics}},  {\em Universe} {\bf 7} (2021), no.~8 273, [\href{http://arxiv.org/abs/2107.00087}{{\tt arXiv:2107.00087}}].

\bibitem{Ooguri:2006in}
H.~Ooguri and C.~Vafa, {\it {On the Geometry of the String Landscape and the Swampland}},  {\em Nucl. Phys. B} {\bf 766} (2007) 21--33, [\href{http://arxiv.org/abs/hep-th/0605264}{{\tt hep-th/0605264}}].

\bibitem{Hamada:2021yxy}
Y.~Hamada, M.~Montero, C.~Vafa, and I.~Valenzuela, {\it {Finiteness and the swampland}},  {\em J. Phys. A} {\bf 55} (2022), no.~22 224005, [\href{http://arxiv.org/abs/2111.00015}{{\tt arXiv:2111.00015}}].

\bibitem{Stout:2021ubb}
J.~Stout, {\it {Infinite Distance Limits and Information Theory}},  \href{http://arxiv.org/abs/2106.11313}{{\tt arXiv:2106.11313}}.

\bibitem{Stout:2022phm}
J.~Stout, {\it {Infinite Distances and Factorization}},  \href{http://arxiv.org/abs/2208.08444}{{\tt arXiv:2208.08444}}.

\bibitem{Calderon-Infante:2023ler}
J.~Calder\'on-Infante, A.~Castellano, A.~Herr\'aez, and L.~E. Ib\'a\~nez, {\it {Entropy bounds and the species scale distance conjecture}},  {\em JHEP} {\bf 01} (2024) 039, [\href{http://arxiv.org/abs/2306.16450}{{\tt arXiv:2306.16450}}].

\bibitem{Lee:2018urn}
S.-J. Lee, W.~Lerche, and T.~Weigand, {\it {Tensionless Strings and the Weak Gravity Conjecture}},  {\em JHEP} {\bf 10} (2018) 164, [\href{http://arxiv.org/abs/1808.05958}{{\tt arXiv:1808.05958}}].

\bibitem{Lee:2019wij}
S.-J. Lee, W.~Lerche, and T.~Weigand, {\it {Emergent strings from infinite distance limits}},  {\em JHEP} {\bf 02} (2022) 190, [\href{http://arxiv.org/abs/1910.01135}{{\tt arXiv:1910.01135}}].

\bibitem{Basile:2023blg}
I.~Basile, D.~L{\"u}st, and C.~Montella, {\it {Shedding black hole light on the emergent string conjecture}},  {\em JHEP} {\bf 07} (2024) 208, [\href{http://arxiv.org/abs/2311.12113}{{\tt arXiv:2311.12113}}].

\bibitem{Bedroya:2024ubj}
A.~Bedroya, R.~K. Mishra, and M.~Wiesner, {\it {Density of states, black holes and the Emergent String Conjecture}},  {\em JHEP} {\bf 01} (2025) 144, [\href{http://arxiv.org/abs/2405.00083}{{\tt arXiv:2405.00083}}].

\bibitem{Heidenreich:2015nta}
B.~Heidenreich, M.~Reece, and T.~Rudelius, {\it {Sharpening the Weak Gravity Conjecture with Dimensional Reduction}},  {\em JHEP} {\bf 02} (2016) 140, [\href{http://arxiv.org/abs/1509.06374}{{\tt arXiv:1509.06374}}].

\bibitem{Blumenhagen:2017cxt}
R.~Blumenhagen, I.~Valenzuela, and F.~Wolf, {\it {The Swampland Conjecture and F-term Axion Monodromy Inflation}},  {\em JHEP} {\bf 07} (2017) 145, [\href{http://arxiv.org/abs/1703.05776}{{\tt arXiv:1703.05776}}].

\bibitem{Grimm:2018ohb}
T.~W. Grimm, E.~Palti, and I.~Valenzuela, {\it {Infinite Distances in Field Space and Massless Towers of States}},  {\em JHEP} {\bf 08} (2018) 143, [\href{http://arxiv.org/abs/1802.08264}{{\tt arXiv:1802.08264}}].

\bibitem{Heidenreich:2018kpg}
B.~Heidenreich, M.~Reece, and T.~Rudelius, {\it {Emergence of Weak Coupling at Large Distance in Quantum Gravity}},  {\em Phys. Rev. Lett.} {\bf 121} (2018), no.~5 051601, [\href{http://arxiv.org/abs/1802.08698}{{\tt arXiv:1802.08698}}].

\bibitem{Blumenhagen:2018nts}
R.~Blumenhagen, D.~Kl\"awer, L.~Schlechter, and F.~Wolf, {\it {The Refined Swampland Distance Conjecture in Calabi-Yau Moduli Spaces}},  {\em JHEP} {\bf 06} (2018) 052, [\href{http://arxiv.org/abs/1803.04989}{{\tt arXiv:1803.04989}}].

\bibitem{Lee:2018spm}
S.-J. Lee, W.~Lerche, and T.~Weigand, {\it {A Stringy Test of the Scalar Weak Gravity Conjecture}},  {\em Nucl. Phys. B} {\bf 938} (2019) 321--350, [\href{http://arxiv.org/abs/1810.05169}{{\tt arXiv:1810.05169}}].

\bibitem{Ooguri:2018wrx}
H.~Ooguri, E.~Palti, G.~Shiu, and C.~Vafa, {\it {Distance and de Sitter Conjectures on the Swampland}},  {\em Phys. Lett. B} {\bf 788} (2019) 180--184, [\href{http://arxiv.org/abs/1810.05506}{{\tt arXiv:1810.05506}}].

\bibitem{Corvilain:2018lgw}
P.~Corvilain, T.~W. Grimm, and I.~Valenzuela, {\it {The Swampland Distance Conjecture for K\"ahler moduli}},  {\em JHEP} {\bf 08} (2019) 075, [\href{http://arxiv.org/abs/1812.07548}{{\tt arXiv:1812.07548}}].

\bibitem{Grimm:2018cpv}
T.~W. Grimm, C.~Li, and E.~Palti, {\it {Infinite Distance Networks in Field Space and Charge Orbits}},  {\em JHEP} {\bf 03} (2019) 016, [\href{http://arxiv.org/abs/1811.02571}{{\tt arXiv:1811.02571}}].

\bibitem{Buratti:2018xjt}
G.~Buratti, J.~Calder\'on, and A.~M. Uranga, {\it {Transplanckian axion monodromy!?}},  {\em JHEP} {\bf 05} (2019) 176, [\href{http://arxiv.org/abs/1812.05016}{{\tt arXiv:1812.05016}}].

\bibitem{Lust:2019zwm}
D.~L\"ust, E.~Palti, and C.~Vafa, {\it {AdS and the Swampland}},  {\em Phys. Lett. B} {\bf 797} (2019) 134867, [\href{http://arxiv.org/abs/1906.05225}{{\tt arXiv:1906.05225}}].

\bibitem{Joshi:2019nzi}
A.~Joshi and A.~Klemm, {\it {Swampland Distance Conjecture for One-Parameter Calabi-Yau Threefolds}},  {\em JHEP} {\bf 08} (2019) 086, [\href{http://arxiv.org/abs/1903.00596}{{\tt arXiv:1903.00596}}].

\bibitem{Erkinger:2019umg}
D.~Erkinger and J.~Knapp, {\it {Refined swampland distance conjecture and exotic hybrid Calabi-Yaus}},  {\em JHEP} {\bf 07} (2019) 029, [\href{http://arxiv.org/abs/1905.05225}{{\tt arXiv:1905.05225}}].

\bibitem{Marchesano:2019ifh}
F.~Marchesano and M.~Wiesner, {\it {Instantons and infinite distances}},  {\em JHEP} {\bf 08} (2019) 088, [\href{http://arxiv.org/abs/1904.04848}{{\tt arXiv:1904.04848}}].

\bibitem{Font:2019cxq}
A.~Font, A.~Herr\'aez, and L.~E. Ib\'a\~nez, {\it {The Swampland Distance Conjecture and Towers of Tensionless Branes}},  {\em JHEP} {\bf 08} (2019) 044, [\href{http://arxiv.org/abs/1904.05379}{{\tt arXiv:1904.05379}}].

\bibitem{Baume:2020dqd}
F.~Baume and J.~Calder\'on~Infante, {\it {Tackling the SDC in AdS with CFTs}},  {\em JHEP} {\bf 08} (2021) 057, [\href{http://arxiv.org/abs/2011.03583}{{\tt arXiv:2011.03583}}].

\bibitem{Perlmutter:2020buo}
E.~Perlmutter, L.~Rastelli, C.~Vafa, and I.~Valenzuela, {\it {A CFT distance conjecture}},  {\em JHEP} {\bf 10} (2021) 070, [\href{http://arxiv.org/abs/2011.10040}{{\tt arXiv:2011.10040}}].

\bibitem{Klaewer:2020lfg}
D.~Klaewer, S.-J. Lee, T.~Weigand, and M.~Wiesner, {\it {Quantum corrections in 4d $N$ = 1 infinite distance limits and the weak gravity conjecture}},  {\em JHEP} {\bf 03} (2021) 252, [\href{http://arxiv.org/abs/2011.00024}{{\tt arXiv:2011.00024}}].

\bibitem{Lanza:2021udy}
S.~Lanza, F.~Marchesano, L.~Martucci, and I.~Valenzuela, {\it {The EFT stringy viewpoint on large distances}},  {\em JHEP} {\bf 09} (2021) 197, [\href{http://arxiv.org/abs/2104.05726}{{\tt arXiv:2104.05726}}].

\bibitem{Lee:2021qkx}
S.-J. Lee and T.~Weigand, {\it {Elliptic K3 surfaces at infinite complex structure and their refined Kulikov models}},  {\em JHEP} {\bf 09} (2022) 143, [\href{http://arxiv.org/abs/2112.07682}{{\tt arXiv:2112.07682}}].

\bibitem{Lee:2021usk}
S.-J. Lee, W.~Lerche, and T.~Weigand, {\it {Physics of infinite complex structure limits in eight dimensions}},  {\em JHEP} {\bf 06} (2022) 042, [\href{http://arxiv.org/abs/2112.08385}{{\tt arXiv:2112.08385}}].

\bibitem{Rudelius:2022gbz}
T.~Rudelius, {\it {Asymptotic scalar field cosmology in string theory}},  {\em JHEP} {\bf 10} (2022) 018, [\href{http://arxiv.org/abs/2208.08989}{{\tt arXiv:2208.08989}}].

\bibitem{Baume:2023msm}
F.~Baume and J.~Calder\'on-Infante, {\it {On higher-spin points and infinite distances in conformal manifolds}},  {\em JHEP} {\bf 12} (2023) 163, [\href{http://arxiv.org/abs/2305.05693}{{\tt arXiv:2305.05693}}].

\bibitem{Rudelius:2023mjy}
T.~Rudelius, {\it {Revisiting the refined Distance Conjecture}},  {\em JHEP} {\bf 09} (2023) 130, [\href{http://arxiv.org/abs/2303.12103}{{\tt arXiv:2303.12103}}].

\bibitem{Alvarez-Garcia:2023gdd}
R.~\'Alvarez-Garc\'\i{}a, S.-J. Lee, and T.~Weigand, {\it {Non-minimal elliptic threefolds at infinite distance. Part I. Log Calabi-Yau resolutions}},  {\em JHEP} {\bf 08} (2024) 240, [\href{http://arxiv.org/abs/2310.07761}{{\tt arXiv:2310.07761}}].

\bibitem{Alvarez-Garcia:2023qqj}
R.~\'Alvarez-Garc\'\i{}a, S.-J. Lee, and T.~Weigand, {\it {Non-minimal elliptic threefolds at infinite distance II: asymptotic physics}},  {\em JHEP} {\bf 01} (2025) 058, [\href{http://arxiv.org/abs/2312.11611}{{\tt arXiv:2312.11611}}].

\bibitem{Castellano:2023jjt}
A.~Castellano, I.~Ruiz, and I.~Valenzuela, {\it {Stringy evidence for a universal pattern at infinite distance}},  {\em JHEP} {\bf 06} (2024) 037, [\href{http://arxiv.org/abs/2311.01536}{{\tt arXiv:2311.01536}}].

\bibitem{Castellano:2023stg}
A.~Castellano, I.~Ruiz, and I.~Valenzuela, {\it {Universal Pattern in Quantum Gravity at Infinite Distance}},  {\em Phys. Rev. Lett.} {\bf 132} (2024), no.~18 181601, [\href{http://arxiv.org/abs/2311.01501}{{\tt arXiv:2311.01501}}].

\bibitem{Ooguri:2024ofs}
H.~Ooguri and Y.~Wang, {\it {Universal Bounds on CFT Distance Conjecture}},  \href{http://arxiv.org/abs/2405.00674}{{\tt arXiv:2405.00674}}.

\bibitem{Calderon-Infante:2024oed}
J.~Calder\'on-Infante and I.~Valenzuela, {\it {Tensionless String Limits in 4d Conformal Manifolds}},  \href{http://arxiv.org/abs/2410.07309}{{\tt arXiv:2410.07309}}.

\bibitem{Ashmore:2015joa}
A.~Ashmore and D.~Waldram, {\it {Exceptional Calabi-Yau spaces: the geometry of $\mathcal{N}=2$ backgrounds with flux}},  {\em Fortsch. Phys.} {\bf 65} (2017), no.~1 1600109, [\href{http://arxiv.org/abs/1510.00022}{{\tt arXiv:1510.00022}}].

\bibitem{Aoufia:2024awo}
C.~Aoufia, I.~Basile, and G.~Leone, {\it {Species scale, worldsheet CFTs and emergent geometry}},  \href{http://arxiv.org/abs/2405.03683}{{\tt arXiv:2405.03683}}.

\bibitem{Friedrich:2025gvs}
B.~Friedrich, J.~Monnee, T.~Weigand, and M.~Wiesner, {\it {Emergent Strings in Type IIB Calabi--Yau Compactifications}},  \href{http://arxiv.org/abs/2504.01066}{{\tt arXiv:2504.01066}}.

\bibitem{Baume:2016psm}
F.~Baume and E.~Palti, {\it {Backreacted Axion Field Ranges in String Theory}},  {\em JHEP} {\bf 08} (2016) 043, [\href{http://arxiv.org/abs/1602.06517}{{\tt arXiv:1602.06517}}].

\bibitem{Klaewer:2016kiy}
D.~Klaewer and E.~Palti, {\it {Super-Planckian Spatial Field Variations and Quantum Gravity}},  {\em JHEP} {\bf 01} (2017) 088, [\href{http://arxiv.org/abs/1610.00010}{{\tt arXiv:1610.00010}}].

\bibitem{Gendler:2020dfp}
N.~Gendler and I.~Valenzuela, {\it {Merging the weak gravity and distance conjectures using BPS extremal black holes}},  {\em JHEP} {\bf 01} (2021) 176, [\href{http://arxiv.org/abs/2004.10768}{{\tt arXiv:2004.10768}}].

\bibitem{Lanza:2020qmt}
S.~Lanza, F.~Marchesano, L.~Martucci, and I.~Valenzuela, {\it {Swampland Conjectures for Strings and Membranes}},  {\em JHEP} {\bf 02} (2021) 006, [\href{http://arxiv.org/abs/2006.15154}{{\tt arXiv:2006.15154}}].

\bibitem{Cribiori:2021gbf}
N.~Cribiori, D.~Lust, and M.~Scalisi, {\it {The gravitino and the swampland}},  {\em JHEP} {\bf 06} (2021) 071, [\href{http://arxiv.org/abs/2104.08288}{{\tt arXiv:2104.08288}}].

\bibitem{Buratti:2021fiv}
G.~Buratti, J.~Calder{\'o}n-Infante, M.~Delgado, and A.~M. Uranga, {\it {Dynamical Cobordism and Swampland Distance Conjectures}},  {\em JHEP} {\bf 10} (2021) 037, [\href{http://arxiv.org/abs/2107.09098}{{\tt arXiv:2107.09098}}].

\bibitem{Basile:2023rvm}
I.~Basile and C.~Montella, {\it {Domain walls and distances in discrete landscapes}},  {\em JHEP} {\bf 02} (2024) 227, [\href{http://arxiv.org/abs/2309.04519}{{\tt arXiv:2309.04519}}].

\bibitem{Castellano:2021yye}
A.~Castellano, A.~Font, A.~Herraez, and L.~E. Ib{\'a}{\~n}ez, {\it {A gravitino distance conjecture}},  {\em JHEP} {\bf 08} (2021) 092, [\href{http://arxiv.org/abs/2104.10181}{{\tt arXiv:2104.10181}}].

\bibitem{Castellano:2022bvr}
A.~Castellano, A.~Herr{\'a}ez, and L.~E. Ib{\'a}{\~n}ez, {\it {The emergence proposal in quantum gravity and the species scale}},  {\em JHEP} {\bf 06} (2023) 047, [\href{http://arxiv.org/abs/2212.03908}{{\tt arXiv:2212.03908}}].

\bibitem{Castellano:2021mmx}
A.~Castellano, A.~Herr{\'a}ez, and L.~E. Ib{\'a}{\~n}ez, {\it {IR/UV mixing, towers of species and swampland conjectures}},  {\em JHEP} {\bf 08} (2022) 217, [\href{http://arxiv.org/abs/2112.10796}{{\tt arXiv:2112.10796}}].

\bibitem{Scalisi:2018eaz}
M.~Scalisi and I.~Valenzuela, {\it {Swampland distance conjecture, inflation and $\alpha$-attractors}},  {\em JHEP} {\bf 08} (2019) 160, [\href{http://arxiv.org/abs/1812.07558}{{\tt arXiv:1812.07558}}].

\bibitem{Bedroya:2019snp}
A.~Bedroya and C.~Vafa, {\it {Trans-Planckian Censorship and the Swampland}},  {\em JHEP} {\bf 09} (2020) 123, [\href{http://arxiv.org/abs/1909.11063}{{\tt arXiv:1909.11063}}].

\bibitem{Andriot:2020lea}
D.~Andriot, N.~Cribiori, and D.~Erkinger, {\it {The web of swampland conjectures and the TCC bound}},  {\em JHEP} {\bf 07} (2020) 162, [\href{http://arxiv.org/abs/2004.00030}{{\tt arXiv:2004.00030}}].

\bibitem{Etheredge:2022opl}
M.~Etheredge, B.~Heidenreich, S.~Kaya, Y.~Qiu, and T.~Rudelius, {\it {Sharpening the Distance Conjecture in diverse dimensions}},  {\em JHEP} {\bf 12} (2022) 114, [\href{http://arxiv.org/abs/2206.04063}{{\tt arXiv:2206.04063}}].

\bibitem{Etheredge:2023odp}
M.~Etheredge, B.~Heidenreich, J.~McNamara, T.~Rudelius, I.~Ruiz, and I.~Valenzuela, {\it {Running decompactification, sliding towers, and the distance conjecture}},  {\em JHEP} {\bf 12} (2023) 182, [\href{http://arxiv.org/abs/2306.16440}{{\tt arXiv:2306.16440}}].

\bibitem{Etheredge:2023zjk}
M.~Etheredge and B.~Heidenreich, {\it {Geodesic Gradient Flows in Moduli Space}},  \href{http://arxiv.org/abs/2311.18693}{{\tt arXiv:2311.18693}}.

\bibitem{Etheredge:2024tok}
M.~Etheredge, B.~Heidenreich, T.~Rudelius, I.~Ruiz, and I.~Valenzuela, {\it {Taxonomy of Infinite Distance Limits}},  \href{http://arxiv.org/abs/2405.20332}{{\tt arXiv:2405.20332}}.

\bibitem{Etheredge:2023usk}
M.~Etheredge, {\it {Dense geodesics, tower alignment, and the Sharpened Distance Conjecture}},  {\em JHEP} {\bf 01} (2024) 122, [\href{http://arxiv.org/abs/2308.01331}{{\tt arXiv:2308.01331}}].

\bibitem{Grieco:2025bjy}
A.~Grieco, I.~Ruiz, and I.~Valenzuela, {\it {EFT strings and dualities in 4d $\mathcal{N}=1$}},  \href{http://arxiv.org/abs/2504.16984}{{\tt arXiv:2504.16984}}.

\bibitem{Etheredge:2024amg}
M.~Etheredge, B.~Heidenreich, and T.~Rudelius, {\it {A Distance Conjecture for Branes}},  \href{http://arxiv.org/abs/2407.20316}{{\tt arXiv:2407.20316}}.

\bibitem{Etheredge:2025ahf}
M.~Etheredge, {\it {Taxonomy of branes in infinite distance limits}},  \href{http://arxiv.org/abs/2505.10615}{{\tt arXiv:2505.10615}}.

\bibitem{Cecotti:2021cvv}
S.~Cecotti, {\it {Swampland geometry and the gauge couplings}},  {\em JHEP} {\bf 09} (2021) 136, [\href{http://arxiv.org/abs/2102.03205}{{\tt arXiv:2102.03205}}].

\bibitem{Farquet:2012cs}
D.~Farquet and C.~A. Scrucca, {\it {Scalar geometry and masses in Calabi-Yau string models}},  {\em JHEP} {\bf 09} (2012) 025, [\href{http://arxiv.org/abs/1205.5728}{{\tt arXiv:1205.5728}}].

\bibitem{Cecotti:2015wqa}
S.~Cecotti, {\em {Supersymmetric Field Theories}: {Geometric Structures and Dualities}}.
\newblock Cambridge University Press, 1, 2015.

\bibitem{Cecotti:2020rjq}
S.~Cecotti, {\it {Special Geometry and the Swampland}},  {\em JHEP} {\bf 09} (2020) 147, [\href{http://arxiv.org/abs/2004.06929}{{\tt arXiv:2004.06929}}].

\bibitem{Mautner-ergodic}
F.~I. Mautner, {\it Geodesic flows on symmetric riemann spaces},  {\em Annals of Mathematics} {\bf 65} (1957), no.~3 416--431.

\bibitem{Moore-ergodic}
C.~C. Moore, {\it Ergodicity of flows on homogeneous spaces},  {\em American Journal of Mathematics} {\bf 88} (1966), no.~1 154--178.

\bibitem{sullivan}
D.~Sullivan, {\it Disjoint spheres, approximation by imaginary quadratic numbers, and the logarithm law for geodesics},  {\em Acta Mathematica} {\bf 149} (1982), no.~1 215--237.

\bibitem{Kleinbock-Margulis}
D.~Y. Kleinbock and G.~A. Margulis, {\it Logarithm laws for flows on homogeneous spaces},  {\em Inventiones mathematicae} {\bf 138} (1999), no.~3 451--494.

\bibitem{Keurentjes:2006cw}
A.~Keurentjes, {\it {Determining the dual}},  \href{http://arxiv.org/abs/hep-th/0607069}{{\tt hep-th/0607069}}.

\bibitem{Link:2008}
G.~Link, ``An introduction to globally symmetric spaces.'' \url{https://metaphor.ethz.ch/x/2023/fs/401-3226-00L/literature/LinkIntroSymSpaces.pdf}, 2008.

\bibitem{Erickson:2008}
J.~Erickson, ``Parabolic geometries for people that like pictures.'' \url{https://www.math.umd.edu/~jwericks/Parabolic%20Geometries%20RIT/Lecture%2010.pdf}.

\bibitem{Borel}
A.~Borel and L.~Ji, {\em Compactifications of Symmetric and Locally Symmetric Spaces}.
\newblock Birkhäuser Boston, MA, 2006.

\bibitem{Ballman}
W.~Ballmann, {\it Nonpositively curved manifolds of higher rank},  {\em Annals of Mathematics} {\bf 122} (1985), no.~3 597--609.

\bibitem{BurnsSpatzier}
K.~Burns and R.~Spatzier, {\it Manifolds of nonpositive curvature and their buildings},  {\em Publications Math{\'e}matiques de l'Institut des Hautes {\'E}tudes Scientifiques} {\bf 65} (1987), no.~1 35--59.

\bibitem{Henneaux:2007ej}
M.~Henneaux, D.~Persson, and P.~Spindel, {\it {Spacelike Singularities and Hidden Symmetries of Gravity}},  {\em Living Rev. Rel.} {\bf 11} (2008) 1, [\href{http://arxiv.org/abs/0710.1818}{{\tt arXiv:0710.1818}}].

\bibitem{Margulis}
G.~A. Margulis, {\it Arithmetic properties of discrete subgroups},  {\em Russ. Math. Surv.} {\bf 29} (1974), no.~1 107--156.

\bibitem{Corlette}
K.~Corlette, {\it Archimedean superrigidity and hyperbolic geometry},  {\em Annals of Mathematics} {\bf 135} (1992), no.~1 165--182.

\bibitem{Israel:2025ouq}
D.~Israel, I.~V. Melnikov, and Y.~Proto, {\it {Shift orbifolds, decompactification limits, and lattices}},  \href{http://arxiv.org/abs/2502.18453}{{\tt arXiv:2502.18453}}.

\bibitem{steinberg}
R.~Steinberg, {\em Lectures on Chevalley Groups}.
\newblock No.~pts. 1-4 in Lectures on Chevalley Groups. Yale University, Department of Mathematics, 1967.

\bibitem{Obers:1998fb}
N.~A. Obers and B.~Pioline, {\it {U duality and M theory}},  {\em Phys. Rept.} {\bf 318} (1999) 113--225, [\href{http://arxiv.org/abs/hep-th/9809039}{{\tt hep-th/9809039}}].

\bibitem{hardy}
G.~H. Hardy and E.~M. Wright, {\em An Introduction to the Theory of Numbers}.
\newblock Oxford, fourth~ed., 1975.

\bibitem{Giveon:1994fu}
A.~Giveon, M.~Porrati, and E.~Rabinovici, {\it {Target space duality in string theory}},  {\em Phys. Rept.} {\bf 244} (1994) 77--202, [\href{http://arxiv.org/abs/hep-th/9401139}{{\tt hep-th/9401139}}].

\bibitem{Coimbra:2011ky}
A.~Coimbra, C.~Strickland-Constable, and D.~Waldram, {\it {$E_{d(d)} \times \mathbb{R}^+$ generalised geometry, connections and M theory}},  {\em JHEP} {\bf 02} (2014) 054, [\href{http://arxiv.org/abs/1112.3989}{{\tt arXiv:1112.3989}}].

\bibitem{deBoer:2001wca}
J.~de~Boer, R.~Dijkgraaf, K.~Hori, A.~Keurentjes, J.~Morgan, D.~R. Morrison, and S.~Sethi, {\it {Triples, fluxes, and strings}},  {\em Adv. Theor. Math. Phys.} {\bf 4} (2002) 995--1186, [\href{http://arxiv.org/abs/hep-th/0103170}{{\tt hep-th/0103170}}].

\bibitem{Fraiman:2021hma}
B.~Fraiman and H.~P. de~Freitas, {\it {Freezing of gauge symmetries in the heterotic string on T$^{4}$}},  {\em JHEP} {\bf 04} (2022) 007, [\href{http://arxiv.org/abs/2111.09966}{{\tt arXiv:2111.09966}}].

\bibitem{Nakajima:2023zsh}
S.~Nakajima, {\it {New non-supersymmetric heterotic string theory with reduced rank and exponential suppression of the cosmological constant}},  \href{http://arxiv.org/abs/2303.04489}{{\tt arXiv:2303.04489}}.

\bibitem{DeFreitas:2024ztt}
H.~P. De~Freitas, {\it {Non-supersymmetric heterotic strings and chiral CFTs}},  {\em JHEP} {\bf 11} (2024) 002, [\href{http://arxiv.org/abs/2402.15562}{{\tt arXiv:2402.15562}}].

\bibitem{Fraiman:2023cpa}
B.~Fraiman, M.~Gra\~na, H.~Parra De~Freitas, and S.~Sethi, {\it {Non-Supersymmetric Heterotic Strings on a Circle}},  \href{http://arxiv.org/abs/2307.13745}{{\tt arXiv:2307.13745}}.

\bibitem{DeFreitas:2024yzr}
H.~Parra~de Freitas, {\it {T-duality for non-critical heterotic strings}},  {\em JHEP} {\bf 01} (2025) 173, [\href{http://arxiv.org/abs/2407.12923}{{\tt arXiv:2407.12923}}].

\bibitem{Fraiman:2025wip}
B.~Fraiman and H.~Parra~de Freitas, ``{Aspects of non-supersymmetric CHL strings (To appear)}.''

\bibitem{Riccioni_2007}
F.~Riccioni and P.~C. West, {\it Thee11origin of all maximal supergravities},  {\em Journal of High Energy Physics} {\bf 2007} (July, 2007) 063–063.

\bibitem{Bergshoeff:2007qi}
E.~A. Bergshoeff, I.~De~Baetselier, and T.~A. Nutma, {\it {E(11) and the embedding tensor}},  {\em JHEP} {\bf 09} (2007) 047, [\href{http://arxiv.org/abs/0705.1304}{{\tt arXiv:0705.1304}}].

\bibitem{Riccioni_2009}
F.~Riccioni, D.~Steele, and P.~West, {\it The e11 origin of all maximal supergravities. the hierarchy of field-strengths},  {\em Journal of High Energy Physics} {\bf 2009} (Sept., 2009) 095–095.

\bibitem{Marrani:2019jvd}
A.~Marrani and L.~Romano, {\it {Orbits in nonsupersymmetric magic theories}},  {\em Int. J. Mod. Phys. A} {\bf 34} (2019), no.~32 1950190, [\href{http://arxiv.org/abs/1906.05830}{{\tt arXiv:1906.05830}}].

\bibitem{Ginsparg:1986bx}
P.~H. Ginsparg, {\it {Comment on Toroidal Compactification of Heterotic Superstrings}},  {\em Phys. Rev. D} {\bf 35} (1987) 648.

\bibitem{Polchinski_1996}
J.~Polchinski and E.~Witten, {\it Evidence for heterotic — type i string duality},  {\em Nuclear Physics B} {\bf 460} (Feb., 1996) 525–540.

\bibitem{Kneiper1997}
G.~Knieper, {\it On the asymptotic geometry of nonpositively curved manifolds},  {\em Geometric and Functional Analysis} {\bf 7} (1997), no.~4 755--782.

\bibitem{Fraczyk23}
M.~Fraczyk and T.~Gelander, {\it Infinite volume and infinite injectivity radius},  {\em Annals of Mathematics} {\bf 197} (2023), no.~1 389--421.

\bibitem{Margulis-arithmetic}
G.~A. Margulis, {\it Arithmeticity of nonuniform lattices in weakly noncompact groups},  {\em Functional Analysis and Its Applications} {\bf 9} (1975), no.~1 31--38.

\bibitem{Margulis-rigidity}
G.~A. Margulis, {\em Discrete Subgroups of Semisimple Lie Groups}, vol.~17 of {\em Ergebnisse der Mathematik und ihrer Grenzgebiete (3) [Results in Mathematics and Related Areas (3)]}.
\newblock Springer-Verlag, Berlin, 1991.

\bibitem{Calderon-Infante:2020dhm}
J.~Calder\'on-Infante, A.~M. Uranga, and I.~Valenzuela, {\it {The Convex Hull Swampland Distance Conjecture and Bounds on Non-geodesics}},  {\em JHEP} {\bf 03} (2021) 299, [\href{http://arxiv.org/abs/2012.00034}{{\tt arXiv:2012.00034}}].

\bibitem{Arkani-Hamed:2006emk}
N.~Arkani-Hamed, L.~Motl, A.~Nicolis, and C.~Vafa, {\it {The String landscape, black holes and gravity as the weakest force}},  {\em JHEP} {\bf 06} (2007) 060, [\href{http://arxiv.org/abs/hep-th/0601001}{{\tt hep-th/0601001}}].

\bibitem{Palti:2017elp}
E.~Palti, {\it {The Weak Gravity Conjecture and Scalar Fields}},  {\em JHEP} {\bf 08} (2017) 034, [\href{http://arxiv.org/abs/1705.04328}{{\tt arXiv:1705.04328}}].

\bibitem{trenner2011asymptotic}
T.~Trenner and P.~Wilson, {\it Asymptotic curvature of moduli spaces for calabi--yau threefolds},  {\em Journal of Geometric Analysis} {\bf 21} (2011), no.~2 409--428.

\bibitem{Raman:2024fcv}
S.~Raman and C.~Vafa, {\it {Swampland and the Geometry of Marked Moduli Spaces}},  \href{http://arxiv.org/abs/2405.11611}{{\tt arXiv:2405.11611}}.

\bibitem{Marchesano:2023thx}
F.~Marchesano, L.~Melotti, and L.~Paoloni, {\it {On the moduli space curvature at infinity}},  {\em JHEP} {\bf 02} (2024) 103, [\href{http://arxiv.org/abs/2311.07979}{{\tt arXiv:2311.07979}}].

\bibitem{Castellano:2024gwi}
A.~Castellano, F.~Marchesano, L.~Melotti, and L.~Paoloni, {\it {The Moduli Space Curvature and the Weak Gravity Conjecture}},  \href{http://arxiv.org/abs/2410.10966}{{\tt arXiv:2410.10966}}.

\bibitem{Samtleben:2008pe}
H.~Samtleben, {\it {Lectures on Gauged Supergravity and Flux Compactifications}},  {\em Class. Quant. Grav.} {\bf 25} (2008) 214002, [\href{http://arxiv.org/abs/0808.4076}{{\tt arXiv:0808.4076}}].

\bibitem{Alekseevsky:2005}
D.~Alekseevsky and V.~Cort{\'e}s, {\it Classification of pseudo-{R}iemannian symmetric spaces of quaternionic {K}\"ahler type},  {\em Amer. Math. Soc. Transl. Ser. 2} {\bf 213} (2005), no.~2.

\bibitem{Cremmer:1984hc}
E.~Cremmer and A.~Van~Proeyen, {\it {Classification of Kahler Manifolds in $N=2$ Vector Multiplet Supergravity Couplings}},  {\em Class. Quant. Grav.} {\bf 2} (1985) 445.

\end{thebibliography}\endgroup

\end{document}